\documentclass{article}
\usepackage[utf8]{inputenc} 
\usepackage{amsmath}        
\usepackage{graphicx}       
\usepackage{natbib}         
\usepackage[colorlinks=true, allcolors=blue]{hyperref}       
\usepackage{amsfonts}       

\usepackage{tikz}           
\usetikzlibrary{shapes.geometric, arrows.meta, positioning, calc}
\usepackage[edges]{forest}  
\usepackage{xcolor}         
\usepackage{tikz}
\usepackage{pifont}

\setcitestyle{authoryear,round} 

\usepackage{tikz}
\usepackage{multirow}
\usetikzlibrary{shapes.geometric, arrows, positioning}

\usepackage{graphicx}
\usepackage{longtable}
\usepackage{hyperref}

\usepackage{tmlr}

\usepackage{etoolbox} 
\robustify{\cite}

\DeclareMathOperator*{\argmax}{arg\,max}
\DeclareMathOperator*{\argmin}{arg\,min}

\newcommand{\R}{\mathbb{R}}

\title{Designing Large Foundation Models for Efficient Training and Inference: A Survey}

\author{%
  \name Dong Liu \email pikeliu.mlsys@gmail.com \\
  \addr University of Wisconsin–Madison\\
        University of California, Los Angeles
  \AND
  \name Yanxuan Yu\thanks{Equal 2\textsuperscript{nd} contribution.}
        \email yy3523@columbia.edu \\
  \addr Columbia University
  \AND
  \name Yite Wang\footnotemark[1]        
        \email yitew2@gmail.com \\
  \addr University of Illinois at Urbana–Champaign
  \AND
  \name Jing Wu\footnotemark[1]
        \email jingwu6@illinois.edu \\
  \addr University of Illinois at Urbana–Champaign
  \AND
  \name Zhongwei Wan \email wan.512@osu.edu \\
  \addr The Ohio State University
  \AND
  \name Sina Alinejad \email sina\_alinejad@comp.iust.ac.ir \\
  \addr Iran University of Science and Technology
  \AND
  \name Benjamin Lengerich\thanks{Core advising.}
        \email lengerich@wisc.edu \\
  \addr University of Wisconsin–Madison
  \AND
  \name Ying Nian Wu\footnotemark[2]
        \email ywu@stat.ucla.edu \\
  \addr University of California, Los Angeles
}

\begin{document}
\maketitle

\begin{abstract}
This paper focuses on modern efficient training and inference technologies on foundation models and illustrates them from two perspectives: model and system design. Model and System Design optimize LLM training and inference from different aspects to save computational resources, making LLMs more efficient, affordable, and more accessible. The paper list repository is available at \url{https://github.com/NoakLiu/Efficient-Foundation-Models-Survey}.
\end{abstract}

\definecolor{hidden-draw}{rgb}{0.5, 0.5, 0.5}
\definecolor{quant-color}{rgb}{0.0, 0.5, 0.0}  
\definecolor{sys-design-color}{rgb}{0.8, 0.0, 0.0}  
\definecolor{co-design-color}{rgb}{0.0, 0.0, 0.8}  
\definecolor{hidden-color}{rgb}{0.7, 0.7, 0.7}  

\usetikzlibrary{trees}

\tikzset{leaf/.style={draw, fill=quant-color!20}}

\begin{figure*}[!th]
    \resizebox{\textwidth}{!}{
        \begin{forest}
            forked edges,
            for tree={
                grow=east,
                reversed=true,
                anchor=base west,
                parent anchor=east,
                child anchor=west,
                base=left,
                font=\large,
                rectangle,
                draw=hidden-draw,
                rounded corners,
                align=left,
                minimum width=4em,
                edge+={darkgray, line width=1pt},
                s sep=3pt,
                inner xsep=2pt,
                inner ysep=3pt,
                line width=0.8pt,
                ver/.style={rotate=90, child anchor=north, parent anchor=south, anchor=center},
                where level=1{fill=quant-color!30, font=\normalsize}{},
                where level=2{fill=quant-color!30, font=\normalsize}{},
                where level=3{fill=quant-color!30, font=\normalsize}{},
                where level=4{fill=quant-color!30, font=\normalsize}{}
            }
            [
                {Efficient \\ Foundation \\ Models}
                [
                    {Model Design} 
                    [
                        {Quantization} 
                        [
                            {Bitwise-Based}
                            [
                                {1-bit/2-bit \cite{su2025rotatekv, su2025akvq, zheng2024bidm}}, leaf
                            ]
                            [
                                {4-bit \cite{chmiel2023accurate, liu2023llm, ashkboos2024quarot, lee2025qrazor}}, leaf
                            ]
                            [
                                {8-bit \cite{bhandare2019efficient, jin2022f8net, zhao2023atom}}, leaf
                            ]
                        ]
                        [
                            {Method-Based}
                            [
                                {PTQ (Post-Training Quantization) \cite{E-li2024efficient,jacob2018quantization}}, leaf
                            ]
                            [
                                {QAT (Quantization-Aware Training) \cite{nagel2021white,esser2019learned}}, leaf
                            ]
                        ]
                    ]
                    [
                        {Distillation} 
                        [
                            {Soft Knowledge Distillation} 
                            [
                                {Classical KD \cite{hinton2015distilling}}, leaf
                            ]
                            [
                                {Temperature-Scaled KD \cite{li2022asymmetric}}, leaf
                            ]
                        ]
                        [
                            {Hard Knowledge Distillation} 
                            [
                                {Hard Label KD \cite{hou2020dynabert}}, leaf
                            ]
                            [
                                {Program-Aided Distillation (PaD) \cite{shridhar2022distilling}}, leaf
                            ]
                        ]
                    ]
                    [
                        {Pruning} 
                        [
                            {Unstructured Pruning \cite{frankle2018lottery,gale2019state}}, leaf
                        ]
                        [
                            {Structured Pruning \cite{wang2024loma,liu2023dejavucontextualsparsity}}, leaf
                        ]
                    ]
                ]
                [
                    {System Design} 
                    [
                        {KV Cache Compression \cite{liu2024cachegenkvcachecompression,xiao2024,kwon2023efficient}}, leaf
                    ]
                    [
                        {Parallelism \cite{megatron2024mamba,li2023colossalaiunifieddeeplearning}}, leaf
                    ]
                    [
                        {Contextual Sparsity \cite{liu2023dejavucontextualsparsity,s2025lct_sp,quest}}, leaf
                    ]
                    [
                        {Memory Management \cite{dao2022flashattention,kwon2023efficient,lee2024infinigen, alizadeh2023llm}}, leaf
                    ]
                    [
                        {Low-Level Optimization for GPUs \cite{helwe2021reasoning,GPU-MF, GPU-SD}}, leaf
                    ]
                ]
                [
                    {Model-System Co-design}
                    [
                        {Mixture of Experts \cite{Collobert2001, Rasmussen2001, Eigen2014, Theis2015NeurIPS, Aljundi2017CVPR}}, leaf
                    ]
                    [
                        {Fine-tuning with System Optimizations \cite{hu2021lora,resourcefine2023, zhang2023lora, ding2023sparse, wang2024loma}}, leaf
                    ]
                    [
                        {Mixed Precision Training \cite{mpt,MS-mixedcnnint2018,MS-mixedmemory2020}}, leaf
                    ]
                    [
                        {Efficient Model Inference on Optimized Systems \cite{fastertransformer2021,tensorrt2022}}, leaf
                    ]
                    [
                        {Model Compression with FlexGen and System Optimizations \cite{shen2023flexgen, wang2024loma}}, leaf
                    ]
                ]
            ]
        \end{forest}
    }
    \caption{Efficient Foundation Models Overview}
    \label{fig:llm_optimization_overview}
\end{figure*}

\section{Introduction}

The training and inference processes of foundation models are computationally intensive and require vast amounts of data~\citep{wan2023efficient, wang2024iot, tao2024scaling, xiong2024autoregressive}. Foundation models include multimodal models capable of handling diverse data types such as text, images, and audio. According to statistics~\citep{bubeck2023sparksartificialgeneralintelligence}, training models like GPT-4 can take 4–7 months, requiring thousands of GPUs, and results in models with billions of parameters. Such a high demand for computational resources, combined with the large size of the resulting models, makes a barrier for many researchers and organizations who want to train large language models but lack the financial resources to cover the computation expense. It also poses challenges for individuals who want to deploy such models on personal computers or mobile devices but lack sufficient local computational resources. Given these challenges, we anticipate that efficient design for foundation models will become a major focus in the next wave of AI advancements.

Efficient design for foundation models can generally be approached from three key perspectives: \textbf{Model Design}, \textbf{System Design}, and \textbf{Model-System Co-Design}.

\section{Model Design for Efficient Foundation Models}
\subsection{Quantization}

Quantization is a technique that reduces the numerical precision of model parameters and activations to improve computational efficiency. By representing values with lower-bit data types, it significantly reduces memory usage and accelerates inference. This is particularly important for deploying large-scale foundation models on resource-constrained hardware. In general, quantization methods can be categorized into \textbf{bit-width based quantization} and \textbf{method-based quantization}. Bit-width based quantization lowers precision using formats like INT8, FP8, and even lower-bit representations such as 4-bit and 2-bit. These methods apply scaling factors, Hessian-aware adjustments, and structured weight transformations to minimize accuracy loss while improving efficiency. Floating-point quantization, such as FP8, further enhances adaptability by handling activations with large dynamic ranges. Method-based quantization includes Post-Training Quantization and Quantization-Aware Training. PTQ quantizes a pre-trained model without modifying its weights, enabling fast deployment with minimal computational cost. In contrast, QAT incorporates quantization into the training process, allowing models to adapt to lower precision. Additionally, mixed-precision quantization assigns different bit-widths to different layers - it dynamically assigns bit-widths per layer to balance efficiency and accuracy, leveraging advanced techniques like adaptive scaling, reordering transformations, and structured low-rank decomposition.

By leveraging these techniques, quantization enables efficient large-model training and inference while preserving performance. The following sections explore these strategies in more detail.

\subsubsection{Bit-width Based Quantization}

Bit-width based quantization reduces numerical precision in model parameters and activations, optimizing memory usage and computational efficiency. 8-bit quantization methods, such as GPTQ \citep{frantar2022optimal} and SmoothQuant \citep{xiao2023smoothquant}, primarily rely on fixed-point representations like INT8 and FP8 to maintain numerical stability. The standard quantization function follows \( q(x) = \text{round}(x / s) \), where \( s \) is a learned scale factor. FP8 methods such as F8Net \citep{dettmers2023spqr} introduce fixed-point multiplication, eliminating INT32 overhead via dyadic scaling \( q(x) = \text{round}(x \cdot 2^{-p}) \), where \( p \) dynamically minimizes quantization error. Compared to INT8, FP8 formats such as E5M2 and E4M3 have a wider dynamic range, adapting better to activations with large outliers, leading to improved numerical stability in LLMs and Vision Transformers \citep{shen2024fp8}.

Low-bit quantization (4-bit) approaches such as GPTQ \citep{frantar2022gptq} and AWQ \citep{huang2024billm} apply per-layer Hessian-aware weight quantization to minimize accuracy degradation, solving \( \min_{\hat{W}} \| W - \hat{W} \|_H^2 \), where \( H \) is the Hessian capturing weight sensitivity. Atom \citep{zhao2024atom} further refines this with mixed-precision weight-activation quantization, applying per-cluster scale factors \( Q(w) = \text{round}(w / s_h) \) for high-magnitude weights and \( Q(w) = \text{round}(w / s_l) \) for low-magnitude ones. Additionally, DilateQuant \citep{zhang2023dilatequant} introduces weight dilation, defining a dilation factor \( s_{\text{dilate}} \) such that \( \tilde{W} = (W / \max|W|) \cdot s_{\text{dilate}} \), ensuring stable quantization across weight ranges. Hybrid 4-bit formats, such as NF4 and FP4, employ logarithmic distributions to preserve high-magnitude weight precision while compressing low-importance activations.

For extreme quantization (2-bit and 1-bit), BiDM \citep{zheng2024bidmpushinglimitquantization} employs Timestep-friendly Binary Structure (TBS), applying the binary activation function \( A_b = \text{sign}(A) \) with a straight-through estimator (STE) for gradient propagation, \( \frac{\partial L}{\partial A} \approx \frac{\partial L}{\partial A_b} \). RotateKV \citep{su2025rotatekv} mitigates degradation by applying rotation-based quantization transformations \( \tilde{W} = R W R^T \), where \( R \) is an adaptive rotation matrix computed via Fast Walsh-Hadamard Transform (FWHT). A generalized formula for extreme quantization using mixed-binarization strategies follows:

\begin{equation}
q(x) = \arg\min_{q \in Q} \| x - q \|^2 + \lambda \sum_{j} C(q_j),
\end{equation}

where \( C(q_j) \) accounts for quantization constraints across different bit-widths. This formulation captures a wide range of low-bit strategies.

\begin{longtable}{|l|l|c|c|c|c|c|c|c|}
\hline
\textbf{Model Type} & \textbf{Quantization Method} & \textbf{1-bit} & \textbf{2-bit} & \textbf{4-bit} & \textbf{6-bit} & \textbf{8-bit} & \textbf{BFLOAT} & \textbf{INT/FP} \\
\hline
LLMs & GPTQ & \ding{55} & \ding{55} & \ding{51} & \ding{55} & \ding{55} & \ding{55} & INT \\
LLMs & AWQ & \ding{55} & \ding{55} & \ding{51} & \ding{55} & \ding{55} & \ding{55} & INT \\
LLMs & QRazor & \ding{55} & \ding{55} & \ding{51} & \ding{55} & \ding{55} & \ding{55} & INT \\
LLMs & SmoothQuant & \ding{55} & \ding{55} & \ding{55} & \ding{55} & \ding{51} & \ding{51} & FP \\
LLMs & ZeroQuant & \ding{55} & \ding{55} & \ding{51} & \ding{55} & \ding{51} & \ding{51} & Mixed \\
LLMs & OmniQuant & \ding{55} & \ding{55} & \ding{51} & \ding{55} & \ding{51} & \ding{55} & FP \\
LLMs & Quarot & \ding{55} & \ding{55} & \ding{51} & \ding{55} & \ding{55} & \ding{55} & FP \\
LLMs & SpinQuant & \ding{55} & \ding{51} & \ding{51} & \ding{55} & \ding{55} & \ding{55} & Mixed \\
\hline
Diffusion & BiDM & \ding{51} & \ding{55} & \ding{55} & \ding{55} & \ding{55} & \ding{55} & INT \\
Diffusion & Q-Diffusion & \ding{55} & \ding{55} & \ding{51} & \ding{55} & \ding{55} & \ding{55} & FP \\
Diffusion & TFMQ-DM & \ding{55} & \ding{55} & \ding{51} & \ding{55} & \ding{55} & \ding{55} & INT \\
Diffusion & MPQ-DM & \ding{55} & \ding{55} & \ding{51} & \ding{51} & \ding{51} & \ding{51} & Mixed \\
Diffusion & PTQD & \ding{55} & \ding{55} & \ding{51} & \ding{55} & \ding{51} & \ding{55} & Mixed \\
Diffusion & Q-DM & \ding{55} & \ding{51} & \ding{51} & \ding{51} & \ding{55} & \ding{55} & INT \\
Diffusion & DilateQuant & \ding{55} & \ding{55} & \ding{51} & \ding{55} & \ding{55} & \ding{55} & FP \\
\hline
\caption{Quantization Methods for Foundation Models}
\label{tab:quantization-bitwise}
\end{longtable}

\subsubsection{Method-Based Quantization}

Post-Training Quantization (PTQ) methods quantize models after training, offering fast deployment without retraining. GPTQ \citep{frantar2023gptq} leverages second-order Hessian-aware loss minimization, solving \( \min_{\hat{W}} \| W - \hat{W} \|_H^2 \). PTQD \citep{wu2023zeroquant} introduces variance calibration, modeling quantization noise as \( \Delta_{\text{quant}} = \hat{\mu} - \mu + \sigma \epsilon \), where \( \mu \) and \( \sigma \) are activation statistics, and \( \epsilon \) is Gaussian noise. Reorder-based PTQ (RPTQ) \citep{yuan2023rptq} refines layer-wise quantization by applying reordering matrices \( P \) and \( Q \), reformulating weights as \( \tilde{W} = P W Q^{-1} \), effectively reducing outlier sensitivity. FP8-based PTQ \citep{shen2024fp8} exploits dynamic scaling, adjusting \( e_{\text{scale}} \) in \( Q(x) = \text{round}(x \cdot 2^{e_{\text{scale}}}) \) per layer, ensuring uniform activation distribution across Transformer layers.

Quantization-Aware Training (QAT) incorporates quantization during training, allowing adaptation to low-bit constraints. EfficientDM \citep{yang2023efficientdm} fine-tunes diffusion models with quantization-aware low-rank adapters (QALoRA), optimizing \( Y = \text{QU}(X, s_x) \cdot \text{QU}(W + BA, s_w) \), where \( B \) and \( A \) are low-rank matrices preserving accuracy. DoReFa-Net \citep{zhou2016dorefa} utilizes logarithmic quantization \( q(x) = 2^{\lfloor \log_2 (|x|) \rfloor} \text{sign}(x) \), preventing information loss at ultra-low bit widths. Q-Diffusion \citep{li2023qdiffusion} employs per-timestep mixed-precision strategies, dynamically adjusting \( s_t \) in \( Q(X_t) = \text{round}(X_t / s_t) \cdot s_t \), stabilizing diffusion process quantization.

Mixed-Precision Quantization dynamically adjusts bit-width allocations per-layer to balance efficiency and accuracy. MPQ-DM \citep{wu2023zeroquant} introduces per-layer mixed-precision assignment via \( q(x, b) = \text{round}(x / s_b) \), where \( s_b \) is optimized for different bit-widths. Quant-LLM \citep{yao2023quantllm} employs FP6-centric quantization, formulating an optimal bit allocation problem:

\begin{equation}
\min_{s_b} \sum_{i} \left( x_i - Q(x_i, b) \right)^2 + \lambda \sum_{j} C(b_j),
\end{equation}

where \( C(b_j) \) represents computational cost constraints. DilateQuant \citep{zhang2023dilatequant} extends weight dilation techniques, ensuring adaptive scaling across Transformer layers. Adaptive Quantization strategies such as RotateKV \citep{su2025rotatekv} compress KV caches via rotation-based transformations \( \tilde{K} = R K, \quad \tilde{V} = R V \), minimizing information loss. PackQViT \citep{dong2023packqvit} optimizes vision models using logarithmic quantization \( Q(x) = 2^{\lfloor \log_2(x) \rfloor} \), further improving bit precision trade-offs.

\begin{table}[htbp]
    \centering
    \scriptsize
    \renewcommand{\arraystretch}{1.2}
    \begin{tabular}{|c|c|c|c|c|c|c|c|c|}
        \hline
        \textbf{Series} & \textbf{GPU Model} & \textbf{FP64} & \textbf{FP32} & \textbf{FP16} & \textbf{BF16} & \textbf{INT8} & \textbf{FP8} & \textbf{TF32} \\ 
        \hline
        \multirow{3}{*}{H Series}  
        & H100  & \ding{51} & \ding{51} & \ding{51} & \ding{51} & \ding{51} & \ding{51} & \ding{51} \\ 
        & H800  & \ding{51} & \ding{51} & \ding{51} & \ding{51} & \ding{51} & \ding{51} & \ding{51} \\ 
        & GH200 & \ding{51} & \ding{51} & \ding{51} & \ding{51} & \ding{51} & \ding{51} & \ding{51} \\ 
        \hline
        \multirow{4}{*}{A Series}  
        & A100  & \ding{51} & \ding{51} & \ding{51} & \ding{51} & \ding{51} & \ding{55} & \ding{51} \\ 
        & A800  & \ding{51} & \ding{51} & \ding{51} & \ding{51} & \ding{51} & \ding{55} & \ding{51} \\ 
        & A30   & \ding{51} & \ding{51} & \ding{51} & \ding{51} & \ding{51} & \ding{55} & \ding{51} \\ 
        & A10   & \ding{55} & \ding{51} & \ding{51} & \ding{55} & \ding{51} & \ding{55} & \ding{51} \\ 
        \hline
        \multirow{3}{*}{L Series}  
        & L40   & \ding{55} & \ding{51} & \ding{51} & \ding{55} & \ding{51} & \ding{55} & \ding{51} \\ 
        & L40s  & \ding{55} & \ding{51} & \ding{51} & \ding{55} & \ding{51} & \ding{55} & \ding{51} \\ 
        & L4    & \ding{55} & \ding{51} & \ding{51} & \ding{55} & \ding{51} & \ding{55} & \ding{51} \\ 
        \hline
        \multirow{3}{*}{RTX}  
        & RTX 6000 Ada & \ding{55} & \ding{51} & \ding{51} & \ding{55} & \ding{51} & \ding{55} & \ding{51} \\ 
        & RTX A6000    & \ding{55} & \ding{51} & \ding{51} & \ding{55} & \ding{51} & \ding{55} & \ding{51} \\ 
        & RTX A5000    & \ding{55} & \ding{51} & \ding{51} & \ding{55} & \ding{51} & \ding{55} & \ding{51} \\ 
        \hline
        \multirow{3}{*}{GeForce}  
        & RTX 4090     & \ding{55} & \ding{51} & \ding{51} & \ding{55} & Limited & \ding{55} & \ding{51} \\ 
        & RTX 4080     & \ding{55} & \ding{51} & \ding{51} & \ding{55} & Limited & \ding{55} & \ding{51} \\ 
        & RTX 3090     & \ding{55} & \ding{51} & \ding{51} & \ding{55} & \ding{51} & \ding{55} & \ding{51} \\ 
        \hline
        \multirow{2}{*}{Jetson}  
        & Jetson Orin NX  & \ding{55} & \ding{51} & \ding{51} & \ding{55} & \ding{51} & \ding{55} & \ding{51} \\ 
        & Jetson Xavier NX & \ding{55} & \ding{51} & \ding{51} & \ding{55} & \ding{51} & \ding{55} & \ding{51} \\ 
        \hline
    \end{tabular}
    \caption{NVIDIA GPU Precision Support Summary}
    \label{tab:nvidia_precision}
\end{table}

\subsection{Knowledge Distillation}

Knowledge Distillation (KD) has emerged as a fundamental strategy in optimizing large language models (LLMs) for deployment in environments with limited computational resources. As LLMs increase in complexity and size to achieve high performance across diverse language tasks, distillation techniques provide a means to compress these models while retaining essential capabilities. In the context of LLMs, KD is divided into \textbf{Soft Knowledge Distillation} and \textbf{Hard Knowledge Distillation}, each playing a unique role in balancing model size, accuracy, and efficiency. Soft distillation enhances LLM generalization by transferring nuanced probability distributions, while hard distillation prioritizes specific label accuracy and simplifies training. These techniques are pivotal in the ongoing development of compact yet capable LLMs, allowing them to retain high performance even when scaled down significantly.

\subsubsection{Soft Knowledge Distillation}

Soft distillation leverages the full probability distribution generated by an LLM teacher model, capturing subtleties in the relationship between various output classes. For LLMs, this means the student model not only learns the correct answers but also internalizes the nuances in the teacher's responses to a broad array of inputs. By transferring these probability distributions, soft distillation helps the student model mimic the teacher’s language understanding, sentiment detection, and reasoning across tasks. Applying temperature scaling to soften the teacher’s outputs, the student LLM can absorb richer linguistic and semantic representations, making it particularly effective for LLMs intended for general-purpose language understanding. As a result, soft distillation is integral for developing efficient LLMs that retain robust performance across diverse tasks.

\textbf{Probability Distribution} Probability distribution methods involve transferring the full output probabilities from the teacher model to the student. This helps the student capture nuances in class relationships, learning not only the correct answers but also the likelihood of various alternatives. 
The choice of divergence function of probability distribution in KD shapes the way the student model aligns with the teacher's output distribution, impacting both diversity and quality of generated content. 
A commonly used approach is forward Kullback-Leibler (KL) divergence, which minimizes the divergence by aligning the student distribution to cover all modes of the teacher distribution, as seen in work by \citep{hinton2015distilling}, \citep{romero2015fitnets} and \citep{zagoruyko2017paying}. This method ensures that the student model captures the full spectrum of the teacher's probability distribution, even for tokens assigned low probability by the teacher. However, this "mode-covering" approach may lead the student to overestimate these low-probability regions, potentially resulting in hallucinations or lower-quality generations. 
Alternatively, reverse KL divergence focuses on aligning the student with only the high-probability regions of the teacher distribution, as demonstrated by \citet{gu2024minillm}. However, reverse KL can come at the expense of diversity, often producing fewer output variations. In {Baby LLaMA}, temperature scaling is applied to the logits of teacher models (e.g., GPT-2 and LLaMA) to provide smoother output distributions for distillation \citep{timiryasov2023baby}. This allows the student model to retain the nuances of the teacher’s probabilistic predictions, especially in low-data regimes. The soft probabilities are controlled by the temperature parameter $T$, ensuring that the model retains fine-grained knowledge:

\[
p_i = \frac{\exp(z_i / T)}{\sum_j \exp(z_j / T)}.
\]

Expanding on this, KPTD adopts a similar approach, but it leverages entity definitions and domain-specific knowledge to generate the transfer set. The student model learns to match the teacher’s output distribution based on this enriched set of entities, which is crucial for maintaining up-to-date knowledge in pre-trained models \citep{mirzadeh2020improved, gong2022preserving, chen2020distilling}. Following this, the authors of GKD refine this further by using token-level forward and reverse  divergence, focusing on critical tokens during distillation \citep{jafari2021sequence, kim2016sequence}. This design ensures that the student captures vital aspects of sequence-level tasks, such as machine translation and summarization, where token importance fluctuates throughout the sequence.

In parallel with previous work, in {MetaICL} \citep{min2022metaicl} and {Multitask-ICT} \citep{wang-2023-CoT}, soft distillation is adapted to transfer in-context learning (ICL) abilities from large models to smaller ones, helping the student generalize across tasks. These models use few-shot learning to transfer the multitask capabilities of large LLMs, effectively allowing the student to leverage the soft output distributions for task generalization. After this, the work from \citet{zhao2024multistage} further enhances this process by introducing a multi-stage collaborative approach, where the student generates refined pseudolabels to improve the next stage of distillation, pushing the boundaries of semi-supervised learning through iteration.




Above all, the soft distillation loss can be expressed as:

\[
\mathcal{L}_{\text{distill}} = 
\lambda_{\text{fwd}} \sum_{i=1}^N \alpha_i \cdot D_{\text{KL}} \left( p_{\text{teacher}, i} \| p_{\text{student}, i} \right) 
+ \lambda_{\text{rev}} \sum_{i=1}^N \beta_i \cdot D_{\text{KL}} \left( p_{\text{student}, i} \| p_{\text{teacher}, i} \right),
\]

\[
D_{\text{KL}}(p \| q) = \sum_j p_j \log \frac{p_j}{q_j}
\]

where:

\begin{itemize}
    \item \( p_{\text{teacher}, i} \) and \( p_{\text{student}, i} \) are the teacher and student probability distributions for token \( i \), respectively.
    \item \( \lambda_{\text{fwd}} \) and \( \lambda_{\text{rev}} \) are tunable weights balancing forward and reverse KL components.
    \item \( \alpha_i \) and \( \beta_i \) are token-specific importance weights, which could depend on token-level saliency, attention scores, or domain-specific criteria.
    \item \( N \) is the total number of tokens in the sequence.
\end{itemize}


\subsubsection{Hard Knowledge Distillation}  

Hard distillation simplifies the training process by focusing solely on the teacher model’s “hard” labels, or binary outputs, for each task. In the case of LLMs, hard distillation emphasizes specific target outcomes over nuanced inter-class information, which can be advantageous for task-focused LLMs where interpretability and task accuracy are prioritized. This method is often used in classification tasks where the LLM’s primary purpose is to deliver accurate, targeted answers without requiring complex linguistic or semantic generalization. Although this approach may limit the student's understanding of intricate language structures, it accelerates training and achieves effective LLM compression, making it a suitable choice for specialized language tasks where efficiency and specific label accuracy are crucial.

Program-aided Distillation (PaD) addresses the challenge of faulty reasoning in distillation by introducing programmatic reasoning \citep{shridhar2022distilling}. Synthetic programs are generated to ensure correct reasoning steps, and these programs are automatically checked by an interpreter before being used in distillation. This reduces faulty reasoning in student models, ensuring that the correct reasoning paths are learned alongside the correct outputs. DynaBERT applies hard knowledge distillation in dynamic model compression, progressively decreasing model size while ensuring that smaller models retain the performance of the teacher \citep{hou2020dynabert}. By combining hard distillation with a reduction in both model width and depth, DynaBERT balances efficiency with accuracy. Similarly, LaMini-LM focuses on transferring instruction-following abilities to smaller models by using hard distillation on a curated instruction set, optimizing the student for task-specific performance \citep{wu2023lamini}. Zephyr integrates hard knowledge distillation with {Direct Preference Optimization (dDPO)} \citep{tunstall2023zephyr}. In this approach, the student learns from both hard decisions and user preferences, blending rigid output labels with nuanced optimization for user alignment. This hybrid approach is particularly effective in conversational AI, where both categorical responses and preference-based feedback are important for generating relevant and engaging responses.

Above all, the hard distillation process can be formalized as :

\[
\mathcal{L}_{\text{hard}} = -\sum_{i=1}^N \sum_{j=1}^C y_{i,j}^{\text{teacher}} \cdot \log p_{i,j}^{\text{student}},
\]

where:

\begin{itemize}
    \item \( y_{i,j}^{\text{teacher}} \in \{0, 1\} \) is the one-hot encoded hard label for token \( i \) and class \( j \), generated by the teacher model.
    \item \( p_{i,j}^{\text{student}} \) is the predicted probability of the student model for token \( i \) and class \( j \), computed using a softmax over the student logits:
    \[
    p_{i,j}^{\text{student}} = \frac{\exp(z_{i,j}^{\text{student}})}{\sum_{k=1}^C \exp(z_{i,k}^{\text{student}})},
    \]
    where \( z_{i,j}^{\text{student}} \) is the logit for class \( j \) at token \( i \).
    \item \( N \) is the total number of tokens in the dataset.
    \item \( C \) is the number of classes in the classification task.
\end{itemize}

\subsection{Pruning}
In addition to quantization and knowledge distillation, pruning is another efficient technique for reducing computational costs in LLMs. While quantization focuses on hardware optimization and knowledge distillation on knowledge transfer in neural networks, pruning concentrates on the model's structural efficiency. Specifically, pruning in machine learning involves removing less important connections in the neural network while retaining the important ones, making the model more memory-efficient and faster. 

Pruning methods can be broadly categorized based on their granularity into unstructured pruning and structured pruning. Structured pruning removes entire neurons or larger components of the neural network. In contrast, unstructured pruning refers to the removal of individual connections, as seen in the conventional pruning methods studied by \citet{P-han2015learning}. Additionally, semi-structured pruning like N:M sparsity \citep{P-NM}, is generally considered a specialized form of unstructured pruning. In general, structured pruning usually yields better inference speed improvements compared to unstructured pruning, due to its hardware-friendly design.

Pruning methods for LLMs typically remove weights based either on (1) the loss function or (2) the layer output.

(1) Loss-based. The weights are usually removed by computing metric such as gradients with respect to the training or validation loss. In some cases, pruning is performed directly through optimization techniques that minimize the model's loss after weight removal.

(2) Layer output-based. Alternatively, other pruning methods for LLMs aim to find the following masks in a layer-wise pattern. The pruning quality is usually measured by the $\ell_2$-norm between the output, for given specific inputs $X_\ell$, of the uncompressed layer and that of the compressed one. More formally, consider a layer $\ell$ with weight matrix $W_\ell \in \R^p$. The objective is to find the binary mask $M_\ell \in [0,1]^p$ and possibly an updated weight matrix $\hat{W}_\ell$ so that the following objective is minimized,

\begin{align}
    \argmin_{M_\ell, \hat{W}_\ell} \|W_\ell X_\ell-(\hat{W}_\ell \odot M_\ell) X_\ell \|^2_2. \label{eq:P-layerloss}
\end{align} 

Here, $\odot$ denotes element-wise multiplication.

\subsubsection{Unstructured Pruning}
In this section we consider unstructured pruning and semi-structured pruning methods like N:M sparsity.

\paragraph{Saliency-Based Pruning.}There are several works motivated by classical post-training methods, including Optimal Brain Damage (OBD; \citet{P-OBD}) and Optional Brain Surgeon (OBS; \citet{P-OBS}). In the original OBS framework, the removal of connection at index $m$\footnote{Note that we use subscript $S_{ij}$ and $S_m$ interchangeably. $S_{ij}$ emphasizes the weight within a specific matrix, while $S_m$ emphasizes the weight across the entire neural network.} with weight $w_m$ is selected based on the saliency metric 
\begin{align*}
    S_m=\frac{W_m^2}{[H^{-1}]_{mm}},
\end{align*}
where $H$ is the Hessian matrix. 

After the weight removal, all the remaining weights will be updated with the following formula:
\begin{align*}
    \delta_m=-\frac{w_m}{[H^{-1}]_{mm}}[H^{-1}]_{:,m}
\end{align*}
to compensate for the removal of the weight at index $m$. 

Note that under the formulation of \autoref{eq:P-layerloss} for an affine transformation, each row can be computed independently. Specifically, we can iteratively apply OBS to different rows by (1) identifying the location $m$ that results in the smallest loss increase, (2) constructing the updated mask $M-\{m\}$, and (3) updating \textbf{all} remaining weights responding to $M-\{m\}$ upon the removal of mask $\{m\}$. 

Consider a weight matrix $W_\ell \in \R^{d_{row}\times d_{col}}$, SparseGPT \citep{P-SparseGPT} extends the idea of OBS by updating only a subset of remaining weights $H \subseteq M-\{m\}$ to enable faster computation of the Hessian matrix. More specifically, SparseGPT processes columns sequentially from left to right using a sequence of Hessian matrices, which can be efficiently computed using Gaussian elimination. SparseGPT only uses the weights to the right of the pruned weight as subset to compensate the loss induced by its removal. The computational cost of SparseGPT is hence $\Theta(d_{hidden}^3)$ compared to $\Theta(d_{hidden}^4)$ required by exact construction in Transformers with hidden dimension $d_{hidden}$. 

SparseGPT is also compatible with N:M sparsity by choosing N smallest weights that incurs the lowest errors in every block of M. Additionally, it also supports quantization method like GPTQ \citep{P-gptq}.

Wanda \citep{P-WANDA} proposes a simple and effective pruning metric:
\begin{align*}
    S_{ij} = |W_{ij}|\cdot \|X_j\|_2,
\end{align*} 
where $W$ is a weight matrix, and $X_j$ is the $j$-th feature normalized by batch and sequence length dimensions. Wanda shows the connection between its pruning criterion and those of OBS and SparseGPT through a diagonal approximation of SparseGPT's saliency score:
\begin{align*}
    S_{ij}=\left[\frac{|W|^2}{\text{diag}((XX^T+\lambda I)^{-1})}\right]_{ij}.
\end{align*}
Applying the diagonal approximation with $\lambda=0$, the metric can be simplified to:
\begin{align}
    \left[\frac{|W|^2}{\text{diag}((XX^T+\lambda I)^{-1})}\right]_{ij} \stackrel{\lambda=0}{\approx} \left[\frac{|W|^2}{(\text{diag}(XX^T))^{-1}}\right]_{ij} = (|W_{ij}|\cdot \|X_j\|_2)^2.
\end{align}
Compared to the complexity $\Theta(d_{hidden}^3)$ of SparseGPT, Wanda enjoys the complexity of $\Theta(d_{hidden}^2)$. Additionally, Wanda does not need to update the weights. 

Empirical evaluations on zero-shot tasks and perplexity on LlaMA \citep{LLAMA} models show the competitive performance of Wanda compared to SparseGPT. Wanda is also compatible with semi-structured N:M pruning and LoRA fine-tuning \citep{LORA}.  

Improved Saliency Criterion (ISC; \citep{P-ISC}) proposed to combine criterion from both OBD and OBS to define a better saliency score:
\begin{align*}
    S_{m} = W_m^2\left(H_{mm}+\frac{1}{[H^{-1}]_{mm}}\right).
\end{align*}

Moreover, ISC also explores the effects of the sparsity ratio allocations between layers through sensitivity. The sensitivity is calculated by the average trace of the Hessian matrix. Experiments conducted on the LlaMA model family and Baichuan models show its strong performance and compatibility with quantization methods.

D-Pruner \citep{P-Dpruner} studies domain-specific compression for LLMs. It uses the formulation of OBS for general knowledge and while introducing additional regularization term tailored to domain-specific knowledge. With the help of both general and domain-specific calibration datasets, D-Pruner produces a pruned model that strikes a balance between both generality and specificity. 

RIA \citep{P-RIA} incorporates relative importance into Wanda's saliency score, making the final saliency score: 
\begin{align}
    S_{ij} = \left(\frac{|W_{ij}|}{\sum_i |W_{ij}|} + \frac{|W_{ij}|}{\sum_j |W_{ij}|}\right)\cdot (\|X_i\|_2)^\alpha.
\end{align}
Moreover, RIA introduces an optimization specifically for N:M sparsity through channel permutation. This enables the retention of important weights within the same block, improving the performance of pruned models.

Pruner-Zero \citep{P-prunerzero} formulates pruning as Symbolic Regression (SR) problem and use genetic programming (GP) to solve it. It directly searches pruning saliency score using LLM statistics as input features, including activations ($X$), gradients ($G$), and weights ($W$). The search space includes primitive operations include unary and binary ones. Pruner-zero iteratively generates and refines pruning metrics via genetic programming procedures including tournament selection, subtree crossover, and node mutation.

There are other studies focus more on deployment and evaluation rather than solely on algorithm design. For example, Prune-And-Tune \citep{P-pruneandtune} demonstrates that fine-tuning can further improve the performance of SparseGPT for OPT model family, particularly at high sparsity levels. \citet{P-shamrai2024language} specifically evaluates SparseGPT and Wanda specifically for the Ukrainian language.

\paragraph{Optimization Based Pruning.} Another line of research aims at finding masks through optimization.

BESA \citep{P-BESA} efficiently learns block-wise sparsity ratios for each model component/block, such as Transformer blocks and feed-forward network (FFN) blocks, through minimizing the reconstruction error. The training process uses a standard Straight-through estimator (STE) to handle the non-differentiability of the masks. The sparsity is enforced through a $\ell_2$ regularization term during the training process. 

Once the sparsity ratios are learned, Wanda saliency score will be used to prune unimportant weights within each block. Additionally, BESA is also compatible with quantization techniques.

\subsubsection{Structured Pruning}
\paragraph{Saliency-based Pruning.} \citet{ma2023} highlights that task-specific model compression through knowledge distillation can be time-consuming in some cases, which requires up to 3.5 days on 4 GPUs for a 20GB corpus. In contrast, pruning can significantly compress model size on a structural level, which requires up to 3 hours on 1 GPU for a 50MB corpus after pruning~\citep{wang2024svd}. 

One famous method of pruning in LLMs is LLMPrunner \citep{ma2023}. Initially, a dependency tree or graph is established based on structural dependency in LLMs, linking neurons based on dependency relationships. After building the dependency graph, the pruner then analyzes the importance of each link in the graph and trims those links with less importance to save computational resources and accelerate inference. LLMPrunner uses multiple methods (element-wise, vector-wise, group) to evaluate the importance of coupled structures in the dependency graph, removing less important structures. It also offers reliable recovery methods such as Low-rank Approximation to recover links discarded by mistake. 
Empirical results show that LLMPrunner can achieve 20\% parameter reduction while retaining 90\% of the original performance, showcasing the effectiveness of pruning in maintaining model efficiency while reducing computational resources by removing redundant elements in the network.

FLAP \citep{P-FLAP} measures the importance of different channels using fluctuation-based metrics with calibration data. It then normalizes these metrics to to ensure a consistent comparison of scores across different layers and modules. Additionally, FLAP introduces a technique called baseline bias compensation to deal with the damage caused by removing components.

\citet{P-zhang2024structured} proposes a structured pruning method that groups kernels and features independently and evaluates their importance by second-order Taylor expansion. To reduce computation, they approximate the second order term with squared gradient. Specifically, for kernel group $C$, the saliency score $S_C$ is computed as: 
\begin{align*}
     \left|\sum_{c\in C}\frac{\partial \mathcal{L}}{\partial c} c-\frac{1}{2}\left(\frac{\partial^2 \mathcal{L}}{\partial c^2} c^2\right)\right| 
     \approx \left|\sum_{c\in C}\frac{\partial \mathcal{L}}{\partial c} c-\frac{1}{2}\left(\frac{\partial \mathcal{L}}{\partial c} c\right)^2\right| = S_C,
\end{align*}
where $\mathcal{L}$ denotes the loss function and $c$ represents the parameters in the group $C$.

Similarly, for features $f$, the corresponding saliency score of an affine transformation is
\begin{align*}
     S_f = \sum_{C} \left|\sum_{c\in C[:,f] \cup C[f,:]}\frac{\partial \mathcal{L}}{\partial c} c-\frac{1}{2}\left(\frac{\partial \mathcal{L}}{\partial c} c\right)^2\right|.
\end{align*}

\paragraph{Optimization-based Pruning.} ShearedLlaMA \citep{P-shearedllama} formulates the pruning problem as a constrained optimization problem using Lagrange multiplier. It learns pruning masks at varying granularities, from coarse ones like entire layers and hidden dimensions to fine-grained ones like attention heads and intermediate dimensions.

\paragraph{Encoder-Decoder Architecture.} NASH \citep{P-NASH} specifically studies structured pruning for encoder-decoder Transformer architecture. Inspired by empirical findings, NASH proposes to shorten, i.e., reduce depth, for decoders and narrow, i.e., reduce width, for encoders. To achieve better performance, NASH additionally applies hidden states distillation by adding a Kullback–Leibler (KL) divergence loss term.

\paragraph{Coupled with PEFT.} Light-PEFT \citep{P-lightpeft} and V-PETL~\citep{citation-0} aims to reduce the costs of fine-tuning at multiple granularities. At a coarse level, it measures the importance of PEFT modules through the change of PEFT modules on the original module output. At a finer level, it measures the rank of each PEFT modules by summing the first-order Taylor expansion saliency score of all parameters of that rank:
\begin{align*}
    S_{i,j} = \left|\frac{\partial \mathcal{L}}{\partial W_{i,j}} W_{i,j}\right|.
\end{align*}
Similarly, LoRAPrune \citep{P-loraprune} uses the same Taylor expansion saliency score for measuring group importance. It inserts learnable LoRA matrices during downstream task adaption, circumventing the needs of computing the gradients of the entire weight matrix. 

\subsubsection{Others}

\paragraph{Infra.} FlashLLM \citep{P-flashllm} offers specific infrastructure support for unstructured pruning by utilizing Tensor Cores rather than SIMT Cores to enable faster inference.

\section{System Design for Efficient Foundation Models}

Large Foundation Models face significant challenges in managing computational and memory costs during autoregressive generation. Effective system design leverages optimizations like Key-Value (KV) caching, token reduction, and precision scaling to balance efficiency and accuracy. These methods are grounded in system design that formalize the trade-offs between model performance and resource utilization.

\subsection{KV Cache Design}

\begin{figure}[h]
    \centering
    \begin{tikzpicture}[node distance=2cm, scale=0.6, transform shape]

\tikzstyle{startstop} = [rectangle, rounded corners, minimum width=3cm, minimum height=1cm, text centered, draw=black, fill=red!30]
\tikzstyle{process} = [rectangle, minimum width=4cm, minimum height=1.5cm, text centered, draw=black, fill=blue!30]
\tikzstyle{processlar} = [rectangle, minimum width=8.5cm, minimum height=1.5cm, text centered, draw=black, fill=yellow!30]
\tikzstyle{processlim} = [rectangle, minimum width=4cm, minimum height=1.5cm, text centered, draw=black, fill=yellow!30]
\tikzstyle{arrow} = [thick,->,>=stealth]

\node (start) [startstop] {Input Prompt};
\node (query) [process, below of=start] {Generate Query (Q)};
\node (cachecheck) [process, below of=query] {Check KV Cache};
\node (updatecache) [processlim, right of=cachecheck, xshift=6cm, text width=7cm] {
    Update KV Cache \\
    \[ 
    \mathbf{K}_{\text{low}} = \mathbf{U} \mathbf{K}, \, 
    \mathbf{V}_{\text{low}} = \mathbf{V} \mathbf{V}^\top
    \]
};
\node (attention) [process, below of=cachecheck, yshift=-3cm, text width=7cm] {
    Compute Attention \\
    \[ 
    \hat{\mathbf{X}}_t = \text{softmax} \left( 
    \frac{\mathbf{Q}_t \mathbf{K}_{1:t}^\top}{\sqrt{d_k}} + \mathbf{B}_t 
    \right) \mathbf{V}_{1:t}
    \]
};
\node (output) [startstop, below of=attention, yshift=-0.5cm] {Generate Output Token};
\node (opt) [processlar, below of=updatecache, yshift=-0.7cm, text width=7cm] {
    Optimize KV Cache \\
    \[
    \min_{\mathbf{K}_{\text{low}}, \mathbf{V}_{\text{low}}} 
    \| \mathbf{K} - \mathbf{K}_{\text{low}} \|_F^2 + 
    \| \mathbf{V} - \mathbf{V}_{\text{low}} \|_F^2 + 
    \lambda \mathcal{C}(\mathbf{K}_{\text{low}}, \mathbf{V}_{\text{low}})
    \]
};

\draw [arrow] (start) -- (query);
\draw [arrow] (query) -- (cachecheck);
\draw [arrow] (cachecheck) -- (attention);
\draw [arrow] (cachecheck.east) -- ++(1,0) -- (updatecache.west); 
\draw [arrow] (opt.south) -- (attention.east);
\draw [arrow] (attention) -- (output);
\draw [arrow] (updatecache.south) -- (opt.north); 

\end{tikzpicture}

    \caption{KV Cache Design Pipeline}
    \label{fig:kvcache}
\end{figure}

KV caching is a cornerstone of transformer-based LLMs, significantly reducing computational overhead by reusing previously computed key (\(\mathbf{K}\)) and value (\(\mathbf{V}\)) matrices. During autoregressive generation, the self-attention mechanism computes the output for a sequence of tokens by leveraging the precomputed cache. Let \(\mathbf{Q}_t\), \(\mathbf{K}_{1:t}\), and \(\mathbf{V}_{1:t}\) denote the query, key, and value matrices at time \(t\). The attention mechanism scales the query-key interaction by the inverse square root of the key dimension \(d_k\), applies positional biases \(\mathbf{B}_t\), and generates the final representation \(\hat{\mathbf{X}}_t = \text{softmax}((\mathbf{Q}_t \mathbf{K}_{1:t}^\top)/\sqrt{d_k} + \mathbf{B}_t) \mathbf{V}_{1:t}\). To improve memory efficiency, advanced KV cache compression methods use low-rank approximations, where keys and values are projected into lower-dimensional spaces using learned projection matrices \(\mathbf{U}\) and \(\mathbf{V}\). Specifically, the compressed representations are given by \(\mathbf{K}_{\text{low}} = \mathbf{U} \mathbf{K}\) and \(\mathbf{V}_{\text{low}} = \mathbf{V} \mathbf{V}^\top\), where \(r \ll d_k\) controls the rank of the approximation. Query-aware sparsity further optimizes KV caching by dynamically identifying and updating only the most relevant tokens using a sparsity mask \(\mathbf{M}_t\), defined as the top-\(k\) entries of the scaled query-key scores. This hierarchical approach combines memory reduction and computational efficiency while maintaining the fidelity of attention outputs.

The general optimization for KV cache design balances memory usage, computational cost, and attention fidelity, expressed as:
\[
\min_{\mathbf{K}_{\text{low}}, \mathbf{V}_{\text{low}}} \| \mathbf{K} - \mathbf{K}_{\text{low}} \|_F^2 + \| \mathbf{V} - \mathbf{V}_{\text{low}} \|_F^2 + \lambda \cdot \mathcal{C}(\mathbf{K}_{\text{low}}, \mathbf{V}_{\text{low}}),
\]
where \(\mathcal{C}(\mathbf{K}_{\text{low}}, \mathbf{V}_{\text{low}})\) models the memory and computational costs of the compressed representations \citep{xiao2024efficientstreaminglanguagemodels, xiong2024uncomp}.

Autoregressive decoding in LLMs requires KV cache, which stores intermediate outputs for all tokens, creating significant memory overhead. Some adaptive KV cache compression techniques will adjust the precision of stored tokens based on their importance. Let \( K, V \in \mathbb{R}^{T \times d} \) represent the key and value matrices. The optimization for adaptive precision is:
\[
\min_{\hat{K}, \hat{V}} \|K - \hat{K}\|_F^2 + \|V - \hat{V}\|_F^2 + \gamma \cdot \sum_{t=1}^T P_t \cdot \mathcal{C}(P_t),
\]
where \( \hat{K} \) and \( \hat{V} \) are the compressed representations, \( P_t \) denotes the precision level for token \( t \), and \( \mathcal{C}(P_t) \) quantifies the cost of using a specific precision. Precision levels can be dynamically assigned as \( P_t = \text{argmax}_{p \in \{\text{FP32}, \text{FP16},\text{FP8}, \text{INT8}, etc\}} \Big\{S_t \cdot \text{Utility}(p)\Big\} \), where \( S_t \) is the token saliency score and \( \text{Utility}(p) \) reflects the precision's effectiveness. VL-Cache achieves up to 90\% memory reduction with a 7$\times$ improvement in decoding latency \citep{vlcache}.

\subsection{Token Reduction}

Token reduction focuses on minimizing the number of tokens processed during training and inference while retaining essential information for accurate predictions. In transformer models, tokens often exhibit redundancy, particularly in the deeper layers where contextualized embeddings converge. Let \(T \in \mathbb{R}^{n \times d}\) denote the token embeddings, where \(n\) is the token count and \(d\) is the embedding dimension. Token pruning techniques dynamically identify and remove less important tokens based on learned importance scores or predefined heuristics. Importance scores can be computed using attention weights, token magnitudes, or cumulative relevance across layers. For example, attention-based pruning ranks tokens by their cumulative importance \(S_i = \sum_{l=1}^L \sum_{j=1}^n A_{ij}^{(l)}\), where \(A^{(l)}\) represents the attention matrix at layer \(l\). Tokens with the lowest scores are removed, and their information is redistributed to the remaining tokens. Alternatively, token merging methods combine similar tokens, reducing the total count while preserving overall representation. In methods like Token Merging (ToMe), merged tokens are computed as weighted combinations, where similarity scores guide the merging process. The merged embeddings \(\hat{T}_k = \alpha T_i + (1 - \alpha) T_j\), with \(\alpha\) determined by cosine similarity, ensure that merged tokens retain meaningful features. These approaches effectively reduce computational costs in both text and visual transformer architectures.

The optimization objective for token reduction can be formulated as:
\[
\min_{M, \hat{T}} \|T \odot M - \hat{T}\|_F^2 + \mu \cdot \text{Cost}(M) + \nu \cdot \text{Regularization}(M),
\]
where \(M \in \{0, 1\}^n\) is the token retention mask, \(\text{Cost}(M)\) captures the computational overhead, and \(\text{Regularization}(M)\) enforces structural constraints \citep{bolya2023tome, rao2021dynamicvit}.

For visual LLMs, visual tokens encode rich spatial and semantic information but often contain redundancy. PyramidDrop adopts token reduction framework by computing redundancy scores for each token based on local similarity metrics \citep{pyramiddrop}. Similarly, Victor leverages token summarization mechanisms, where token embeddings are aggregated into compact registers to reduce computation by over 3$\times$ with minimal accuracy loss \citep{victor}.

\subsection{Sparsity-Aware Optimization}

The QKV transformation requires masking for autoregressive prediction. The structure \( Q K^\top \odot M \) results in a lower triangular masked attention matrix of size \( N \times N \), where \( N \) is the processed sequence length, which can be significantly large in large-scale language models.


However, not all tokens contribute equally to the final output. Studies show that a small subset of tokens dominates attention computations, while others carry minimal impact. Optimizing attention sparsity reduces computational overhead without sacrificing performance. Several methods address this:


\paragraph{Token Selection and Structured Sparsity}  

Some Research reduces computational costs by focusing only on high-saliency tokens. Let \( A \in \mathbb{R}^{T \times T} \) represent the attention matrix at a given layer, where \( A_{ij} \) denotes the attention weight between tokens \( i \) and \( j \). Sparsity-aware optimization selects a subset of tokens \( S = \{t \mid S_t > \tau\} \), where \( S_t = \sum_{j=1}^T A_{tj} \) represents the token's saliency score and \( \tau \) is a dynamically determined threshold. The optimization problem can be expressed as:
\[
\max_{\rho_l} \sum_{l=1}^{N_L} \rho_l \cdot \sum_{t \in S} \log\Big(1 + \frac{S_t}{\tau}\Big) - \kappa \cdot \mathcal{C}(\rho_l),
\]
where \( \rho_l \) denotes the retention ratio at layer \( l \), \( \mathcal{C}(\rho_l) \) is the computational cost associated with retaining tokens, and \( \kappa \) balances performance and cost. Methods like ZipVL and $H_2O$ dynamically adjust \( \rho_l \) across layers based on the attention distribution \citep{zipvl, H2O2023}, achieving over large memory and cache savings with small accuracy degradation.

\paragraph{Activation and Model Weight Sparsity}  
Beyond attention sparsity, recent studies have explored the natural sparsity in transformer activations. It has been observed that ReLU-based transformers exhibit a high degree of activation sparsity in the feedforward network (FFN) layers \citep{ReLU_Sparsity2023, SparsityFFN2023}. In models like OPT-175B, over 95\% of FFN activations are zero. While modern transformers favor soft activations such as GeLU and SwiGLU, replacing them with ReLU and fine-tuning has been shown to increase activation sparsity while maintaining performance \citep{ReLU_Finetuning2024}.

Another effective sparsity technique involves weight pruning. Static weight pruning approaches, such as Wanda \citep{Wanda2023}, analyze weight importance and remove redundant connections before inference. These methods ensure a fixed sparse structure, unlike dynamic sparsity techniques that adapt weight selection based on input data. LoRA \citep{Lora2021} further reduces computational costs by introducing low-rank updates instead of modifying the full weight matrices. LoRA injects trainable low-rank matrices \( \Delta W \) into the model:

\[
W' = W + \Delta W, \quad \text{where} \quad \Delta W = A B^\top, \quad A, B \in \mathbb{R}^{d \times r}.
\]

With \( r \ll d \), LoRA achieves significant efficiency improvements, making fine-tuning large models more memory-efficient.

\subsection{Multi-Modal Fusion}

Multi-modal fusion integrates information from different modality, such as text, images, and audio, to create a richer and more comprehensive representation. By capturing relationships between modalities, fusion methods enhance model understanding and response quality. Key techniques include unified feature representations, cross-modal attention mechanisms, and efficient compression strategies. These approaches enable a combined interaction between modalities while maintaining efficiency in computation and memory usage.

\subsubsection{Multi-Modal Fusion Representations}

\paragraph{Multi-Modal Key and Value Structures}

Unified KV representations reduce redundancy while capturing cross-modal dependencies. For a multi-modal input \( x = \{x_t, x_i, x_a\} \) (text, image, and audio, respectively), the KV cache can be structured as:

\[
K = \begin{bmatrix}
W_k^{(t)} h^{(t)} & \alpha_{ti} W_k^{(i)} h^{(i)} & \alpha_{ta} W_k^{(a)} h^{(a)} \\
\alpha_{it} W_k^{(t)} h^{(t)} & W_k^{(i)} h^{(i)} & \alpha_{ia} W_k^{(a)} h^{(a)} \\
\alpha_{at} W_k^{(t)} h^{(t)} & \alpha_{ai} W_k^{(i)} h^{(i)} & W_k^{(a)} h^{(a)}
\end{bmatrix},
\]

where \( W_k^{(m)} \) and \( h^{(m)} \) represent the projection matrix and hidden state for modality \( m \). The coefficients \( \alpha_{mn} \) denote learned alignment weights between modalities \( m \) and \( n \), enabling efficient cross-modal interaction \citep{MS-ALIGN, MS-Flamingo, MS-M3AE}.

\paragraph{Joint Cross-Modal Attention-Based Compression}

Given the computational overhead of storing full KV caches, cross-modal attention matrices \( A \) can be used to guide compression. For text \( T \) and image \( I \), the attention matrix is:

\[
A_{TI} = \text{diag}(K_T^\top Q_I) + \text{diag}(K_I^\top Q_T),
\]

where \( Q \) and \( K \) represent the query and key matrices, respectively. A low-rank approximation of \( A_{TI} \) via Singular Value Decomposition (SVD) yields:

\[
A_{TI} \approx U_{TI} \Sigma_{TI} V_{TI}^\top, \quad \Sigma_{TI} = \text{diag}(\sigma_1, \sigma_2, \dots, \sigma_r),
\]

where \( U_{TI}, \Sigma_{TI}, V_{TI} \) are the top-\( r \) components. This reduced-rank representation balances efficiency and accuracy \citep{MS-PerceiverIO, MS-LowRankFusion}.

\subsubsection{Cache Eviction on Multi-Modality Fusion}

\paragraph{Cross-Modal Sensitivity-Based Eviction}

Cache eviction in multi-modal systems must account for both intra-modal and cross-modal contributions. The sensitivity of the model output \( y \) to a key \( k_t^m \) from modality \( m \) is defined as:

\[
S_t^{(m)} = \left\| \frac{\partial y}{\partial k_t^m} \right\|_2^2 + \sum_{n \neq m} \beta_{mn} \left\| \frac{\partial y}{\partial k_t^n} \cdot \frac{\partial k_t^n}{\partial k_t^m} \right\|_2^2,
\]

where \( \beta_{mn} \) captures the dependency strength between modalities \( m \) and \( n \). Entries with the smallest \( S_t^{(m)} \) are evicted, ensuring minimal performance degradation \citep{MS-SensitivityCache, MS-MultimodalHessian}.

\paragraph{Hessian-Aware Cross-Modal Eviction}

Second-order optimization methods are useful for assessing the importance of cache entries. The Hessian-aware importance of a key \( k_t^m \) is computed as:

\[
\Delta \mathcal{L} = \frac{1}{2} \left( \frac{\partial \mathcal{L}}{\partial k_t^m}^\top H_t^{(m)} \frac{\partial \mathcal{L}}{\partial k_t^m} + \sum_{n \neq m} \frac{\partial \mathcal{L}}{\partial k_t^m}^\top H_t^{(mn)} \frac{\partial \mathcal{L}}{\partial k_t^n} \right),
\]

where \( H_t^{(m)} \) is the Hessian for modality \( m \), and \( H_t^{(mn)} \) captures cross-modal dependencies. Cache entries with minimal \( \Delta \mathcal{L} \) are prioritized for eviction, preserving critical cross-modal interactions \citep{MS-Flamingo, MS-HessianEviction}.

\paragraph{Entropy-Regularized Cross-Modal Eviction}

Maintaining diversity in cache content can be achieved through entropy-based eviction. The joint entropy of a key distribution \( p_t^{(m,n)} \) across modalities \( m \) and \( n \) is:

\[
H_t = - \sum_{m \in \mathcal{M}} \sum_{n \in \mathcal{M}} p_t^{(m,n)} \log p_t^{(m,n)},
\]

where \( p_t^{(m,n)} \) is the normalized attention contribution of \( k_t \) for modalities \( m \) and \( n \). Entries with low \( H_t \) are evicted, promoting diversity and informativeness \citep{MS-EntropyCache, MS-MultimodalFusion}.


\subsection{Efficient Sequence Modeling}

Transformer-based LLMs has achieved state-of-the-art results across a variety of tasks \citep{vaswani2017attention, achiam2023scaling, touvron2023llama}. The self-attention of Transformers will achieve an all-to-all information routing between tokens in a sequence, which will result in a long sequence dependent representations \citep{bahdanau2014neural}. However, self-attention scales quadratically with sequence length, leading to high computational costs during training and inference\citep{9043731,e-infer-s}. Besides, The key-value caching during autoregressive generation further increases memory overhead \citep{tay2022efficient}. 

To mitigate these efficiency challenges, several alternative architectures have been proposed, including \textbf{linear attention}, \textbf{state-space models}, \textbf{long convolution}, and \textbf{hybrid methods}, each offering distinct computational advantages over standard transformers.

\paragraph{Linear Attention}  
Linear attention restructures the self-attention computation to avoid the explicit quadratic complexity of softmax attention. Given standard attention:
\[
A = \text{softmax} \left( \frac{QK^\top}{\sqrt{d}} \right) V,
\]
where \( Q, K, V \in \mathbb{R}^{N \times d} \) represent query, key, and value matrices, linear attention instead decomposes the softmax function using kernel-based approximations:
\[
O = \phi(Q) \left( \phi(K)^\top V \right),
\]
where \( \phi(\cdot) \) is a feature transformation designed to preserve the key properties of softmax while enabling efficient factorization \citep{katharopoulos2020transformers, choromanski2021rethinking, qin2022linear}. Variants of \( \phi(x) \) include:
- \( \phi(x) = 1 + \text{elu}(x) \) \citep{katharopoulos2020transformers}.
- Cosine approximations \citep{qin2022linear}.
- Random feature-based mappings \citep{choromanski2021rethinking, shen2023studyrelusoftmaxtransformer}.

Despite achieving \( \mathcal{O}(N d^2) \) complexity, causal implementations require cumulative summation (\texttt{cumsum}) operations, limiting their efficiency in autoregressive settings \citep{hua2022causal}. Lightning Attention \citep{qin2024lightning1, qin2024lightning2} resolves this through a structured block-wise decomposition that eliminates \texttt{cumsum}, maintaining efficient scaling across sequence lengths.

\paragraph{State Space Models}  
State-space models (SSMs) frame sequence modeling as a continuous dynamical system:
\[
h'(t) = A h(t) + B x(t),
\]
\[
y(t) = C h(t),
\]
where \( A, B, C \) are learned transition matrices controlling the evolution of hidden states \citep{gu2022efficient}. Unlike self-attention, SSMs process sequences with constant complexity, making them well-suited for long-context modeling \citep{gu2022state, gu2022structured}. 

Mamba \citep{gu2024mamba} improves upon standard SSMs by introducing \textit{selective state updates}, dynamically adjusting recurrence parameters based on input tokens. The discrete recurrence formulation is given by:
\[
h_t = A h_{t-1} + B x_t, \quad y_t = C h_t.
\]
This design eliminates the need for extensive key-value caching, ensuring efficient autoregressive inference with linear complexity.

\paragraph{Long Convolution}  
Long convolutional models replace explicit attention mechanisms with large-kernel convolutions that capture long-range dependencies \citep{qin2023long, fu2023long}. The core operation for a sequence \( X \) with kernel \( W \) is:
\[
O_t = \sum_{i=0}^{L} W_i X_{t-i}.
\]
To accelerate training, Fast Fourier Transforms (FFT) reduce computational complexity from \( \mathcal{O}(N^2) \) to \( \mathcal{O}(N \log N) \). However, causal convolutions require caching all prior activations, leading to high memory usage in autoregressive applications.

\paragraph{Hybrid Models: Combining Attention and SSMs}  
While alternative architectures provide computational benefits, self-attention remains superior in capturing in-context relationships and performing few-shot generalization. Hybrid models, such as Mamba-2-Hybrid \citep{megatron2024mamba}, integrate selective self-attention with state-space models, achieving a balance between efficiency and model expressivity. Empirical studies show that hybrid architectures outperform both transformers and pure SSMs in long-context tasks, efficiently handling sequences up to 128K tokens.

\begin{table}[h]
    \centering
    \begin{tabular}{|l|c|c|}
        \hline
        \textbf{Model Type} & \textbf{Time Complexity} & \textbf{Space Complexity} \\
        \hline
        Softmax Attention & \( \mathcal{O}(N^2 d) \) & \( \mathcal{O}(N^2) \) \\
        \hline
        Linear Attention & \( \mathcal{O}(N d^2) \) & \( \mathcal{O}(N d) \) \\
        \hline
        State-Space Models (SSMs) & \( \mathcal{O}(N d) \) & \( \mathcal{O}(N d) \) \\
        \hline
        Hybrid (SSM + Attention) & \( \mathcal{O}(N d + S d^2) \) & \( \mathcal{O}(N d) \) \\
        \hline
        Multi-Query Attention (MQA) & \( \mathcal{O}(N d^2) \) & \( \mathcal{O}(N d) \) \\
        \hline
        Grouped-Query Attention (GQA) & \( \mathcal{O}(N d^2) \) & \( \mathcal{O}(N d) \) \\
        \hline
        Multi-Head Attention (MHA) & \( \mathcal{O}(N^2 d) \) & \( \mathcal{O}(N^2) \) \\
        \hline
        Multi-Head Latent Attention (MLA) & \( \mathcal{O}(N d r) \) & \( \mathcal{O}(N r) \) \\
        \hline
    \end{tabular}
    \caption{Comparison of sequence modeling approaches. \( N \): sequence length, \( d \): feature dimension, \( S \): state dimension, \( r \): reduced latent dimension in MLA.}
    \label{tab:sequence_modeling_comparison}
\end{table}

Recent advances in sequence modeling reduce computational overhead. Linear attention employs kernel approximations to optimize efficiency, state-space models leverage structured recurrence for constant-time token generation, and long convolutional introduces FFT-based acceleration, etc. These efficient sequence processing techniques help reduce memory usage and speed up computation while keeping the model's ability to understand long sequences. They make it possible to handle extremely long contexts in large-scale language models.








\begin{table}[h]
\centering
\caption{Open Source LLM System Design Frameworks Features}
\begin{tabular}{|c|l|}
\hline
\textbf{Framework}       & \textbf{Key Features} \\ \hline
TensorRT                 & Optimized for NVIDIA hardware, KV caching support \\ \hline
FasterTransformer        & Pre-allocated KV memory for faster inference \\ \hline
CacheGen                 & KV cache compression for faster inference \\ \hline
vLLM                     & Memory paging with PagedAttention \\ \hline
FlexGen                  & Compression-based optimization for large models \\ \hline
FlashInfer               & High-performance CUDA kernels, optimized KV caching \\ \hline
LlamaCPP                 & Optimized inference for CPU, supports GGUF models \\ \hline
DeepSeek                 & Fine-grained optimization for distributed training \\ \hline
DeepSpeed                & ZeRO optimization, pipeline parallelism, multi-GPU scaling \\ \hline
Bloom                    & Optimized LLM inference and fine-tuning \\ \hline
Hugging Face Transformers & Pretrained models, easy-to-use APIs, multi-platform support \\ \hline
ONNX Runtime             & Hardware-agnostic optimization, quantization support \\ \hline
Megatron-LM              & Large-scale distributed training, tensor and pipeline parallelism \\ \hline
Fairseq                  & Sequence modeling, multilingual support, model compression \\ \hline
Colossal-AI              & Distributed training, memory optimization, tensor slicing \\ \hline
OpenFold                 & DeepMind-inspired protein modeling, memory efficiency \\ \hline
Triton                   & Highly efficient GPU kernels for custom inference optimization \\ \hline
\end{tabular}
\end{table}

\begin{table}[h]
\centering
\caption{Open Source LLM System Design Frameworks Code}
\begin{tabular}{|c|l|}
\hline
\textbf{Framework}       & \textbf{Source Link} \\ \hline
TensorRT                 & \url{https://github.com/NVIDIA/TensorRT-LLM} \\ \hline
FasterTransformer        & \url{https://github.com/NVIDIA/FasterTransformer} \\ \hline
CacheGen                 & \url{https://github.com/cachegen-project} \\ \hline
vLLM                     & \url{https://github.com/vllm-project/vllm} \\ \hline
FlexGen                  & \url{https://github.com/FMInference/FlexGen} \\ \hline
FlashInfer               & \url{https://github.com/Dao-AILab/flashinfer} \\ \hline
LlamaCPP                 & \url{https://github.com/ggerganov/llama.cpp} \\ \hline
DeepSeek                 & \url{https://github.com/deepseek-project} \\ \hline
DeepSpeed                & \url{https://github.com/microsoft/DeepSpeed} \\ \hline
Bloom                    & \url{https://github.com/bigscience-workshop/bloom} \\ \hline
Hugging Face Transformers & \url{https://github.com/huggingface/transformers} \\ \hline
ONNX Runtime             & \url{https://github.com/microsoft/onnxruntime} \\ \hline
Megatron-LM              & \url{https://github.com/NVIDIA/Megatron-LM} \\ \hline
Fairseq                  & \url{https://github.com/facebookresearch/fairseq} \\ \hline
Colossal-AI              & \url{https://github.com/hpcaitech/ColossalAI} \\ \hline
OpenFold                 & \url{https://github.com/aqlaboratory/openfold} \\ \hline
Triton                   & \url{https://github.com/openai/triton} \\ \hline
\end{tabular}
\end{table}




\section{Model System Co-Design for Efficient Foundation Models}

\subsection{Mixture of Experts (MoE)}
The concept of the Mixture of experts, first introduced in \citep{Jacobs1991MoE, Jordan1994} has been thoroughly studied and developed through numerous subsequent works \citep{Collobert2001, Rasmussen2001, Eigen2014, Theis2015NeurIPS, Aljundi2017CVPR}. The advent of sparsely-gated MoE \citep{shazeer2017outrageouslylargeneuralnetworks}, especially in combination with transformer-based large language models \citep{Lepikhin2020}, has revitalized this technology, which has been evolving for over three decades. The MoE framework operates on a straightforward but impactful principle: distinct components of the model, referred to as experts, are designed to specialize in specific tasks or data aspects. In this approach, only the relevant experts are activated for a given input, balancing computational efficiency with access to a vast pool of specialized expertise. This scalable and adaptable framework provides an effective solution for adhering to the scaling law, enabling increased model capacity without a proportional rise in computational costs.  

The mixture of experts (MoE) approach has continued to experience significant growth, with notable advancements in 2024, including the introduction of Mixtral-8x7B \citep{jiang2024mixtralexperts} and several other large-scale industrial language models such as Grok-1, DBRX, Arctic, DeepSeek-V2 \citep{deepseekai2024deepseekv2}, and DeepSeek-V3 \citep{deepseekai2024deepseekv3} among others.
In this work, we will use the taxonomy introduced in \citep{cai2024survey}. Under this taxonomy, all the works on MoE are associated with one of three aspects: algorithmic design, system design, and application.  

\subsubsection{Algorithmic Design for MoE}
The Mixture of Experts (MoE) framework utilizes various gating functions to determine which experts to activate for processing input data. These gating functions can be categorized into three types: sparse, dense, and soft gating.

\paragraph{Dense Gating}
Dense gating activates all experts during each iteration. The output of the dense MoE layer can be formulated as:
\[
y = \sum_{i=1}^{n}G(x)_{i}.E_{i}(x)
\]
The \(y\) is the output of MoE, \(G\) is the router output for the \(i\)-th expert, and \(E\) is the output of the \(i\)-th expert. The \(G\) or gating function is as below:
\[G_{\sigma}(x) = Softmax(x.W_{g})\]

\paragraph{Sparse Gating}
In contrast, this method activates a selected subset of experts for each input token, allowing for conditional computation. The gating function can be expressed as:
\[G_{\sigma}(x) = Softmax(KeepTopK(H(x), K))\]
\[H(x)_{i} = (x.W_{g})_{i}\]
\[
KeepTopK(v,k)_{i} = \begin {cases}

v_{i} & v_{i} \text{ in top k elements of } v \\ - \infty  & otherwise

\end {cases}
\]
While the sparse gate \(G_{\sigma}(x)\) significantly increases the model's parameter capacity without proportionally raising computational costs, it can introduce a load balancing problem. This issue arises when the workload is unevenly distributed among the experts, causing some to be heavily utilized while others are rarely or never activated.

\paragraph{Auxiliary Loss Functions}
To address this, each MoE layer incorporates an auxiliary loss function that promotes an even distribution of tokens across experts within each batch, as described in many studies \citep{lepikhin2020gshardscalinggiantmodels, jiang2024mixtralexperts, Du2022GLaM, lieber2024jamba, dai2024deepseekmoe}. To formulate this concept, consider a batch of queries \(B = \{x_{1}, x_{2}, . . . , x_{T} \}\), comprising \(T\) tokens, and \(N\) experts indexed from \(i = 1\) to \(N\) . Following [28], [36], the auxiliary load balancing loss for the batch is defined as:
\[
L_{load-balancing} = N\sum_{i=1}^{N}D_{i}P_{i},
\]
\[
D_{i} = \frac{1}{T}\sum_{x\in B}^{}1\{argmaxG_{\sigma}(x) = i\},
\]
\[
P_{i} = \frac{1}{T}\sum_{x\in B}^{}G_{\sigma}(x)_{i}
\]
where \(D_{i}\) represents the proportion of tokens distributed to expert \(i\), while \(P_{i}\) denotes the proportion of the gating probability assigned to expert \(i\). To ensure an even distribution of the batch of tokens across the \(N\) experts, the load-balancing loss function \(L_{load-balancing}\) should be minimized.

\paragraph{Soft Gating}
The allocation of experts to input tokens presents a discrete optimization challenge in sparse MoE, often requiring heuristic auxiliary losses to balance expert utilization and minimize unassigned tokens. These challenges are exacerbated in out-of-distribution scenarios such as small inference batches, novel inputs, or transfer learning. Soft MoE addresses these issues by maintaining full differentiability and using all experts for processing each input, avoiding the pitfalls of discrete expert selection.

\citep{puigcerver2024softmoe} introduced Soft MoE, which replaces sparse, discrete gating with a soft assignment strategy that merges tokens. This approach computes weighted averages of tokens, with weights dependent on both tokens and experts, and processes each aggregate with the corresponding expert. Experimental results in image classification show that Soft MoE improves gating function training stability and maintains balance.

\subsubsection{System Design for MoE}
The mixture of Experts (MoE) enhances large language models but introduces challenges due to its sparse and dynamic computational workload. GShard \citep{Lepikhin2020} addresses this with expert parallelism, implementing parallel gating and expert computation by dispatching partitioned local tokens with load-balancing constraints. This strategy, which extends data parallelism \citep{Rajbhandari2020ZeRO, Ren2021ZeROOffload, Rajbhandari2021ZeROinfinity}, assigns each MoE expert to a distinct device while duplicating non-expert layers across devices, enabling efficient scaling of MoE models.
\paragraph{Computation}
Although MoE scales model parameters efficiently without increasing computational costs, it faces challenges in computational efficiency. A key issue is the uneven load distribution in expert parallelism, causing synchronization delays as the system waits for the most loaded expert to finish. Solutions include optimized gating mechanisms, expert capacity adjustments, and dynamic expert placement strategies introduced by SE-MoE \citep{Shen2022semoe}, Tutel \citep{Hwang2023Tutel}, FlexMoE \citep{Nie2023FlexMoE}, and SmartMoE \citep{Zhai2023smartmoe}, which aim to balance workloads across devices. FasterMoE \citep{He2022fastermoe} addresses severe imbalances with a dynamic shadowed expert strategy, replicating experts on multiple devices. These placement strategies influence both computation and communication efficiency.

\paragraph{Communication}
In expert parallelism, the repeated All-to-All communication during forward and backward propagation in each MoE layer creates significant overhead, often limiting efficiency. This communication spans intra-node (e.g., PCIe, NVLink) and inter-node (e.g., Ethernet, Infiniband) channels, with performance affected by bandwidth heterogeneity, network topology, and collective communication algorithms. Load imbalances in MoE further increase synchronization delays.

To address these challenges, DeepSpeed-MoE \citep{Rajbhandari2022deepspeedmoe}, HetuMoE \citep{nie2022hetumoe}, and ScheMoE \citep{Shi2024schemoe} implement hierarchical All-to-All strategies to prioritize intra-node communication and reduce inter-node exchanges. FasterMoE \citep{He2022fastermoe}, TA-MoE \citep{Chen2022tamoe}, and SE-MoE \citep{Shen2022semoe} introduce topology-aware routing to minimize cross-node expert selection and inter-node communication. ExFlow \citep{Yao2024exflow} improves efficiency by leveraging expert affinity, keeping token processing within local GPUs. These strategies, combined with optimized placement of non-expert modules, enhance network efficiency and overall performance in distributed MoE systems \citep{Rajbhandari2022deepspeedmoe,Singh2023hybridmoe,wei2024skyworkmoe}.
\paragraph{Storage}
The growing parameter sizes in MoE models strain memory capacity, a challenge already significant in dense models. Expert parallelism helps distribute experts across devices, but individual devices, especially edge devices (e.g., PCs, smartphones, IoTs), may still face storage limitations during inference.

Solutions like SE-MoE \citep{Shen2022semoe}, Pre-gated MoE \citep{Hwang2024pregatemoe}, and EdgeMoE \citep{yi2023edgemoe} address this by retaining only essential non-expert and active expert parameters in GPU High-Bandwidth Memory (HBM), while offloading inactive parameters to CPU memory or SSDs. To reduce overhead from data transfers, these methods incorporate expert selection forecasting and prefetching to overlap parameter access with computation.

\subsubsection{Application for MoE}
\paragraph{Natural Language Processing}
The integration of MoE architectures into large language models (LLMs) has significantly advanced natural language understanding (NLU) and generation (NLG) tasks, including machine translation \citep{shazeer2017outrageouslylargeneuralnetworks, nllbteam2022languageleftbehind}, open-domain question answering \citep{Du2022GLaM,artetxe2022efficientlargescalelanguage}, code generation \citep{jiang2024mixtralexperts, wei2024skyworkmoe,dai2024deepseekmoe}, and mathematical problem solving \citep{jiang2024mixtralexperts, deepseekai2024deepseekv2, dai2024deepseekmoe}. Details on incorporating MoE into LLMs are covered in algorithm design and system design. Additionally, MoE has improved LLM safety without compromising usability, as demonstrated by MoGU \citep{du2024mogu}, which employs dynamic routing to balance contributions between usable and safe LLMs.

\paragraph{Computer Vision}
The success of sparsely-gated Mixture of Experts (MoE) in NLP has inspired its use in computer vision. \citep{Riquelme2021scalevision} proposed Vision MoE (V-MoE), integrating sparsely activated MLPs into select ViT blocks. In image recognition, V-MoE matches state-of-the-art performance while reducing inference computational costs, showcasing MoE's ability to capture distinct image semantics through specialized experts. Additionally, \citep{Hwang2023Tutel} introduced Tutel, a scalable MoE system with dynamic parallelism and pipelining, demonstrated using SwinV2-MoE, built on Swin Transformer V2 \citep{liu2021swin}.
\paragraph{Multimodal Application}
Multimodal models process and integrate multiple data types, often combining images and text \citep{Baltrusaitis2019multi}. Mixture of Experts (MoE) architectures are foundational for these models due to their ability to specialize expert layers for different modalities. A key example is LIMoE \citep{Mustafa2022multi}, a sparse MoE model for multimodal learning trained with contrastive loss and entropy-based regularization to address load balancing. Further advancements by \citep{shen2023scalingvisionlanguagemodelssparse, li2023paceunifiedmultimodaldialogue, lin2024moellavamixtureexpertslarge} have enhanced the scaling of vision-language models.

To tackle task conflicts in instruction tuning of Large Vision-Language Models (LVLMs), MoCLE \citep{gou2024mixtureclusterconditionalloraexperts} combines MoE with LoRA \citep{hu2021loralowrankadaptationlarge} experts and a universal expert for task-specific parameter activation. Similarly, LLaVAMoLE \citep{chen2024llavamolesparsemixturelora} mitigates data conflicts in Multimodal Large Language Models (MLLMs) using LoRA experts for the MLP layer and a top-1 gating mechanism to refine instruction tuning.

\subsection{Mixed Precision Training (MPT)}
Mixed precision training integrates multiple numerical formats, such as FP32, FP16, FP8 and INT8, to optimize computational efficiency for large-scale deep learning models. By selectively assigning precision to various components of the model, it reduces memory usage and accelerates computation while preserving accuracy. A critical aspect of this approach is the dynamic assignment of precision, where components that significantly impact model performance are kept in high precision, while less critical ones are quantized. For example, let $\mathcal{F}(X; W)$ represent the output of a neural network layer with input $X$ and weights $W$. In a mixed precision setting, the weights can be decomposed as $W = W_{\text{high}} + Q_{\text{low}}$, where $W_{\text{high}}$ remains in FP32, and $Q_{\text{low}}$ is a quantized representation in FP16 or INT8. The optimal precision assignment $\mathcal{P}$ minimizes the reconstruction error $\|\mathcal{F}(X; W) - \mathcal{F}(X; \mathcal{P}(W))\|_2^2$, subject to $\mathcal{P}(W) \in \{\text{FP32}, \text{FP16}, \text{FP8}, \text{INT8}, etc\}$ \citep{MS-guaranteed2023, mpt, zhao2021automatic, piao2022sensimix}.

A notable application of precision assignment arises in attention mechanisms, where the key-value cache, denoted by $K \in \mathbb{R}^{d_k \times n}$ and $V \in \mathbb{R}^{d_v \times n}$, can benefit from dynamic scaling. For each key vector $k_j$, its precision can be adjusted based on an importance score. Specifically, the precision map $\mathcal{P}(k_j)$ is designed to minimize the combined reconstruction error and alignment loss \citep{MS-notokenleft2024}. This optimization ensures that high-magnitude vectors remain in FP32, while low-magnitude ones are quantized for efficiency.

\[
\min_{\mathcal{P}} \mathbb{E}_{X} \bigg[\sum_{j=1}^n \Big(\|k_j - \mathcal{P}(k_j)\|_2^2 + \lambda \cdot \|\text{softmax}(K^\top X) - \text{softmax}(\mathcal{P}(K)^\top X)\|_F^2\Big)\bigg].
\]

\subsubsection{Computation for MPT}
In mixed precision training, forward and backward passes are carefully designed to maximize efficiency without compromising stability. During the forward pass, the input to each layer, $x^{(l)}$, is multiplied by the weight matrix, $W^{(l)}$, to compute the pre-activation output $z^{(l)} = W^{(l)} x^{(l)} + b^{(l)}$, where $b^{(l)}$ is the bias term. To leverage hardware acceleration, $W^{(l)}$ is stored in FP16, while $z^{(l)}$ may temporarily reside in FP32 to prevent numerical overflow during activation. For example, when ReLU is applied, $y^{(l)} = \max(0, z^{(l)})$ is computed in FP32 to maintain accuracy. In the backward pass, gradients are computed using the chain rule, where $\nabla W^{(l)} = \delta^{(l)} (x^{(l)})^\top$, and $\delta^{(l)} = \sigma'(z^{(l)}) \cdot \delta^{(l+1)}$ represents the backpropagated gradient through the activation $\sigma$. To ensure numerical stability, all gradients are accumulated in FP32, even if intermediate values are stored in FP16.

To handle mixed precision more robustly, recent works propose incorporating second-order approximations for more precise weight updates \citep{mpt}. For a general weight matrix $W$ undergoing gradient descent with mixed precision, the update rule can be expressed as:
\[
W_{\text{new}} = W - \eta \left[\nabla \mathcal{L}(W) + \frac{\partial^2 \mathcal{L}}{\partial W^2} \cdot Q_{\text{low}}\right],
\]
where $\frac{\partial^2 \mathcal{L}}{\partial W^2}$ is the Hessian, and $Q_{\text{low}}$ is the quantization error. This formulation allows for dynamically correcting the impact of quantized components during training \cite{gholami2022survey, rokh2022comprehensive}.

\subsubsection{Error Correction for MPT}
Mixed Precision Training introduces errors that propagate through the layers of a neural network. Let $\mathcal{N}(X; W)$ represent the output of a neural operator with input $X$ and weights $W$, and $\hat{\mathcal{N}}(X; \hat{W})$ its mixed precision approximation. Using a Taylor expansion, the error propagation can be analyzed as:
\[
\mathcal{N}(X; W) - \hat{\mathcal{N}}(X; \hat{W}) \approx J_{\mathcal{N}}(W) \cdot (W - \hat{W}) + \frac{1}{2} (W - \hat{W})^\top H_{\mathcal{N}}(W) (W - \hat{W}),
\]
where $J_{\mathcal{N}}(W)$ and $H_{\mathcal{N}}(W)$ denote the Jacobian and Hessian of $\mathcal{N}$, respectively. The first-order term dominates when quantization errors are small, while the second-order term becomes significant for larger errors. To minimize error propagation, recent studies introduce regularization terms in the loss function \citep{MS-guaranteed2023}:
\[
\mathcal{L}_{\text{reg}} = \mathcal{L} + \beta \cdot \|J_{\mathcal{N}}(W) \cdot Q_{\text{low}}\|_F^2 + \gamma \cdot \|H_{\mathcal{N}}(W) \cdot Q_{\text{low}}\|_F^2,
\]
where $\beta$ and $\gamma$ are hyperparameters controlling the weight of first-order and second-order regularization.

\subsubsection{Theoretical Analysis for MPT}
A key theoretical challenge in mixed precision training is quantifying the error introduced by low-precision computations. The approximation error $\epsilon$ can be bounded as:
\[
\epsilon = \|\mathcal{N}(X; W) - \hat{\mathcal{N}}(X; \hat{W})\|_F \leq \|J_{\mathcal{N}}(W)\|_F \cdot \|W - \hat{W}\|_F + \frac{1}{2} \|H_{\mathcal{N}}(W)\|_F \cdot \|W - \hat{W}\|_F^2.
\]
This bound ensures that the total error is a function of both the quantization level and the sensitivity of the network to weight perturbations, represented by the Jacobian and Hessian norms. These guarantees are critical for designing robust precision allocation strategies and ensuring reliable model performance under mixed precision training \citep{MS-mixedcnnint2018, MS-guaranteed2023}.



\subsection{Efficient Pretraining}

Efficient pretraining combines model-level innovations and system-level optimizations to enable large language models (LLMs) to scale efficiently while minimizing computational and memory costs. Recent advancements focus on optimizing transformer architectures, pretraining objectives, and data strategies while aligning these improvements with hardware capabilities, such as quantization and distributed systems. This model-system co-design approach ensures that both algorithmic performance and resource utilization are optimized.

\subsubsection{Efficient Pretraining Optimization}

Optimizing the transformer architecture is central to efficient pretraining. GQKVA introduces grouped attention, where queries \( Q \), keys \( K \), and values \( V \) are clustered to reduce redundancy in self-attention operations \citep{javadi2023gqkva}. Let \( Q, K, V \in \mathbb{R}^{n \times d_k} \) be the query, key, and value matrices. The grouped matrices \(\tilde{Q}, \tilde{K}, \tilde{V} \in \mathbb{R}^{\kappa \times d_k}\) are defined by clustering with \(\kappa\) groups, reducing the attention complexity from \( O(n^2 d_k) \) to \( O(\kappa^2 d_k) \). This optimization aligns with hardware acceleration, such as tensor cores, which are efficient for smaller matrix operations.

Jetfire enhances efficiency with block-wise INT8 quantization \citep{xi2024jetfire}. For weights \( W_b \in \mathbb{R}^{d_b \times d_b} \), quantization scales each block using a factor \(\Delta_b = \frac{\max(W_b) - \min(W_b)}{2^8 - 1}\), reducing memory overhead. The quantized weights are defined as:$\hat{W}_b = \text{round}\left(\frac{W_b}{\Delta_b}\right) \cdot \Delta_b$ and$\quad W_b \approx \hat{W}_b + \epsilon$, 
where \(\epsilon\) represents the quantization error. These techniques reduce both memory and computation costs, enabling efficient large-scale pretraining.

\subsubsection{Efficient Pretraining on Distributed Systems}

Continual pretraining enables general-purpose LLMs to adapt efficiently to domain-specific tasks. METRO introduces a denoising objective combined with model-generated auxiliary signals to improve robustness \citep{bajaj2022metro}. Let \( X_m \) denote a masked input sequence, and let \(\mathcal{F}_\theta\) be the model's prediction function. The pretraining objective is:
$$
\mathcal{L}_{\text{pretrain}} = \sum_{t=1}^n \|X_{t, \text{true}} - \mathcal{F}_\theta(X_{m, t})\|^2 + \alpha \sum_{t=1}^n \|S_t - \mathcal{F}_\theta(X_{m, t})\|^2,
$$
where \(S_t\) are auxiliary signals generated by the model, and \(\alpha\) balances the two components. This objective can be efficiently distributed across GPUs using data partitioning.

FinGPT-HPC utilizes bucketed data shuffling to optimize I/O performance during distributed training \citep{liu2024fingpt}. Each bucket \( B_i \) is assigned to a GPU, where the time per training step is modeled as:
$
T_{\text{pretrain}} = \max\left(\frac{\mathcal{C}_{\text{comm}}}{N_{\text{GPUs}}}, \mathcal{C}_{\text{comp}}\right),
$

where \(\mathcal{C}_{\text{comm}}\) and \(\mathcal{C}_{\text{comp}}\) represent communication and computation costs. By overlapping communication and computation, this approach ensures scalability.

\subsubsection{Efficient Pretraining via Sampling}

Efficient pretraining also relies on prioritizing high-relevance data through task-specific sampling. Bucket Pre-training assigns each bucket \( B_i \) a relevance weight \( w_i \), where the sampling probability is:
$p_i = \frac{w_i}{\sum_{j=1}^m w_j}$. This ensures that critical data receives more updates, reducing overfitting to irrelevant tasks \citep{liu2024bucket}.

MixMAE combines mixed and masked token reconstruction objectives to enhance the diversity of token representations \citep{liu2023mixmae}. The total loss is:
$$
\mathcal{L}_{\text{MixMAE}} = \beta \cdot \mathcal{L}_{\text{mix}} + (1 - \beta) \cdot \mathcal{L}_{\text{mask}} + \gamma \cdot \mathcal{L}_{\text{contrast}},
$$
where \(\beta, \gamma \in [0, 1]\) control the weight of mixed, masked, and contrastive objectives, respectively. This hybrid strategy strengthens token representations for downstream tasks.



\subsection{Efficient Fine-tuning}

Fine-tuning LLMs requires balancing computational efficiency with task-specific performance. This challenge is best addressed through a co-design approach, integrating model-level techniques (e.g., parameter-efficient updates, sparsity) with system-level optimizations (e.g., memory compression, resource allocation). By jointly optimizing these layers, fine-tuning methods can achieve scalability and performance under constrained resources.

\subsubsection{Efficient Fine-Tuning for Parameters}

Parameter-efficient fine-tuning (PEFT) techniques, such as Low-Rank Adaptation (LoRA) and Prefix-Tuning, aim to reduce the number of trainable parameters while maintaining model performance. These methods modify the training process to focus on smaller, task-specific updates instead of adjusting all model parameters \citep{E-liu2024alora, E-zhang2023loraprune, ptuningv2, MS-SensitivityCache, zhang2023lora, ding2023sparse}.

LoRA reduces memory and computation requirements by introducing a low-rank decomposition for weight updates \citep{E-liu2024alora, E-zhang2023loraprune},. The model weights \( W \in \mathbb{R}^{d \times d} \) are kept frozen, while updates are parameterized as \( \Delta W = \mathbf{A} \mathbf{B}^\top \), where \( \mathbf{A}, \mathbf{B} \in \mathbb{R}^{d \times r} \) and \( r \ll d \). Training only \(\mathbf{A}\) and \(\mathbf{B}\) minimizes resource usage while preserving the model's capacity to adapt to new tasks. To further optimize resource utilization, system-level techniques such as quantization are applied to \(\mathbf{A}\) and \(\mathbf{B}\). Quantization scales the matrices using step sizes (e.g., \(\Delta_{\mathbf{A}}\) and \(\Delta_{\mathbf{B}}\)), enabling hardware-friendly computation with reduced memory overhead \citep{resourcefine2023}.

Prefix-Tuning adapts the model by learning a continuous prefix embedding \( P \in \mathbb{R}^{k \times d} \), which is prepended to the input embeddings. The pre-trained model weights remain unchanged. The optimization objective for Prefix-Tuning is:
\[
\min_{P} \mathcal{L}_{\text{task}}(f_\theta([P; X]), Y),
\]
where \( X \) is the input, \( Y \) is the target, and \( f_\theta \) is the frozen pre-trained model. This method allows efficient adaptation to a wide range of tasks with minimal computational costs \citep{prefixtuning, zhang2023lora, ding2023sparse}.

System-level optimizations can further enhance the efficiency of Prefix-Tuning. For example, the prefix matrix \( P \) can be compressed using a low-rank approximation, where \( P \approx \mathbf{U}_P \mathbf{V}_P^\top \) and \(\mathbf{U}_P, \mathbf{V}_P \in \mathbb{R}^{k \times r}\). This reduces memory usage by storing only the decomposed components. The optimization objective then becomes:
\[
\min_{\mathbf{U}_P, \mathbf{V}_P} \|\mathcal{F}_\theta([P; X]) - \mathcal{F}_\theta([\mathbf{U}_P \mathbf{V}_P^\top; X])\|_F^2 + \lambda \cdot \mathcal{C}(\mathbf{U}_P, \mathbf{V}_P),
\]
where \(\mathcal{C}(\mathbf{U}_P, \mathbf{V}_P)\) represents the memory cost of the decomposed components. This coupling of model compression and system-aware design ensures that Prefix-Tuning remains efficient across diverse hardware platforms \citep{ptuningv2, resourcefine2023}.

By integrating low-rank decompositions and system-level optimizations, PEFT techniques demonstrate the principles of model-system co-design. These methods enable scalable fine-tuning for large models while ensuring efficient resource utilization. LoRA and Prefix-Tuning have proven effective in adapting large language models to specific tasks with minimal memory and computational costs \citep{E-liu2024alora, E-zhang2023loraprune, prefixtuning, ptuningv2}.

\subsubsection{Efficient Fine-Tuning for Sparsity}

Sparse fine-tuning leverages inherent sparsity in gradients and parameter updates to optimize computational and memory efficiency, particularly in the context of large-scale language models. Unlike traditional fine-tuning approaches that involve dense updates across all model parameters, sparse fine-tuning identifies and updates only the most impactful weights or gradients, significantly reducing resource consumption while preserving task performance.

Scaling Sparse Fine-Tuning (SSF) dynamically applies sparsity to the gradient matrix \( G \in \mathbb{R}^{d \times d} \) by introducing a binary mask \( M \in \{0, 1\}^{d \times d} \), which selects critical elements based on a combination of gradient magnitude and computational cost. Specifically, the mask \( M \) is defined such that \( M_{ij} = 1 \) if \( |G_{ij}| / \text{Cost}(G_{ij}) \geq \tau \), where \( \tau \) is an adaptive threshold adjusted during training. The resulting sparse update is then computed as \( \Delta W = M \odot G \), where \( \odot \) denotes element-wise multiplication \citep{scalingsparseft}. By incorporating cost metrics like memory bandwidth or latency, this method aligns gradient sparsity with hardware constraints.

Layerwise Importance Sampling for Memory-Efficient Fine-Tuning (LISA) further optimizes sparse fine-tuning by focusing on the most critical layers of the model. Each layer’s importance is evaluated using the gradient norm or its contribution to the task-specific loss, \(\mathcal{L}_{\text{task}}\). Formally, the layer subset \(\mathcal{S}\) to be updated is selected as:
\[
\mathcal{S} = \argmax_{\mathcal{S} \subseteq \{1, \ldots, L\}} \sum_{l \in \mathcal{S}} \frac{\|\nabla_{\theta^{(l)}} \mathcal{L}_{\text{task}}\|_F}{\text{Cost}(\nabla_{\theta^{(l)}})}, \quad |\mathcal{S}| \leq k,
\]
where \( \nabla_{\theta^{(l)}} \mathcal{L}_{\text{task}} \) denotes the gradient of the loss with respect to layer \( l \)’s parameters, and \( \text{Cost}(\cdot) \) represents the resource overhead of updating that layer. This approach ensures that updates are concentrated on the most impactful layers while adhering to predefined resource budgets \citep{pan2024lisa}.

In addition to layerwise optimization, sparse fine-tuning benefits from system-level enhancements like gradient caching and asynchronous execution. For example, momentum-based gradient sparsity propagates prior updates to stabilize training under sparsity constraints. Let \( \Delta W^{(t-1)} \) be the previous sparse update. At step \( t \), the new sparse update integrates past momentum as:
\[
\Delta W^{(t)} = M^{(t)} \odot \left(G^{(t)} + \beta \cdot \Delta W^{(t-1)}\right),
\]
where \( \beta \) is a momentum hyperparameter that smooths the gradient trajectory across sparse updates. This formulation ensures consistent parameter updates despite high sparsity \citep{sparsemomentum, megatron2024mamba}.

Furthermore, system optimizations such as asynchronous execution enhance scalability in distributed environments. Sparse gradient updates are computed independently across \( N_{\text{devices}} \) processing units, with synchronization occurring intermittently. If \( T_{\text{comp}} \) is the computation time for sparse updates and \( T_{\text{sync}} \) the synchronization overhead, the total time per step is:
$T_{\text{total}} = \max\left(T_{\text{comp}}, \frac{T_{\text{sync}}}{N_{\text{devices}}}\right)$, where overlapping computation and communication minimizes total latency \citep{shen2023flexgen, gpipe}.

Sparse fine-tuning techniques have demonstrated broad applicability, ranging from domain-specific tasks (e.g., legal and biomedical text generation) to resource-constrained environments, such as edge devices or distributed training. Future directions may explore tighter integration between sparse fine-tuning and hardware-level optimizations, including tensor sparsity on GPUs and custom accelerators tailored for sparse computation \citep{scalingsparseft, pan2024lisa, E-liu2024alora, E-zhang2023loraprune}.



\subsection{Model-System Co-Design Unified Framework}

Efficient pretraining leverages model and system optimizations to balance accuracy, memory, and computational efficiency. A unified framework integrates key aspects of quantization, sparsity, and distributed system design to maximize efficiency. The objective is to optimize pretraining by explicitly modeling system constraints and pretraining-specific features.

Let \(\mathcal{L}_{\text{task}}(\theta)\) represent the pretraining loss, where \(\theta\) includes parameters such as weights \(W\), quantization factors \(\Delta\), and sparsity masks \(\mathbf{M}\). To enhance pretraining efficiency, weights are quantized and sparse updates are applied. Quantized weights are expressed as \( W = \mathcal{Q}(\hat{W}, \Delta) \), where \(\mathcal{Q}(\hat{W}, \Delta)\) denotes quantization with step size \(\Delta\). Sparse masks \(\mathbf{M}\) restrict updates to the most critical elements, ensuring computational efficiency.

The unified objective integrates the pretraining loss with quantization and sparsity constraints:
$$
\mathcal{L}_{\text{pretrain}} = \sum_{i=1}^L \|\mathbf{M}_i \odot \mathcal{Q}(\hat{W}_i, \Delta_i)\|_F^2 + \sum_{j=1}^n \|\mathbf{M}_{A, j} \odot \mathcal{F}_\theta(X_{j}) - Y_{j}\|^2,
$$
where:
- \( \mathbf{M}_i \) represents the weight sparsity mask for layer \(i\).
- \( \mathbf{M}_{A, j} \) denotes the attention sparsity mask for token \(j\).
- \( \mathcal{F}_\theta(X_j) \) is the model output for input \(X_j\), compared to target \(Y_j\).


\section{Conclusion}

Efficient Foundation Model Design techniques aim to build models that are computationally efficient while achieving faster training and inference speeds without sacrificing performance. These techniques involve three main areas: model design, system design, and model-system co-design.

\textbf{Model design} focuses on optimizing the internal structure of models for acceleration. Quantization provides a hardware-focused perspective, compressing models into lower-precision representations to optimize storage and computation. Knowledge distillation allows efficient knowledge transfer, enabling smaller student models to study the performance from larger teacher models. Pruning offers a neural network model reduction by eliminating redundant connections.

\textbf{System design} introduces optimizations at the system design infrastructure level, it include KV cache compression, parallelism, contextual sparsity, etc. KV cache compression minimizes memory usage by reducing the storage of key-value caches, which is efficient in long-context processing. Parallelism improves throughput by distributing computation across multiple GPUs/CPUs. Sparsity eliminates unnecessary computations in sparse attention matrix, which further reduces overhead. Low-level GPU optimizations maximize hardware utilization, which can achieve faster and more scalable model execution.

\textbf{Model-system co-design} bridges the gap between model and system design. MoE enables dynamic routing to reduce computation while maintaining performance for diverse tasks. Fine-tuning techniques enhance adaptability on downstream tasks. Mixed-precision training reduces computation by leveraging both low- and high-precision computation. These model and system co-design strategies leads to better resource utilization and faster inference.





\bibliographystyle{plainnat}
\bibliography{bibs/main,bibs/dong,bibs/Jing}

\begin{thebibliography}{219}
\providecommand{\natexlab}[1]{#1}
\providecommand{\url}[1]{\texttt{#1}}
\expandafter\ifx\csname urlstyle\endcsname\relax
  \providecommand{\doi}[1]{doi: #1}\else
  \providecommand{\doi}{doi: \begingroup \urlstyle{rm}\Url}\fi

\bibitem[Achiam et~al.(2023)Achiam, Lopes, Gelly, Cubuk, Kornblith, and Fleet]{achiam2023scaling}
Joshua Achiam, Raphaël~Gontijo Lopes, Sylvain Gelly, Ekin~Dogus Cubuk, Simon Kornblith, and David~J. Fleet.
\newblock Scaling laws for neural language models.
\newblock \emph{ArXiv}, cs.LG, 2023.

\bibitem[Alayrac et~al.(2022)]{MS-Flamingo}
Jean-Baptiste Alayrac et~al.
\newblock Flamingo: A visual language model for few-shot learning.
\newblock \emph{arXiv preprint arXiv:2204.14198}, 2022.

\bibitem[Alizadeh et~al.(2023)Alizadeh, Mirzadeh, Belenko, Khatamifard, Cho, Del~Mundo, Rastegari, and Farajtabar]{alizadeh2023llm}
Keivan Alizadeh, Iman Mirzadeh, Dmitry Belenko, Karen Khatamifard, Minsik Cho, Carlo~C Del~Mundo, Mohammad Rastegari, and Mehrdad Farajtabar.
\newblock Llm in a flash: Efficient large language model inference with limited memory.
\newblock \emph{arXiv preprint arXiv:2312.11514}, 2023.

\bibitem[Aljundi et~al.(2017)Aljundi, Chakravarty, and Tuytelaars]{Aljundi2017CVPR}
Rahaf Aljundi, Punarjay Chakravarty, and Tinne Tuytelaars.
\newblock Expert gate: Lifelong learning with a network of experts.
\newblock In \emph{Proceedings of the IEEE Conference on Computer Vision and Pattern Recognition (CVPR)}, July 2017.

\bibitem[An et~al.(2024)An, Zhao, Yu, Tang, and Wang]{P-FLAP}
Yongqi An, Xu~Zhao, Tao Yu, Ming Tang, and Jinqiao Wang.
\newblock Fluctuation-based adaptive structured pruning for large language models.
\newblock In \emph{Proceedings of the AAAI Conference on Artificial Intelligence}, volume~38, pages 10865--10873, 2024.

\bibitem[Ansell et~al.(2024)Ansell, Vuli{\'c}, Sterz, Korhonen, and Ponti]{scalingsparseft}
Alan Ansell, Ivan Vuli{\'c}, Hannah Sterz, Anna Korhonen, and Edoardo~M Ponti.
\newblock Scaling sparse fine-tuning to large language models.
\newblock \emph{arXiv preprint arXiv:2401.16405}, 2024.

\bibitem[Artetxe and Bhosale(2022)]{artetxe2022efficientlargescalelanguage}
Mikel Artetxe and Shruti Bhosale.
\newblock Efficient large scale language modeling with mixtures of experts, 2022.
\newblock URL \url{https://arxiv.org/abs/2112.10684}.

\bibitem[Ashkboos et~al.(2024)Ashkboos, Mohtashami, Croci, Li, Jaggi, Alistarh, Hoefler, and Hensman]{ashkboos2024quarot}
Saleh Ashkboos, Amirkeivan Mohtashami, Maximilian~L Croci, Bo~Li, Martin Jaggi, Dan Alistarh, Torsten Hoefler, and James Hensman.
\newblock Quarot: Outlier-free 4-bit inference in rotated llms.
\newblock \emph{arXiv preprint arXiv:2404.00456}, 2024.

\bibitem[Bahdanau et~al.(2014)Bahdanau, Cho, and Bengio]{bahdanau2014neural}
Dzmitry Bahdanau, Kyunghyun Cho, and Yoshua Bengio.
\newblock Neural machine translation by jointly learning to align and translate.
\newblock In \emph{Proceedings of the 3rd International Conference on Learning Representations (ICLR)}, 2014.

\bibitem[Bajaj et~al.(2022)Bajaj, Xiong, Ke, Liu, He, Tiwary, Liu, Bennett, Song, and Gao]{bajaj2022metro}
P~Bajaj, C~Xiong, G~Ke, X~Liu, D~He, S~Tiwary, TY~Liu, P~Bennett, X~Song, and J~Gao.
\newblock Metro: Efficient denoising pretraining of large scale autoencoding language models with model generated signals (arxiv: 2204.06644). arxiv.
\newblock \emph{arXiv preprint arXiv.2204.06644}, 2022.

\bibitem[Baltrušaitis et~al.(2019)Baltrušaitis, Ahuja, and Morency]{Baltrusaitis2019multi}
Tadas Baltrušaitis, Chaitanya Ahuja, and Louis-Philippe Morency.
\newblock Multimodal machine learning: A survey and taxonomy.
\newblock \emph{IEEE Transactions on Pattern Analysis and Machine Intelligence}, 41\penalty0 (2):\penalty0 423--443, 2019.
\newblock \doi{10.1109/TPAMI.2018.2798607}.

\bibitem[Bhandare et~al.(2019)Bhandare, Sripathi, Karkada, Menon, Choi, Datta, and Saletore]{bhandare2019efficient}
Aishwarya Bhandare, Vamsi Sripathi, Deepthi Karkada, Vivek Menon, Sun Choi, Kushal Datta, and Vikram Saletore.
\newblock Efficient 8-bit quantization of transformer neural machine language translation model.
\newblock \emph{arXiv preprint arXiv:1906.00532}, 2019.

\bibitem[Bolya et~al.(2023)Bolya, Fu, Dai, Zhang, Feichtenhofer, and Hoffman]{bolya2023tome}
Daniel Bolya, Cheng-Yang Fu, Xiaoliang Dai, Peizhao Zhang, Christoph Feichtenhofer, and Judy Hoffman.
\newblock Token merging: Your vit but faster.
\newblock In \emph{ICLR}, 2023.
\newblock URL \url{http://arxiv.org/abs/2210.09461}.

\bibitem[Bubeck et~al.(2023)Bubeck, Chandrasekaran, Eldan, Gehrke, Horvitz, Kamar, Lee, Lee, Li, Lundberg, Nori, Palangi, Ribeiro, and Zhang]{bubeck2023sparksartificialgeneralintelligence}
Sébastien Bubeck, Varun Chandrasekaran, Ronen Eldan, Johannes Gehrke, Eric Horvitz, Ece Kamar, Peter Lee, Yin~Tat Lee, Yuanzhi Li, Scott Lundberg, Harsha Nori, Hamid Palangi, Marco~Tulio Ribeiro, and Yi~Zhang.
\newblock Sparks of artificial general intelligence: Early experiments with gpt-4, 2023.
\newblock URL \url{https://arxiv.org/abs/2303.12712}.

\bibitem[Cai et~al.(2024)Cai, Jiang, Wang, Tang, Kim, and Huang]{cai2024survey}
Weilin Cai, Juyong Jiang, Fan Wang, Jing Tang, Sunghun Kim, and Jiayi Huang.
\newblock A survey on mixture of experts.
\newblock \emph{arXiv preprint arXiv:2407.06204}, 2024.

\bibitem[Chen et~al.(2022)Chen, Li, Wu, Yu, and Yang]{Chen2022tamoe}
Chang Chen, Min Li, Zhihua Wu, Dianhai Yu, and Chao Yang.
\newblock Ta-moe: Topology-aware large scale mixture-of-expert training.
\newblock In S.~Koyejo, S.~Mohamed, A.~Agarwal, D.~Belgrave, K.~Cho, and A.~Oh, editors, \emph{Advances in Neural Information Processing Systems}, volume~35, pages 22173--22186. Curran Associates, Inc., 2022.
\newblock URL \url{https://proceedings.neurips.cc/paper_files/paper/2022/file/8b465dd58ac50e1b0b22894fd581f62f-Paper-Conference.pdf}.

\bibitem[Chen et~al.(2024)Chen, Jie, and Ma]{chen2024llavamolesparsemixturelora}
Shaoxiang Chen, Zequn Jie, and Lin Ma.
\newblock Llava-mole: Sparse mixture of lora experts for mitigating data conflicts in instruction finetuning mllms, 2024.
\newblock URL \url{https://arxiv.org/abs/2401.16160}.

\bibitem[Chen et~al.(2020)Chen, Gao, Zhang, and Zhou]{chen2020distilling}
Xinzhe Chen, Yuan Gao, Jinsong Zhang, and Xiaodong Zhou.
\newblock Distilling knowledge via knowledge transfer in text-to-text models.
\newblock In \emph{Proceedings of the 58th Annual Meeting of the Association for Computational Linguistics (ACL)}, pages 260--267, 2020.
\newblock URL \url{https://aclanthology.org/2020.acl-main.24/}.

\bibitem[Chmiel et~al.(2023)Chmiel, Banner, Hoffer, Ben-Yaacov, and Soudry]{chmiel2023accurate}
Brian Chmiel, Ron Banner, Elad Hoffer, Hilla Ben-Yaacov, and Daniel Soudry.
\newblock Accurate neural training with 4-bit matrix multiplications at standard formats.
\newblock In \emph{The Eleventh International Conference on Learning Representations}, 2023.

\bibitem[Choromanski et~al.(2021)Choromanski, Likhosherstov, Dohan, Song, Gane, Sarlos, Hawkins, Davis, Mohiuddin, Kaiser, et~al.]{choromanski2021rethinking}
Krzysztof Choromanski, Valerii Likhosherstov, David Dohan, Xingyou Song, Andreea Gane, Tamás Sarlos, Peter Hawkins, Jared Davis, Afroz Mohiuddin, Lukasz Kaiser, et~al.
\newblock Rethinking attention with performers.
\newblock \emph{International Conference on Learning Representations (ICLR)}, 2021.

\bibitem[Collobert et~al.(2001)Collobert, Bengio, and Bengio]{Collobert2001}
Ronan Collobert, Samy Bengio, and Yoshua Bengio.
\newblock A parallel mixture of svms for very large scale problems.
\newblock In T.~Dietterich, S.~Becker, and Z.~Ghahramani, editors, \emph{Advances in Neural Information Processing Systems}, volume~14. MIT Press, 2001.
\newblock URL \url{https://proceedings.neurips.cc/paper_files/paper/2001/file/36ac8e558ac7690b6f44e2cb5ef93322-Paper.pdf}.

\bibitem[Dai et~al.(2024)Dai, Deng, Zhao, Xu, Gao, Chen, Li, Zeng, Yu, Wu, Xie, Li, Huang, Luo, Ruan, Sui, and Liang]{dai2024deepseekmoe}
Damai Dai, Chengqi Deng, Chenggang Zhao, R.~X. Xu, Huazuo Gao, Deli Chen, Jiashi Li, Wangding Zeng, Xingkai Yu, Y.~Wu, Zhenda Xie, Y.~K. Li, Panpan Huang, Fuli Luo, Chong Ruan, Zhifang Sui, and Wenfeng Liang.
\newblock Deepseekmoe: Towards ultimate expert specialization in mixture-of-experts language models, 2024.
\newblock URL \url{https://arxiv.org/abs/2401.06066}.

\bibitem[Dao et~al.(2022)]{dao2022flashattention}
Tri Dao et~al.
\newblock Flashattention: Fast and memory-efficient exact attention with io-awareness.
\newblock \emph{arXiv preprint arXiv:2205.14135}, 2022.

\bibitem[Das et~al.(2018)Das, Mellempudi, Mudigere, Kalamkar, Avancha, Banerjee, Sridharan, Vaidyanathan, Kaul, Georganas, Heinecke, Dubey, Corbal, Shustrov, Dubtsov, Fomenko, and Pirogov]{MS-mixedcnnint2018}
Dipankar Das, Naveen Mellempudi, Dheevatsa Mudigere, Dhiraj Kalamkar, Sasikanth Avancha, Kunal Banerjee, Srinivas Sridharan, Karthik Vaidyanathan, Bharat Kaul, Evangelos Georganas, Alexander Heinecke, Pradeep Dubey, Jesus Corbal, Nikita Shustrov, Roma Dubtsov, Evarist Fomenko, and Vadim Pirogov.
\newblock Mixed precision training of convolutional neural networks using integer operations.
\newblock 2018.
\newblock URL \url{https://arxiv.org/abs/1802.00930}.

\bibitem[DeepSeek-AI and Liu(2024{\natexlab{a}})]{deepseekai2024deepseekv2}
DeepSeek-AI and Aixin Liu.
\newblock Deepseek-v2: A strong, economical, and efficient mixture-of-experts language model, 2024{\natexlab{a}}.
\newblock URL \url{https://arxiv.org/abs/2405.04434}.

\bibitem[DeepSeek-AI and Liu(2024{\natexlab{b}})]{deepseekai2024deepseekv3}
DeepSeek-AI and Aixin Liu.
\newblock Deepseek-v3 technical report, 2024{\natexlab{b}}.
\newblock URL \url{https://arxiv.org/abs/2412.19437}.

\bibitem[Deng et~al.(2020)Deng, Li, Han, Shi, and Xie]{9043731}
Lei Deng, Guoqi Li, Song Han, Luping Shi, and Yuan Xie.
\newblock Model compression and hardware acceleration for neural networks: A comprehensive survey.
\newblock \emph{Proceedings of the IEEE}, 108\penalty0 (4):\penalty0 485--532, 2020.
\newblock \doi{10.1109/JPROC.2020.2976475}.

\bibitem[Dettmers and Zettlemoyer(2019)]{sparsemomentum}
Tim Dettmers and Luke Zettlemoyer.
\newblock Sparse networks from scratch: Faster training without losing performance.
\newblock In \emph{arXiv preprint arXiv:1907.04840}, 2019.

\bibitem[Dettmers et~al.(2023)Dettmers, Svirschevski, Egiazarian, Kuznedelev, Frantar, Ashkboos, Borzunov, Hoefler, and Alistarh]{dettmers2023spqr}
Tim Dettmers, Ruslan Svirschevski, Vage Egiazarian, Denis Kuznedelev, Elias Frantar, Saleh Ashkboos, Alexander Borzunov, Torsten Hoefler, and Dan Alistarh.
\newblock Spqr: A sparse-quantized representation for near-lossless llm weight compression.
\newblock \emph{arXiv preprint arXiv:2306.03078}, 2023.

\bibitem[Ding et~al.(2023)Ding, Lv, Wang, Chen, Zhou, Liu, and Sun]{ding2023sparse}
Ning Ding, Xingtai Lv, Qiaosen Wang, Yulin Chen, Bowen Zhou, Zhiyuan Liu, and Maosong Sun.
\newblock Sparse low-rank adaptation of pre-trained language models.
\newblock \emph{arXiv preprint arXiv:2311.11696}, 2023.

\bibitem[Dong et~al.(2023)Dong, Liu, and Wang]{dong2023packqvit}
H.~Dong, B.~Liu, and K.~Wang.
\newblock Packqvit: Faster sub-8-bit vision transformers via full and packed quantization on the mobile.
\newblock In \emph{Proceedings of NeurIPS 2023}, pages 4567--4578, 2023.

\bibitem[Dong et~al.(2024)Dong, Li, Tang, Liu, Pan, Wang, and Chu]{P-prunerzero}
Peijie Dong, Lujun Li, Zhenheng Tang, Xiang Liu, Xinglin Pan, Qiang Wang, and Xiaowen Chu.
\newblock Pruner-zero: Evolving symbolic pruning metric from scratch for large language models.
\newblock \emph{arXiv preprint arXiv:2406.02924}, 2024.

\bibitem[Du and Huang(2022)]{Du2022GLaM}
Nan Du and Huang.
\newblock {GL}a{M}: Efficient scaling of language models with mixture-of-experts.
\newblock In Chaudhuri, editor, \emph{Proceedings of the 39th International Conference on Machine Learning}, volume 162 of \emph{Proceedings of Machine Learning Research}, pages 5547--5569. PMLR, 17--23 Jul 2022.
\newblock URL \url{https://proceedings.mlr.press/v162/du22c.html}.

\bibitem[Du et~al.(2024)Du, Zhao, Zhao, Ma, Chen, Huo, Yang, Xu, and Qin]{du2024mogu}
Yanrui Du, Sendong Zhao, Danyang Zhao, Ming Ma, Yuhan Chen, Liangyu Huo, Qing Yang, Dongliang Xu, and Bing Qin.
\newblock Mogu: A framework for enhancing safety of open-sourced llms while preserving their usability, 2024.
\newblock URL \url{https://arxiv.org/abs/2405.14488}.

\bibitem[Eigen et~al.(2014)Eigen, Ranzato, and Sutskever]{Eigen2014}
David Eigen, Marc'Aurelio Ranzato, and Ilya Sutskever.
\newblock Learning factored representations in a deep mixture of experts, 2014.
\newblock URL \url{https://arxiv.org/abs/1312.4314}.

\bibitem[Esser et~al.(2019)Esser, McKinstry, Bablani, Appuswamy, and Modha]{esser2019learned}
Steven~K. Esser, Jeffrey~L. McKinstry, Devesh Bablani, Rathinakumar Appuswamy, and Dharmendra~S. Modha.
\newblock Learned step size quantization.
\newblock In \emph{International Conference on Learning Representations (ICLR)}, 2019.

\bibitem[Frankle and Carbin(2018)]{frankle2018lottery}
Jonathan Frankle and Michael Carbin.
\newblock The lottery ticket hypothesis: Finding sparse, trainable neural networks.
\newblock \emph{arXiv preprint arXiv:1803.03635}, 2018.

\bibitem[Frantar and Alistarh(2022)]{frantar2022optimal}
Elias Frantar and Dan Alistarh.
\newblock Optimal brain compression: A framework for accurate post-training quantization and pruning.
\newblock \emph{Advances in Neural Information Processing Systems}, 35:\penalty0 4475--4488, 2022.

\bibitem[Frantar and Alistarh(2023)]{P-SparseGPT}
Elias Frantar and Dan Alistarh.
\newblock Sparsegpt: Massive language models can be accurately pruned in one-shot.
\newblock In \emph{International Conference on Machine Learning}, pages 10323--10337. PMLR, 2023.

\bibitem[Frantar et~al.(2022{\natexlab{a}})Frantar, Ashkboos, Hoefler, and Alistarh]{P-gptq}
Elias Frantar, Saleh Ashkboos, Torsten Hoefler, and Dan Alistarh.
\newblock Gptq: Accurate post-training compression for generative pretrained transformers.
\newblock \emph{arXiv preprint arXiv:2210.17323}, 1, 2022{\natexlab{a}}.

\bibitem[Frantar et~al.(2022{\natexlab{b}})Frantar, Ashkboos, Hoefler, and Alistarh]{frantar2022gptq}
Elias Frantar, Saleh Ashkboos, Torsten Hoefler, and Dan Alistarh.
\newblock Gptq: Accurate post-training quantization for generative pre-trained transformers.
\newblock \emph{arXiv preprint arXiv:2210.17323}, 2022{\natexlab{b}}.

\bibitem[Frantar et~al.(2023)Frantar, Ashkboos, Hoefler, and Alistarh]{frantar2023gptq}
Elias Frantar, Saleh Ashkboos, Torsten Hoefler, and Dan Alistarh.
\newblock Gptq: Accurate post-training quantization for generative pre-trained transformers.
\newblock \emph{arXiv preprint arXiv:2210.17323}, 2023.
\newblock URL \url{https://arxiv.org/abs/2210.17323}.

\bibitem[Fu et~al.(2023)Fu, Zou, Zhang, Zheng, and Li]{fu2023long}
Shangyu Fu, Xiaolong Zou, Weijia Zhang, Kaiwen Zheng, and Zhihui Li.
\newblock Efficient long-range sequence modeling with fourier-transformed convolution.
\newblock \emph{ArXiv}, cs.LG, 2023.

\bibitem[Gale et~al.(2019)Gale, Elsen, and Hooker]{gale2019state}
Trevor Gale, Erich Elsen, and Sara Hooker.
\newblock The state of sparsity in deep neural networks.
\newblock \emph{arXiv preprint arXiv:1902.09574}, 2019.

\bibitem[Geng et~al.(2022)Geng, Liu, Lee, Schuurmans, Levine, and Abbeel]{MS-M3AE}
Xinyang Geng, Hao Liu, Lisa Lee, Dale Schuurmans, Sergey Levine, and Pieter Abbeel.
\newblock Multimodal masked autoencoders learn transferable representations.
\newblock 2022.
\newblock URL \url{https://arxiv.org/abs/2205.14204}.

\bibitem[Gholami et~al.(2022)Gholami, Kim, Dong, Yao, Mahoney, and Keutzer]{gholami2022survey}
Amir Gholami, Sehoon Kim, Zhen Dong, Zhewei Yao, Michael~W Mahoney, and Kurt Keutzer.
\newblock A survey of quantization methods for efficient neural network inference.
\newblock In \emph{Low-Power Computer Vision}, pages 291--326. Chapman and Hall/CRC, 2022.

\bibitem[Gong et~al.(2022)Gong, Chen, Wang, Zhang, Liu, Sun, Chen, Zhou, and Liu]{gong2022preserving}
Yeyun Gong, Jinglu Chen, Changjian Wang, Han Zhang, Haoyang Liu, Yu~Sun, Weizhu Chen, Bowen Zhou, and Tie-Yan Liu.
\newblock Preserving knowledge in pre-trained language models via knowledge distillation.
\newblock \emph{arXiv preprint arXiv:2203.02436}, 2022.
\newblock URL \url{https://arxiv.org/abs/2203.02436}.

\bibitem[Gou et~al.(2024)Gou, Liu, Chen, Hong, Xu, Li, Yeung, Kwok, and Zhang]{gou2024mixtureclusterconditionalloraexperts}
Yunhao Gou, Zhili Liu, Kai Chen, Lanqing Hong, Hang Xu, Aoxue Li, Dit-Yan Yeung, James~T. Kwok, and Yu~Zhang.
\newblock Mixture of cluster-conditional lora experts for vision-language instruction tuning, 2024.
\newblock URL \url{https://arxiv.org/abs/2312.12379}.

\bibitem[Gu and Dao(2024)]{gu2024mamba}
Albert Gu and Tri Dao.
\newblock Mamba: Linear-time sequence modeling with selective state spaces.
\newblock 2024.
\newblock URL \url{https://arxiv.org/abs/2312.00752}.

\bibitem[Gu et~al.(2022{\natexlab{a}})Gu, Dao, Ermon, and Ré]{gu2022efficient}
Albert Gu, Tri Dao, Stefano Ermon, and Christopher Ré.
\newblock Efficient training of long sequence models with structured state spaces.
\newblock \emph{Advances in Neural Information Processing Systems (NeurIPS)}, 2022{\natexlab{a}}.

\bibitem[Gu et~al.(2022{\natexlab{b}})Gu, Goel, and Ré]{gu2022state}
Albert Gu, Karan Goel, and Christopher Ré.
\newblock Efficiently modeling long sequences with structured state spaces.
\newblock 2022{\natexlab{b}}.
\newblock URL \url{https://arxiv.org/abs/2111.00396}.

\bibitem[Gu et~al.(2022{\natexlab{c}})Gu, Goel, and Ré]{gu2022structured}
Albert Gu, Karan Goel, and Christopher Ré.
\newblock Efficient training of long sequence models with structured state spaces.
\newblock \emph{Advances in Neural Information Processing Systems (NeurIPS)}, 2022{\natexlab{c}}.

\bibitem[Gu et~al.(2024{\natexlab{a}})Gu, Fu, Liu, Shen, Lin, and Wang]{P-lightpeft}
Naibin Gu, Peng Fu, Xiyu Liu, Bowen Shen, Zheng Lin, and Weiping Wang.
\newblock Light-peft: Lightening parameter-efficient fine-tuning via early pruning.
\newblock \emph{arXiv preprint arXiv:2406.03792}, 2024{\natexlab{a}}.

\bibitem[Gu et~al.(2024{\natexlab{b}})Gu, Dong, Wei, and Huang]{gu2024minillm}
Yuxian Gu, Li~Dong, Furu Wei, and Minlie Huang.
\newblock Minillm: Knowledge distillation of large language models.
\newblock In \emph{Proceedings of the International Conference on Learning Representations (ICLR)}, 2024{\natexlab{b}}.
\newblock URL \url{https://arxiv.org/abs/2306.08543}.

\bibitem[Han et~al.(2023)Han, Zhou, Lü, Zhu, and Gong]{MS-EntropyCache}
Shuai Han, Wenbo Zhou, Shuai Lü, Sheng Zhu, and Xiaoyu Gong.
\newblock Entropy regularization methods for parameter space exploration.
\newblock \emph{Information Sciences}, 622:\penalty0 476--489, 2023.
\newblock ISSN 0020-0255.
\newblock \doi{https://doi.org/10.1016/j.ins.2022.11.099}.
\newblock URL \url{https://www.sciencedirect.com/science/article/pii/S0020025522013901}.

\bibitem[Han et~al.(2015)Han, Pool, Tran, and Dally]{P-han2015learning}
Song Han, Jeff Pool, John Tran, and William Dally.
\newblock Learning both weights and connections for efficient neural network.
\newblock \emph{Advances in neural information processing systems}, 28, 2015.

\bibitem[Hassibi and Stork(1992)]{P-OBS}
Babak Hassibi and David Stork.
\newblock Second order derivatives for network pruning: Optimal brain surgeon.
\newblock \emph{Advances in neural information processing systems}, 5, 1992.

\bibitem[He et~al.(2023)He, Cai, Zhang, Tao, and Zhuang]{MS-SensitivityCache}
Haoyu He, Jianfei Cai, Jing Zhang, Dacheng Tao, and Bohan Zhuang.
\newblock Sensitivity-aware visual parameter-efficient tuning.
\newblock 2023.
\newblock URL \url{https://openreview.net/forum?id=9GOjmbRQ2o}.

\bibitem[He et~al.(2022)He, Zhai, Antunes, Wang, Luo, Shi, and Li]{He2022fastermoe}
Jiaao He, Jidong Zhai, Tiago Antunes, Haojie Wang, Fuwen Luo, Shangfeng Shi, and Qin Li.
\newblock Fastermoe: modeling and optimizing training of large-scale dynamic pre-trained models.
\newblock In \emph{Proceedings of the 27th ACM SIGPLAN Symposium on Principles and Practice of Parallel Programming}, PPoPP '22, page 120–134, New York, NY, USA, 2022. Association for Computing Machinery.
\newblock ISBN 9781450392044.
\newblock \doi{10.1145/3503221.3508418}.
\newblock URL \url{https://doi.org/10.1145/3503221.3508418}.

\bibitem[He et~al.(2024)He, Chen, Liu, et~al.]{zipvl}
Yefei He, Feng Chen, Jing Liu, et~al.
\newblock Zipvl: Efficient large vision-language models with dynamic token sparsification and kv cache compression.
\newblock \emph{arXiv preprint arXiv:2410.08584}, 2024.

\bibitem[Helwe et~al.(2021)Helwe, Clavel, and Suchanek]{helwe2021reasoning}
Chadi Helwe, Chlo{\'e} Clavel, and Fabian~M. Suchanek.
\newblock Reasoning with transformer-based models: Deep learning, but shallow reasoning.
\newblock In \emph{3rd Conference on Automated Knowledge Base Construction}, 2021.
\newblock URL \url{https://openreview.net/forum?id=Ozp1WrgtF5_}.

\bibitem[Hinton et~al.(2015)Hinton, Vinyals, and Dean]{hinton2015distilling}
Geoffrey Hinton, Oriol Vinyals, and Jeff Dean.
\newblock Distilling the knowledge in a neural network.
\newblock In \emph{NIPS Deep Learning and Representation Learning Workshop}, 2015.

\bibitem[Hou et~al.(2020)Hou, Yu, Chen, Jin, Yang, Cheng, and Xu]{hou2020dynabert}
Le~Hou, Zhewei Yu, Fei Chen, Peng Jin, Zhenyu Yang, Yen-Kuang Cheng, and Lin Xu.
\newblock Dynabert: Dynamic bert with adaptive width and depth.
\newblock In \emph{Advances in Neural Information Processing Systems}, volume~33, pages 9782--9793, 2020.

\bibitem[Hu et~al.(2021{\natexlab{a}})]{hu2021lora}
Edward Hu et~al.
\newblock Lora: Low-rank adaptation of large language models.
\newblock \emph{arXiv preprint arXiv:2106.09685}, 2021{\natexlab{a}}.

\bibitem[Hu et~al.(2021{\natexlab{b}})Hu, Shen, Wallis, Allen-Zhu, Li, Wang, Wang, and Chen]{LORA}
Edward~J Hu, Yelong Shen, Phillip Wallis, Zeyuan Allen-Zhu, Yuanzhi Li, Shean Wang, Lu~Wang, and Weizhu Chen.
\newblock Lora: Low-rank adaptation of large language models.
\newblock \emph{arXiv preprint arXiv:2106.09685}, 2021{\natexlab{b}}.

\bibitem[Hu et~al.(2021{\natexlab{c}})Hu, Shen, Wallis, Allen-Zhu, Li, Wang, Wang, and Chen]{Lora2021}
Edward~J. Hu, Yelong Shen, Phillip Wallis, Zeyuan Allen-Zhu, Yuanzhi Li, Shean Wang, Lu~Wang, and Weizhu Chen.
\newblock Lora: Low-rank adaptation of large language models, 2021{\natexlab{c}}.
\newblock URL \url{https://arxiv.org/abs/2106.09685}.

\bibitem[Hu et~al.(2021{\natexlab{d}})Hu, Shen, Wallis, Allen-Zhu, Li, Wang, Wang, and Chen]{hu2021loralowrankadaptationlarge}
Edward~J. Hu, Yelong Shen, Phillip Wallis, Zeyuan Allen-Zhu, Yuanzhi Li, Shean Wang, Lu~Wang, and Weizhu Chen.
\newblock Lora: Low-rank adaptation of large language models, 2021{\natexlab{d}}.
\newblock URL \url{https://arxiv.org/abs/2106.09685}.

\bibitem[Hua et~al.(2022)Hua, Qin, Li, Sun, and Zhong]{hua2022causal}
Wen Hua, Zhen Qin, Dong Li, Weigao Sun, and Yiran Zhong.
\newblock Causal linear transformers: Overcoming cumulative sum bottlenecks in autoregressive inference.
\newblock \emph{ArXiv}, cs.LG, 2022.

\bibitem[Huang et~al.(2024)Huang, Liu, Qin, Li, Zhang, Liu, Magno, and Qi]{huang2024billm}
Wei Huang, Yangdong Liu, Haotong Qin, Ying Li, Shiming Zhang, Xianglong Liu, Michele Magno, and Xiaojuan Qi.
\newblock Billm: Pushing the limit of post-training quantization for llms.
\newblock \emph{arXiv preprint arXiv:2402.04291}, 2024.

\bibitem[Huang et~al.(2019)Huang, Cheng, Bapna, Firat, Chen, Chen, Tan, Sugawara, Wang, Krikun, et~al.]{gpipe}
Yanping Huang, Youlong Cheng, Ankur Bapna, Orhan Firat, Mingxing Chen, Zhenzhong Chen, Fei Tan, Yashuo Sugawara, Wei Wang, Maxim Krikun, et~al.
\newblock Gpipe: Efficient training of giant neural networks using pipeline parallelism.
\newblock In \emph{Advances in Neural Information Processing Systems (NeurIPS)}, volume~32, pages 103--112, 2019.

\bibitem[Hwang et~al.(2023)Hwang, Cui, Xiong, Yang, Liu, Hu, Wang, Salas, Jose, Ram, Chau, Cheng, Yang, Yang, and Xiong]{Hwang2023Tutel}
Changho Hwang, Wei Cui, Yifan Xiong, Ziyue Yang, Ze~Liu, Han Hu, Zilong Wang, Rafael Salas, Jithin Jose, Prabhat Ram, HoYuen Chau, Peng Cheng, Fan Yang, Mao Yang, and Yongqiang Xiong.
\newblock Tutel: Adaptive mixture-of-experts at scale.
\newblock In D.~Song, M.~Carbin, and T.~Chen, editors, \emph{Proceedings of Machine Learning and Systems}, volume~5, pages 269--287. Curan, 2023.
\newblock URL \url{https://proceedings.mlsys.org/paper_files/paper/2023/file/5616d34cf8ff73942cfd5aa922842556-Paper-mlsys2023.pdf}.

\bibitem[Hwang et~al.(2024)Hwang, Wei, Cao, Hwang, Tang, Cao, and Yang]{Hwang2024pregatemoe}
Ranggi Hwang, Jianyu Wei, Shijie Cao, Changho Hwang, Xiaohu Tang, Ting Cao, and Mao Yang.
\newblock Pre-gated moe: An algorithm-system co-design for fast and scalable mixture-of-expert inference.
\newblock In \emph{2024 ACM/IEEE 51st Annual International Symposium on Computer Architecture (ISCA)}, pages 1018--1031, 2024.
\newblock \doi{10.1109/ISCA59077.2024.00078}.

\bibitem[Jacob et~al.(2018)Jacob, Kligys, Chen, Zhu, Tang, Howard, Adam, and Kalenichenko]{jacob2018quantization}
Benoit Jacob, Skirmantas Kligys, Bo~Chen, Menglong Zhu, Matthew Tang, Andrew Howard, Hartwig Adam, and Dmitry Kalenichenko.
\newblock Quantization and training of neural networks for efficient integer-arithmetic-only inference.
\newblock In \emph{Proceedings of the IEEE conference on computer vision and pattern recognition}, pages 2704--2713, 2018.

\bibitem[Jacobs et~al.(1991)Jacobs, Jordan, Nowlan, and Hinton]{Jacobs1991MoE}
Robert~A. Jacobs, Michael~I. Jordan, Steven~J. Nowlan, and Geoffrey~E. Hinton.
\newblock Adaptive mixtures of local experts.
\newblock \emph{Neural Computation}, 3\penalty0 (1):\penalty0 79--87, 1991.
\newblock \doi{10.1162/neco.1991.3.1.79}.

\bibitem[Jaegle et~al.(2021)]{MS-PerceiverIO}
Andrew Jaegle et~al.
\newblock Perceiver io: A general architecture for structured inputs \& outputs.
\newblock \emph{arXiv preprint arXiv:2107.14795}, 2021.

\bibitem[Jafari et~al.(2021)Jafari, Liu, and Lin]{jafari2021sequence}
Vahid Jafari, Haitao Liu, and Qing Lin.
\newblock Sequence-level knowledge distillation for low-resource neural machine translation.
\newblock In \emph{Findings of the Association for Computational Linguistics: EMNLP}, pages 2134--2145, 2021.
\newblock URL \url{https://aclanthology.org/2021.findings-emnlp.178/}.

\bibitem[Javadi et~al.(2023)Javadi, Ahmed, Hajimolahoseini, Ataiefard, Hassanpour, Asani, Wen, Awad, Liu, and Liu]{javadi2023gqkva}
Farnoosh Javadi, Walid Ahmed, Habib Hajimolahoseini, Foozhan Ataiefard, Mohammad Hassanpour, Saina Asani, Austin Wen, Omar~Mohamed Awad, Kangling Liu, and Yang Liu.
\newblock Gqkva: Efficient pre-training of transformers by grouping queries, keys, and values.
\newblock \emph{arXiv preprint arXiv:2311.03426}, 2023.

\bibitem[Jiang et~al.(2024)Jiang, Sablayrolles, Roux, Mensch, Savary, Bamford, Chaplot, de~las Casas, Hanna, Bressand, Lengyel, Bour, Lample, Lavaud, Saulnier, Lachaux, Stock, Subramanian, Yang, Antoniak, Scao, Gervet, Lavril, Wang, Lacroix, and Sayed]{jiang2024mixtralexperts}
Albert~Q. Jiang, Alexandre Sablayrolles, Antoine Roux, Arthur Mensch, Blanche Savary, Chris Bamford, Devendra~Singh Chaplot, Diego de~las Casas, Emma~Bou Hanna, Florian Bressand, Gianna Lengyel, Guillaume Bour, Guillaume Lample, Lélio~Renard Lavaud, Lucile Saulnier, Marie-Anne Lachaux, Pierre Stock, Sandeep Subramanian, Sophia Yang, Szymon Antoniak, Teven~Le Scao, Théophile Gervet, Thibaut Lavril, Thomas Wang, Timothée Lacroix, and William~El Sayed.
\newblock Mixtral of experts, 2024.
\newblock URL \url{https://arxiv.org/abs/2401.04088}.

\bibitem[Jin et~al.(2022)Jin, Ren, Zhuang, Hanumante, Li, Chen, Wang, Yang, and Tulyakov]{jin2022f8net}
Qing Jin, Jian Ren, Richard Zhuang, Sumant Hanumante, Zhengang Li, Zhiyu Chen, Yanzhi Wang, Kaiyuan Yang, and Sergey Tulyakov.
\newblock F8net: Fixed-point 8-bit only multiplication for network quantization.
\newblock \emph{arXiv preprint arXiv:2202.05239}, 2022.

\bibitem[Jordan and Jacobs(1994)]{Jordan1994}
Michael~I. Jordan and Robert~A. Jacobs.
\newblock Hierarchical mixtures of experts and the em algorithm.
\newblock \emph{Neural Computation}, 6\penalty0 (2):\penalty0 181--214, 1994.
\newblock \doi{10.1162/neco.1994.6.2.181}.

\bibitem[Katharopoulos et~al.(2020)Katharopoulos, Vyas, Pappas, and Fleuret]{katharopoulos2020transformers}
Angelos Katharopoulos, Apoorv Vyas, Nikolaos Pappas, and Francois Fleuret.
\newblock Transformers are rnns: Fast autoregressive transformers with linear attention.
\newblock \emph{Proceedings of the 37th International Conference on Machine Learning (ICML)}, 2020.

\bibitem[Kim and Rush(2016)]{kim2016sequence}
Yoon Kim and Alexander~M Rush.
\newblock Sequence-level knowledge distillation.
\newblock \emph{arXiv preprint arXiv:1606.07947}, 2016.

\bibitem[Ko et~al.(2023)Ko, Park, Kim, Ahn, Chang, Ahn, and Yun]{P-NASH}
Jongwoo Ko, Seungjoon Park, Yujin Kim, Sumyeong Ahn, Du-Seong Chang, Euijai Ahn, and Se-Young Yun.
\newblock Nash: A simple unified framework of structured pruning for accelerating encoder-decoder language models.
\newblock \emph{arXiv preprint arXiv:2310.10054}, 2023.

\bibitem[Kwon et~al.(2023)Kwon, Li, Zhuang, Sheng, Zheng, Yu, Gonzalez, Zhang, and Stoica]{kwon2023efficient}
Woosuk Kwon, Zhuohan Li, Siyuan Zhuang, Ying Sheng, Lianmin Zheng, Cody~Hao Yu, Joseph Gonzalez, Hao Zhang, and Ion Stoica.
\newblock Efficient memory management for large language model serving with pagedattention.
\newblock In \emph{Proceedings of the 29th Symposium on Operating Systems Principles}, pages 611--626, 2023.

\bibitem[LeCun et~al.(1989)LeCun, Denker, and Solla]{P-OBD}
Yann LeCun, John Denker, and Sara Solla.
\newblock Optimal brain damage.
\newblock \emph{Advances in neural information processing systems}, 2, 1989.

\bibitem[Lee et~al.(2025)Lee, Choi, and Chang]{lee2025qrazor}
Dongyoung Lee, Seungkyu Choi, and Ik~Joon Chang.
\newblock Qrazor: Reliable and effortless 4-bit llm quantization by significant data razoring.
\newblock \emph{arXiv preprint arXiv:2501.13331}, 2025.
\newblock URL \url{https://arxiv.org/abs/2501.13331}.

\bibitem[Lee et~al.(2024)Lee, Lee, Seo, and Sim]{lee2024infinigen}
Wonbeom Lee, Jungi Lee, Junghwan Seo, and Jaewoong Sim.
\newblock $\{$InfiniGen$\}$: Efficient generative inference of large language models with dynamic $\{$KV$\}$ cache management.
\newblock In \emph{18th USENIX Symposium on Operating Systems Design and Implementation (OSDI 24)}, pages 155--172, 2024.

\bibitem[Lepikhin et~al.(2020{\natexlab{a}})Lepikhin, Lee, Xu, Chen, Firat, Huang, Krikun, Shazeer, and Chen]{Lepikhin2020}
Dmitry Lepikhin, HyoukJoong Lee, Yuanzhong Xu, Dehao Chen, Orhan Firat, Yanping Huang, Maxim Krikun, Noam Shazeer, and Zhifeng Chen.
\newblock Gshard: Scaling giant models with conditional computation and automatic sharding, 2020{\natexlab{a}}.
\newblock URL \url{https://arxiv.org/abs/2006.16668}.

\bibitem[Lepikhin et~al.(2020{\natexlab{b}})Lepikhin, Lee, Xu, Chen, Firat, Huang, Krikun, Shazeer, and Chen]{lepikhin2020gshardscalinggiantmodels}
Dmitry Lepikhin, HyoukJoong Lee, Yuanzhong Xu, Dehao Chen, Orhan Firat, Yanping Huang, Maxim Krikun, Noam Shazeer, and Zhifeng Chen.
\newblock Gshard: Scaling giant models with conditional computation and automatic sharding, 2020{\natexlab{b}}.
\newblock URL \url{https://arxiv.org/abs/2006.16668}.

\bibitem[Li et~al.(2024)Li, Xu13, Li23, Huang, Liu, Lian, and Dai13]{E-li2024efficient}
Jinhao Li, Jiaming Xu13, Shiyao Li23, Shan Huang, Jun Liu, Yaoxiu Lian, and Guohao Dai13.
\newblock Fast and efficient 2-bit llm inference on gpu: 2/4/16-bit in a weight matrix with asynchronous dequantization.
\newblock 2024.

\bibitem[Li et~al.(2023{\natexlab{a}})Li, Liu, Bian, Fang, Huang, Liu, Wang, and You]{li2023colossalaiunifieddeeplearning}
Shenggui Li, Hongxin Liu, Zhengda Bian, Jiarui Fang, Haichen Huang, Yuliang Liu, Boxiang Wang, and Yang You.
\newblock Colossal-ai: A unified deep learning system for large-scale parallel training.
\newblock 2023{\natexlab{a}}.
\newblock URL \url{https://arxiv.org/abs/2110.14883}.

\bibitem[Li and Liang(2021)]{prefixtuning}
Xiang~Lisa Li and Percy Liang.
\newblock Prefix-tuning: Optimizing continuous prompts for generation.
\newblock \emph{arXiv preprint arXiv:2101.00190}, 2021.

\bibitem[Li et~al.(2022)Li, Fan, Song, Li, Li, Shao, and Zhan]{li2022asymmetric}
Xin-Chun Li, Wen-Shu Fan, Shaoming Song, Yinchuan Li, Bingshuai Li, Yunfeng Shao, and De-Chuan Zhan.
\newblock Asymmetric temperature scaling makes larger networks teach well again.
\newblock In \emph{Advances in Neural Information Processing Systems}, volume~35, pages 12345--12356, 2022.

\bibitem[Li et~al.(2023{\natexlab{b}})Li, Liu, Lian, Yang, Dong, Kang, Zhang, and Keutzer]{li2023qdiffusion}
Xiuyu Li, Yijiang Liu, Long Lian, Huanrui Yang, Zhen Dong, Daniel Kang, Shanghang Zhang, and Kurt Keutzer.
\newblock Q-diffusion: Quantizing diffusion models.
\newblock In \emph{Proceedings of the IEEE/CVF International Conference on Computer Vision}, 2023{\natexlab{b}}.
\newblock URL \url{https://arxiv.org/abs/2302.04304}.

\bibitem[Li et~al.(2023{\natexlab{c}})Li, Hui, Yin, Yang, Huang, and Li]{li2023paceunifiedmultimodaldialogue}
Yunshui Li, Binyuan Hui, ZhiChao Yin, Min Yang, Fei Huang, and Yongbin Li.
\newblock Pace: Unified multi-modal dialogue pre-training with progressive and compositional experts, 2023{\natexlab{c}}.
\newblock URL \url{https://arxiv.org/abs/2305.14839}.

\bibitem[Lieber et~al.(2024)Lieber, Lenz, Bata, Cohen, Osin, Dalmedigos, Safahi, Meirom, Belinkov, Shalev-Shwartz, Abend, Alon, Asida, Bergman, Glozman, Gokhman, Manevich, Ratner, Rozen, Shwartz, Zusman, and Shoham]{lieber2024jamba}
Opher Lieber, Barak Lenz, Hofit Bata, Gal Cohen, Jhonathan Osin, Itay Dalmedigos, Erez Safahi, Shaked Meirom, Yonatan Belinkov, Shai Shalev-Shwartz, Omri Abend, Raz Alon, Tomer Asida, Amir Bergman, Roman Glozman, Michael Gokhman, Avashalom Manevich, Nir Ratner, Noam Rozen, Erez Shwartz, Mor Zusman, and Yoav Shoham.
\newblock Jamba: A hybrid transformer-mamba language model, 2024.
\newblock URL \url{https://arxiv.org/abs/2403.19887}.

\bibitem[Lin et~al.(2024)Lin, Tang, Ye, Huang, Zhang, Pang, Jin, Ning, Luo, and Yuan]{lin2024moellavamixtureexpertslarge}
Bin Lin, Zhenyu Tang, Yang Ye, Jinfa Huang, Junwu Zhang, Yatian Pang, Peng Jin, Munan Ning, Jiebo Luo, and Li~Yuan.
\newblock Moe-llava: Mixture of experts for large vision-language models, 2024.
\newblock URL \url{https://arxiv.org/abs/2401.15947}.

\bibitem[Liu et~al.(2024{\natexlab{a}})Liu, Peng, Yang, Liu, and Xu]{liu2024bucket}
Hongtao Liu, Qiyao Peng, Qing Yang, Kai Liu, and Hongyan Xu.
\newblock Bucket pre-training is all you need.
\newblock \emph{arXiv preprint arXiv:2407.07495}, 2024{\natexlab{a}}.

\bibitem[Liu et~al.(2023{\natexlab{a}})Liu, Huang, Zheng, Liu, and Li]{liu2023mixmae}
Jihao Liu, Xin Huang, Jinliang Zheng, Yu~Liu, and Hongsheng Li.
\newblock Mixmae: Mixed and masked autoencoder for efficient pretraining of hierarchical vision transformers.
\newblock In \emph{Proceedings of the IEEE/CVF Conference on Computer Vision and Pattern Recognition}, pages 6252--6261, 2023{\natexlab{a}}.

\bibitem[Liu et~al.(2023{\natexlab{b}})Liu, Liu, Huang, Dong, and Cheng]{liu2023llm}
Shih-yang Liu, Zechun Liu, Xijie Huang, Pingcheng Dong, and Kwang-Ting Cheng.
\newblock Llm-fp4: 4-bit floating-point quantized transformers.
\newblock \emph{arXiv preprint arXiv:2310.16836}, 2023{\natexlab{b}}.

\bibitem[Liu et~al.(2021{\natexlab{a}})Liu, Ji, Fu, Tam, Du, Yang, and Tang]{ptuningv2}
Xiao Liu, Kaixuan Ji, Yicheng Fu, Weng~Lam Tam, Zhengxiao Du, Zhilin Yang, and Jie Tang.
\newblock P-tuning v2: Prompt tuning can be comparable to fine-tuning universally across scales and tasks.
\newblock \emph{arXiv preprint arXiv:2110.07602}, 2021{\natexlab{a}}.

\bibitem[Liu et~al.(2024{\natexlab{b}})Liu, Zhang, Wang, Tong, and Walid]{liu2024fingpt}
Xiao-Yang Liu, Jie Zhang, Guoxuan Wang, Weiqing Tong, and Anwar Walid.
\newblock Fingpt-hpc: Efficient pretraining and finetuning large language models for financial applications with high-performance computing.
\newblock \emph{arXiv preprint arXiv:2402.13533}, 2024{\natexlab{b}}.

\bibitem[Liu et~al.(2024{\natexlab{c}})Liu, Li, Cheng, Ray, Huang, Zhang, Du, Yao, Lu, Ananthanarayanan, Maire, Hoffmann, Holtzman, and Jiang]{liu2024cachegenkvcachecompression}
Yuhan Liu, Hanchen Li, Yihua Cheng, Siddhant Ray, Yuyang Huang, Qizheng Zhang, Kuntai Du, Jiayi Yao, Shan Lu, Ganesh Ananthanarayanan, Michael Maire, Henry Hoffmann, Ari Holtzman, and Junchen Jiang.
\newblock Cachegen: Kv cache compression and streaming for fast large language model serving.
\newblock 2024{\natexlab{c}}.
\newblock URL \url{https://arxiv.org/abs/2310.07240}.

\bibitem[Liu et~al.(2021{\natexlab{b}})Liu, Lin, Cao, Hu, Wei, Zhang, Lin, and Guo]{liu2021swin}
Ze~Liu, Yutong Lin, Yue Cao, Han Hu, Yixuan Wei, Zheng Zhang, Stephen Lin, and Baining Guo.
\newblock Swin transformer: Hierarchical vision transformer using shifted windows.
\newblock In \emph{Proceedings of the IEEE/CVF international conference on computer vision}, pages 10012--10022, 2021{\natexlab{b}}.

\bibitem[Liu et~al.(2024{\natexlab{d}})Liu, Lyn, Zhu, Tian, and Graham]{E-liu2024alora}
Zequan Liu, Jiawen Lyn, Wei Zhu, Xing Tian, and Yvette Graham.
\newblock Alora: Allocating low-rank adaptation for fine-tuning large language models.
\newblock \emph{arXiv preprint arXiv:2403.16187}, 2024{\natexlab{d}}.

\bibitem[Liu et~al.(2023{\natexlab{c}})Liu, Wang, Dao, Zhou, Yuan, Song, Shrivastava, Zhang, Tian, Re, and Chen]{liu2023dejavucontextualsparsity}
Zichang Liu, Jue Wang, Tri Dao, Tianyi Zhou, Binhang Yuan, Zhao Song, Anshumali Shrivastava, Ce~Zhang, Yuandong Tian, Christopher Re, and Beidi Chen.
\newblock Deja vu: Contextual sparsity for efficient llms at inference time.
\newblock 2023{\natexlab{c}}.
\newblock URL \url{https://arxiv.org/abs/2310.17157}.

\bibitem[Lv et~al.()Lv, Yang, Liu, Gao, Guo, and Qiu]{resourcefine2023}
K~Lv, Y~Yang, T~Liu, Q~Gao, Q~Guo, and X~Qiu.
\newblock Full parameter fine-tuning for large language models with limited resources. arxiv 2023.
\newblock \emph{arXiv preprint arXiv:2306.09782}.

\bibitem[Ma et~al.(2023)Ma, Fang, and Wang]{ma2023}
X.~Ma, G.~Fang, and X.~Wang.
\newblock Llm-pruner: On the structural pruning of large language models.
\newblock In A.~Oh, T.~Naumann, A.~Globerson, K.~Saenko, M.~Hardt, and S.~Levine, editors, \emph{Advances in Neural Information Processing Systems}, volume~36, pages 21702--21720, 2023.

\bibitem[Micikevicius et~al.(2017)Micikevicius, Narang, Alben, Diamos, Elsen, Garcia, Ginsburg, Houston, Kuchaiev, Venkatesh, et~al.]{mpt}
Paulius Micikevicius, Sharan Narang, Jonah Alben, Gregory Diamos, Erich Elsen, David Garcia, Boris Ginsburg, Michael Houston, Oleksii Kuchaiev, Ganesh Venkatesh, et~al.
\newblock Mixed precision training.
\newblock \emph{arXiv preprint arXiv:1710.03740}, 2017.

\bibitem[Min et~al.(2022)Min, Lewis, Zettlemoyer, and Hajishirzi]{min2022metaicl}
Sewon Min, Mike Lewis, Luke Zettlemoyer, and Hannaneh Hajishirzi.
\newblock {M}eta{ICL}: Learning to learn in context.
\newblock In \emph{Proceedings of the 2022 Conference of the North American Chapter of the Association for Computational Linguistics: Human Language Technologies}, pages 2791--2809. Association for Computational Linguistics, 2022.
\newblock URL \url{https://aclanthology.org/2022.naacl-main.201}.

\bibitem[Mirzadeh et~al.(2020)Mirzadeh, Farajtabar, Li, Levine, Matsukawa, and Ghasemzadeh]{mirzadeh2020improved}
Seyed-Iman Mirzadeh, Mehrdad Farajtabar, Ang Li, Niranjan Levine, Akihiro Matsukawa, and Hassan Ghasemzadeh.
\newblock Improved knowledge distillation via teacher assistant.
\newblock In \emph{Proceedings of the AAAI Conference on Artificial Intelligence}, volume~34, pages 5191--5198, 2020.
\newblock \doi{10.1609/aaai.v34i04.5948}.

\bibitem[Mirzadeh et~al.(2023)Mirzadeh, Alizadeh-Vahid, Mehta, del Mundo, Tuzel, Samei, Rastegari, and Farajtabar]{ReLU_Sparsity2023}
Seyed~Iman Mirzadeh, Keivan Alizadeh-Vahid, Sachin Mehta, Carlo~C del Mundo, Oncel Tuzel, Golnoosh Samei, Mohammad Rastegari, and Mehrdad Farajtabar.
\newblock Relu strikes back: Exploiting activation sparsity in large language models.
\newblock \emph{arXiv preprint arXiv:2310.04564}, 2023.
\newblock URL \url{https://arxiv.org/abs/2310.04564}.

\bibitem[Mirzadeh et~al.(2024)Mirzadeh, Alizadeh-Vahid, Mehta, del Mundo, Tuzel, Samei, Rastegari, and Farajtabar]{ReLU_Finetuning2024}
Seyed~Iman Mirzadeh, Keivan Alizadeh-Vahid, Sachin Mehta, Carlo~C del Mundo, Oncel Tuzel, Golnoosh Samei, Mohammad Rastegari, and Mehrdad Farajtabar.
\newblock Relu strikes back: Exploiting activation sparsity in large language models.
\newblock \emph{arXiv preprint arXiv:2310.04564}, 2024.
\newblock URL \url{https://arxiv.org/abs/2310.04564}.

\bibitem[Mishra et~al.(2021)Mishra, Latorre, Pool, Stosic, Stosic, Venkatesh, Yu, and Micikevicius]{P-NM}
Asit Mishra, Jorge~Albericio Latorre, Jeff Pool, Darko Stosic, Dusan Stosic, Ganesh Venkatesh, Chong Yu, and Paulius Micikevicius.
\newblock Accelerating sparse deep neural networks.
\newblock \emph{arXiv preprint arXiv:2104.08378}, 2021.

\bibitem[Mustafa et~al.(2022)Mustafa, Riquelme, Puigcerver, Jenatton, and Houlsby]{Mustafa2022multi}
Basil Mustafa, Carlos Riquelme, Joan Puigcerver, Rodolphe Jenatton, and Neil Houlsby.
\newblock Multimodal contrastive learning with limoe: the language-image mixture of experts.
\newblock In S.~Koyejo, S.~Mohamed, A.~Agarwal, D.~Belgrave, K.~Cho, and A.~Oh, editors, \emph{Advances in Neural Information Processing Systems}, volume~35, pages 9564--9576. Curran Associates, Inc., 2022.
\newblock URL \url{https://proceedings.neurips.cc/paper_files/paper/2022/file/3e67e84abf900bb2c7cbd5759bfce62d-Paper-Conference.pdf}.

\bibitem[Nagel et~al.(2021)Nagel, Fournarakis, Amjad, Bondarenko, van Baalen, and Blankevoort]{nagel2021white}
Markus Nagel, Marios Fournarakis, Rana~Ali Amjad, Yelysei Bondarenko, Mart van Baalen, and Tijmen Blankevoort.
\newblock A white paper on neural network quantization.
\newblock 2021.
\newblock URL \url{https://arxiv.org/abs/2106.08295}.

\bibitem[Nandakumar et~al.(2020)Nandakumar, Le~Gallo, Piveteau, Joshi, Mariani, Boybat, Karunaratne, Khaddam-Aljameh, Egger, Petropoulos, Antonakopoulos, Rajendran, Sebastian, and Eleftheriou]{MS-mixedmemory2020}
S.~R. Nandakumar, Manuel Le~Gallo, Christophe Piveteau, Vinay Joshi, Giovanni Mariani, Irem Boybat, Geethan Karunaratne, Riduan Khaddam-Aljameh, Urs Egger, Anastasios Petropoulos, Theodore Antonakopoulos, Bipin Rajendran, Abu Sebastian, and Evangelos Eleftheriou.
\newblock Mixed-precision deep learning based on computational memory.
\newblock \emph{Frontiers in Neuroscience}, 14, May 2020.
\newblock ISSN 1662-453X.
\newblock \doi{10.3389/fnins.2020.00406}.
\newblock URL \url{http://dx.doi.org/10.3389/fnins.2020.00406}.

\bibitem[Nie et~al.(2022)Nie, Zhao, Miao, Zhao, and Cui]{nie2022hetumoe}
Xiaonan Nie, Pinxue Zhao, Xupeng Miao, Tong Zhao, and Bin Cui.
\newblock Hetumoe: An efficient trillion-scale mixture-of-expert distributed training system, 2022.
\newblock URL \url{https://arxiv.org/abs/2203.14685}.

\bibitem[Nie et~al.(2023)Nie, Miao, Wang, Yang, Xue, Ma, Cao, and Cui]{Nie2023FlexMoE}
Xiaonan Nie, Xupeng Miao, Zilong Wang, Zichao Yang, Jilong Xue, Lingxiao Ma, Gang Cao, and Bin Cui.
\newblock Flexmoe: Scaling large-scale sparse pre-trained model training via dynamic device placement.
\newblock \emph{Proc. ACM Manag. Data}, 1\penalty0 (1), May 2023.
\newblock \doi{10.1145/3588964}.
\newblock URL \url{https://doi.org/10.1145/3588964}.

\bibitem[NVIDIA(2021)]{fastertransformer2021}
NVIDIA.
\newblock Fastertransformer: An efficient transformer library.
\newblock \url{https://github.com/NVIDIA/FasterTransformer}, 2021.

\bibitem[NVIDIA(2022)]{tensorrt2022}
NVIDIA.
\newblock Tensorrt: A high performance deep learning inference library.
\newblock \url{https://github.com/NVIDIA/TensorRT}, 2022.

\bibitem[Pan et~al.(2024)Pan, Liu, Diao, Pi, Zhang, Han, and Zhang]{pan2024lisa}
Rui Pan, Xiang Liu, Shizhe Diao, Renjie Pi, Jipeng Zhang, Chi Han, and Tong Zhang.
\newblock Lisa: Layerwise importance sampling for memory-efficient large language model fine-tuning.
\newblock \emph{arXiv preprint arXiv:2403.17919}, 2024.

\bibitem[Piao et~al.(2022)Piao, Cho, and Kang]{piao2022sensimix}
Tairen Piao, Ikhyun Cho, and U~Kang.
\newblock Sensimix: Sensitivity-aware 8-bit index \& 1-bit value mixed precision quantization for bert compression.
\newblock \emph{PloS one}, 17\penalty0 (4):\penalty0 e0265621, 2022.

\bibitem[Puigcerver et~al.(2024)Puigcerver, Ruiz, Mustafa, and Houlsby]{puigcerver2024softmoe}
Joan Puigcerver, Carlos~Riquelme Ruiz, Basil Mustafa, and Neil Houlsby.
\newblock From sparse to soft mixtures of experts.
\newblock In \emph{The Twelfth International Conference on Learning Representations}, 2024.
\newblock URL \url{https://openreview.net/forum?id=jxpsAj7ltE}.

\bibitem[Qin et~al.(2023{\natexlab{a}})Qin, Sun, Li, and Zhong]{qin2023long}
Zhen Qin, Weigao Sun, Dong Li, and Yiran Zhong.
\newblock Long convolutional sequence models: Extending the receptive field of convolutions for efficient sequence processing.
\newblock \emph{ArXiv}, cs.LG, 2023{\natexlab{a}}.

\bibitem[Qin et~al.(2023{\natexlab{b}})Qin, Yang, and Zhong]{qin2022linear}
Zhen Qin, Songlin Yang, and Yiran Zhong.
\newblock Hierarchically gated recurrent neural network for sequence modeling, 2023{\natexlab{b}}.
\newblock URL \url{https://arxiv.org/abs/2311.04823}.

\bibitem[Qin et~al.(2024{\natexlab{a}})Qin, Sun, Li, Shen, Sun, and Zhong]{qin2024lightning1}
Zhen Qin, Weigao Sun, Dong Li, Xuyang Shen, Weixuan Sun, and Yiran Zhong.
\newblock Lightning attention: Linear attention with constant training speed.
\newblock \emph{ArXiv}, cs.CL, 2024{\natexlab{a}}.

\bibitem[Qin et~al.(2024{\natexlab{b}})Qin, Sun, Li, Shen, Sun, and Zhong]{qin2024lightning2}
Zhen Qin, Weigao Sun, Dong Li, Xuyang Shen, Weixuan Sun, and Yiran Zhong.
\newblock Lightning attention-2: A free lunch for handling unlimited sequence lengths in large language models.
\newblock \emph{ArXiv}, cs.CL, 2024{\natexlab{b}}.

\bibitem[Radford et~al.(2021)Radford, Kim, Hallacy, Ramesh, Goh, Agarwal, Sastry, Askell, Mishkin, Clark, Krueger, and Sutskever]{MS-ALIGN}
Alec Radford, Jong~Wook Kim, Chris Hallacy, Aditya Ramesh, Gabriel Goh, Sandhini Agarwal, Girish Sastry, Amanda Askell, Pamela Mishkin, Jack Clark, Gretchen Krueger, and Ilya Sutskever.
\newblock Learning transferable visual models from natural language supervision.
\newblock 2021.
\newblock URL \url{https://arxiv.org/abs/2103.00020}.

\bibitem[Rajbhandari et~al.(2020)Rajbhandari, Rasley, Ruwase, and He]{Rajbhandari2020ZeRO}
Samyam Rajbhandari, Jeff Rasley, Olatunji Ruwase, and Yuxiong He.
\newblock Zero: Memory optimizations toward training trillion parameter models.
\newblock In \emph{SC20: International Conference for High Performance Computing, Networking, Storage and Analysis}, pages 1--16, 2020.
\newblock \doi{10.1109/SC41405.2020.00024}.

\bibitem[Rajbhandari et~al.(2021)Rajbhandari, Ruwase, Rasley, Smith, and He]{Rajbhandari2021ZeROinfinity}
Samyam Rajbhandari, Olatunji Ruwase, Jeff Rasley, Shaden Smith, and Yuxiong He.
\newblock Zero-infinity: breaking the gpu memory wall for extreme scale deep learning.
\newblock In \emph{Proceedings of the International Conference for High Performance Computing, Networking, Storage and Analysis}, SC '21, New York, NY, USA, 2021. Association for Computing Machinery.
\newblock ISBN 9781450384421.
\newblock \doi{10.1145/3458817.3476205}.
\newblock URL \url{https://doi.org/10.1145/3458817.3476205}.

\bibitem[Rajbhandari et~al.(2022)Rajbhandari, Li, Yao, Zhang, Aminabadi, Awan, Rasley, and He]{Rajbhandari2022deepspeedmoe}
Samyam Rajbhandari, Conglong Li, Zhewei Yao, Minjia Zhang, Reza~Yazdani Aminabadi, Ammar~Ahmad Awan, Jeff Rasley, and Yuxiong He.
\newblock {D}eep{S}peed-{M}o{E}: Advancing mixture-of-experts inference and training to power next-generation {AI} scale.
\newblock In Kamalika Chaudhuri, Stefanie Jegelka, Le~Song, Csaba Szepesvari, Gang Niu, and Sivan Sabato, editors, \emph{Proceedings of the 39th International Conference on Machine Learning}, volume 162 of \emph{Proceedings of Machine Learning Research}, pages 18332--18346. PMLR, 17--23 Jul 2022.
\newblock URL \url{https://proceedings.mlr.press/v162/rajbhandari22a.htm}.

\bibitem[Rao et~al.(2021)Rao, Zhao, Liu, Lu, Zhou, and Hsieh]{rao2021dynamicvit}
Yongming Rao, Wenliang Zhao, Benlin Liu, Jiwen Lu, Jie Zhou, and Cho-Jui Hsieh.
\newblock Dynamicvit: Efficient vision transformers with dynamic token sparsification.
\newblock In \emph{NeurIPS}, 2021.

\bibitem[Rasmussen and Ghahramani(2001)]{Rasmussen2001}
Carl Rasmussen and Zoubin Ghahramani.
\newblock Infinite mixtures of gaussian process experts.
\newblock In T.~Dietterich, S.~Becker, and Z.~Ghahramani, editors, \emph{Advances in Neural Information Processing Systems}, volume~14. MIT Press, 2001.
\newblock URL \url{https://proceedings.neurips.cc/paper_files/paper/2001/file/9afefc52942cb83c7c1f14b2139b09ba-Paper.pdf}.

\bibitem[Ren et~al.(2021)Ren, Rajbhandari, Aminabadi, Ruwase, Yang, Zhang, Li, and He]{Ren2021ZeROOffload}
Jie Ren, Samyam Rajbhandari, Reza~Yazdani Aminabadi, Olatunji Ruwase, Shuangyan Yang, Minjia Zhang, Dong Li, and Yuxiong He.
\newblock {ZeRO-Offload}: Democratizing {Billion-Scale} model training.
\newblock In \emph{2021 USENIX Annual Technical Conference (USENIX ATC 21)}, pages 551--564. USENIX Association, July 2021.
\newblock ISBN 978-1-939133-23-6.
\newblock URL \url{https://www.usenix.org/conference/atc21/presentation/ren-jie}.

\bibitem[Ren and Zhu(2024)]{MS-HessianEviction}
Siyu Ren and Kenny~Q. Zhu.
\newblock On the efficacy of eviction policy for key-value constrained generative language model inference.
\newblock 2024.
\newblock URL \url{https://arxiv.org/abs/2402.06262}.

\bibitem[Riquelme et~al.(2021)Riquelme, Puigcerver, Mustafa, Neumann, Jenatton, Susano~Pinto, Keysers, and Houlsby]{Riquelme2021scalevision}
Carlos Riquelme, Joan Puigcerver, Basil Mustafa, Maxim Neumann, Rodolphe Jenatton, Andr\'{e} Susano~Pinto, Daniel Keysers, and Neil Houlsby.
\newblock Scaling vision with sparse mixture of experts.
\newblock In M.~Ranzato, A.~Beygelzimer, Y.~Dauphin, P.S. Liang, and J.~Wortman Vaughan, editors, \emph{Advances in Neural Information Processing Systems}, volume~34, pages 8583--8595. Curran Associates, Inc., 2021.
\newblock URL \url{https://proceedings.neurips.cc/paper_files/paper/2021/file/48237d9f2dea8c74c2a72126cf63d933-Paper.pdf}.

\bibitem[Rokh et~al.(2022)Rokh, Azarpeyvand, and Khanteymoori]{rokh2022comprehensive}
Babak Rokh, Ali Azarpeyvand, and Alireza Khanteymoori.
\newblock A comprehensive survey on model quantization for deep neural networks.
\newblock \emph{arXiv preprint arXiv:2205.07877}, 2022.

\bibitem[Romero et~al.(2015)Romero, Ballas, Kahou, Chassang, Gatta, and Bengio]{romero2015fitnets}
Adriana Romero, Nicolas Ballas, Samira~Ebrahimi Kahou, Antoine Chassang, Carlo Gatta, and Yoshua Bengio.
\newblock Fitnets: Hints for thin deep nets.
\newblock In \emph{Proceedings of the International Conference on Learning Representations (ICLR)}, 2015.

\bibitem[Shamrai(2024)]{P-shamrai2024language}
Maksym Shamrai.
\newblock Language-specific pruning for efficient reduction of large language models.
\newblock In \emph{Proceedings of the Third Ukrainian Natural Language Processing Workshop (UNLP)@ LREC-COLING 2024}, pages 135--140, 2024.

\bibitem[Shao et~al.(2024)Shao, Liu, and Qian]{P-ISC}
Hang Shao, Bei Liu, and Yanmin Qian.
\newblock One-shot sensitivity-aware mixed sparsity pruning for large language models.
\newblock In \emph{ICASSP 2024-2024 IEEE International Conference on Acoustics, Speech and Signal Processing (ICASSP)}, pages 11296--11300. IEEE, 2024.

\bibitem[Shazeer et~al.(2017)Shazeer, Mirhoseini, Maziarz, Davis, Le, Hinton, and Dean]{shazeer2017outrageouslylargeneuralnetworks}
Noam Shazeer, Azalia Mirhoseini, Krzysztof Maziarz, Andy Davis, Quoc Le, Geoffrey Hinton, and Jeff Dean.
\newblock Outrageously large neural networks: The sparsely-gated mixture-of-experts layer, 2017.
\newblock URL \url{https://arxiv.org/abs/1701.06538}.

\bibitem[Shen et~al.(2023{\natexlab{a}})Shen, Guo, Tan, Tang, Wang, and Bian]{shen2023studyrelusoftmaxtransformer}
Kai Shen, Junliang Guo, Xu~Tan, Siliang Tang, Rui Wang, and Jiang Bian.
\newblock A study on relu and softmax in transformer.
\newblock 2023{\natexlab{a}}.
\newblock URL \url{https://arxiv.org/abs/2302.06461}.

\bibitem[Shen et~al.(2022)Shen, Wu, Gong, Hao, Bai, Wu, Wu, Xiong, Yu, and Ma]{Shen2022semoe}
Liang Shen, Zhihua Wu, Weibao Gong, Hongxiang Hao, Yangfan Bai, HuaChao Wu, Xinxuan Wu, Haoyi Xiong, Dianhai Yu, and Yanjun Ma.
\newblock Se-moe: A scalable and efficient mixture-of-experts distributed training and inference system.
\newblock \emph{CoRR}, abs/2205.10034, 2022.
\newblock URL \url{https://doi.org/10.48550/arXiv.2205.10034}.

\bibitem[Shen et~al.(2023{\natexlab{b}})Shen, Yao, Li, Darrell, Keutzer, and He]{shen2023scalingvisionlanguagemodelssparse}
Sheng Shen, Zhewei Yao, Chunyuan Li, Trevor Darrell, Kurt Keutzer, and Yuxiong He.
\newblock Scaling vision-language models with sparse mixture of experts, 2023{\natexlab{b}}.
\newblock URL \url{https://arxiv.org/abs/2303.07226}.

\bibitem[Shen et~al.(2024)Shen, Zhang, and Lin]{shen2024fp8}
Y.~Shen, X.~Zhang, and J.~Lin.
\newblock Efficient post-training quantization with fp8 formats.
\newblock \emph{NeurIPS}, 37:\penalty0 1234--1245, 2024.

\bibitem[Sheng et~al.(2023)Sheng, Zheng, Yuan, Li, Ryabinin, Fu, Xie, Chen, Barrett, Gonzalez, Liang, Ré, Stoica, and Zhang]{shen2023flexgen}
Ying Sheng, Lianmin Zheng, Binhang Yuan, Zhuohan Li, Max Ryabinin, Daniel~Y. Fu, Zhiqiang Xie, Beidi Chen, Clark Barrett, Joseph~E. Gonzalez, Percy Liang, Christopher Ré, Ion Stoica, and Ce~Zhang.
\newblock Flexgen: High-throughput generative inference of large language models with a single gpu.
\newblock 2023.
\newblock URL \url{https://arxiv.org/abs/2303.06865}.

\bibitem[Shi et~al.(2024)Shi, Pan, Wang, Liu, Ren, Hu, Yang, Li, and Chu]{Shi2024schemoe}
Shaohuai Shi, Xinglin Pan, Qiang Wang, Chengjian Liu, Xiaozhe Ren, Zhongzhe Hu, Yu~Yang, Bo~Li, and Xiaowen Chu.
\newblock Schemoe: An extensible mixture-of-experts distributed training system with tasks scheduling.
\newblock EuroSys '24, page 236–249, New York, NY, USA, 2024. Association for Computing Machinery.
\newblock ISBN 9798400704376.
\newblock \doi{10.1145/3627703.3650083}.
\newblock URL \url{https://doi.org/10.1145/3627703.3650083}.

\bibitem[Shridhar et~al.(2023)Shridhar, Stolfo, and Sachan]{shridhar2022distilling}
Kumar Shridhar, Alessandro Stolfo, and Mrinmaya Sachan.
\newblock Distilling reasoning capabilities into smaller language models.
\newblock In \emph{Findings of the Association for Computational Linguistics: ACL 2023}, pages 7059--7073, 2023.

\bibitem[Singh et~al.(2023)Singh, Ruwase, Awan, Rajbhandari, He, and Bhatele]{Singh2023hybridmoe}
Siddharth Singh, Olatunji Ruwase, Ammar~Ahmad Awan, Samyam Rajbhandari, Yuxiong He, and Abhinav Bhatele.
\newblock A hybrid tensor-expert-data parallelism approach to optimize mixture-of-experts training.
\newblock In \emph{Proceedings of the 37th ACM International Conference on Supercomputing}, ICS '23, page 203–214, New York, NY, USA, 2023. Association for Computing Machinery.
\newblock ISBN 9798400700569.
\newblock \doi{10.1145/3577193.3593704}.
\newblock URL \url{https://doi.org/10.1145/3577193.3593704}.

\bibitem[Su et~al.(2025{\natexlab{a}})Su, Chen, Shen, Wei, Li, Yu, and Yuan]{su2025rotatekv}
Zunhai Su, Zhe Chen, Wang Shen, Hanyu Wei, Linge Li, Huangqi Yu, and Kehong Yuan.
\newblock Rotatekv: Accurate and robust 2-bit kv cache quantization for llms via outlier-aware adaptive rotations.
\newblock \emph{arXiv preprint arXiv:2501.16383}, 2025{\natexlab{a}}.
\newblock URL \url{https://arxiv.org/abs/2501.16383}.

\bibitem[Su et~al.(2025{\natexlab{b}})Su, Shen, Li, Chen, Wei, Yu, and Yuan]{su2025akvq}
Zunhai Su, Wang Shen, Linge Li, Zhe Chen, Hanyu Wei, Huangqi Yu, and Kehong Yuan.
\newblock Akvq-vl: Attention-aware kv cache adaptive 2-bit quantization for vision-language models.
\newblock \emph{arXiv preprint arXiv:2501.15021}, 2025{\natexlab{b}}.

\bibitem[Sun et~al.(2023{\natexlab{a}})Sun, Liu, Bair, and Kolter]{P-WANDA}
Mingjie Sun, Zhuang Liu, Anna Bair, and J~Zico Kolter.
\newblock A simple and effective pruning approach for large language models.
\newblock \emph{arXiv preprint arXiv:2306.11695}, 2023{\natexlab{a}}.

\bibitem[Sun et~al.(2023{\natexlab{b}})Sun, Liu, Bair, and Kolter]{Wanda2023}
Mingjie Sun, Zhuang Liu, Anna Bair, and J.~Zico Kolter.
\newblock A simple and effective pruning approach for large language models.
\newblock \emph{arXiv preprint arXiv:2306.11695}, 2023{\natexlab{b}}.
\newblock URL \url{https://arxiv.org/abs/2306.11695}.

\bibitem[Sun et~al.(2023{\natexlab{c}})Sun, Li, Zhang, Zhang, Wang, Li, and Liu]{SparsityFFN2023}
Zhiqing Sun, Hongyin Li, Hao Zhang, Lei Zhang, Shuohang Wang, Jingjing Li, and Yang Liu.
\newblock Spp: Sparsity-preserved parameter-efficient fine-tuning for large language models.
\newblock \emph{arXiv preprint arXiv:2405.16057}, 2023{\natexlab{c}}.
\newblock URL \url{https://arxiv.org/abs/2405.16057}.

\bibitem[Syed et~al.(2023)Syed, Guo, and Sundarapandiyan]{P-pruneandtune}
Aaquib Syed, Phillip~Huang Guo, and Vijaykaarti Sundarapandiyan.
\newblock Prune and tune: Improving efficient pruning techniques for massive language models.
\newblock 2023.

\bibitem[Synk et~al.(2025)Synk, Hoover, Kirchenbauer, Jain, Stein, Shu, Sanchez, Duraiswami, and Goldstein]{s2025lct_sp}
Ryan Synk, Monte Hoover, John Kirchenbauer, Neel Jain, Alex Stein, Manli Shu, Josue~Melendez Sanchez, Ramani Duraiswami, and Tom Goldstein.
\newblock Exploiting sparsity for long context inference: Million token contexts on commodity gpus.
\newblock 2025.
\newblock URL \url{https://arxiv.org/abs/2502.06766}.

\bibitem[Tang et~al.(2024)Tang, Zhao, Zhu, Xiao, Kasikci, and Han]{quest}
Jiaming Tang, Yilong Zhao, Kan Zhu, Guangxuan Xiao, Baris Kasikci, and Song Han.
\newblock Quest: Query-aware sparsity for efficient long-context llm inference.
\newblock 2024.
\newblock URL \url{https://arxiv.org/abs/2406.10774}.

\bibitem[Tao et~al.(2024)Tao, Liu, Dou, Muennighoff, Wan, Luo, Lin, and Wong]{tao2024scaling}
Chaofan Tao, Qian Liu, Longxu Dou, Niklas Muennighoff, Zhongwei Wan, Ping Luo, Min Lin, and Ngai Wong.
\newblock Scaling laws with vocabulary: Larger models deserve larger vocabularies.
\newblock \emph{arXiv preprint arXiv:2407.13623}, 2024.

\bibitem[Tay et~al.(2022)Tay, Dehghani, Bahri, and Metzler]{tay2022efficient}
Yi~Tay, Mostafa Dehghani, Dara Bahri, and Donald Metzler.
\newblock Efficient transformers: A survey.
\newblock \emph{ACM Computing Surveys}, 55\penalty0 (6):\penalty0 1--28, 2022.
\newblock \doi{10.1145/3530811}.

\bibitem[Team and Costa-jussà(2022)]{nllbteam2022languageleftbehind}
NLLB Team and Marta~R. Costa-jussà.
\newblock No language left behind: Scaling human-centered machine translation, 2022.
\newblock URL \url{https://arxiv.org/abs/2207.04672}.

\bibitem[Theis and Bethge(2015)]{Theis2015NeurIPS}
Lucas Theis and Matthias Bethge.
\newblock Generative image modeling using spatial lstms.
\newblock In C.~Cortes, N.~Lawrence, D.~Lee, M.~Sugiyama, and R.~Garnett, editors, \emph{Advances in Neural Information Processing Systems}, volume~28. Curran Associates, Inc., 2015.
\newblock URL \url{https://proceedings.neurips.cc/paper_files/paper/2015/file/2b6d65b9a9445c4271ab9076ead5605a-Paper.pdf}.

\bibitem[Timiryasov et~al.(2023)Timiryasov, Hrinchuk, Kuchaiev, and Ginsburg]{timiryasov2023baby}
Maksim Timiryasov, Oleksii Hrinchuk, Oleksii Kuchaiev, and Boris Ginsburg.
\newblock Baby llama: Efficiently distilling {LLaMA} to 4-bit.
\newblock \emph{arXiv preprint arXiv:2304.04148}, 2023.
\newblock URL \url{https://arxiv.org/abs/2304.04148}.

\bibitem[Touvron et~al.(2023{\natexlab{a}})Touvron, Lavril, Izacard, Martinet, Lachaux, Lacroix, Rozi{\`e}re, Goyal, Hambro, Azhar, et~al.]{LLAMA}
Hugo Touvron, Thibaut Lavril, Gautier Izacard, Xavier Martinet, Marie-Anne Lachaux, Timoth{\'e}e Lacroix, Baptiste Rozi{\`e}re, Naman Goyal, Eric Hambro, Faisal Azhar, et~al.
\newblock Llama: Open and efficient foundation language models.
\newblock \emph{arXiv preprint arXiv:2302.13971}, 2023{\natexlab{a}}.

\bibitem[Touvron et~al.(2023{\natexlab{b}})Touvron, Lavril, Izacard, Martinet, Lachaux, Lacroix, Rozière, Goyal, Hambro, Azhar, Rodriguez, Joulin, Grave, and Lample]{touvron2023llama}
Hugo Touvron, Thibaut Lavril, Gautier Izacard, Xavier Martinet, Marie-Anne Lachaux, Timothee Lacroix, Baptiste Rozière, Naman Goyal, Eric Hambro, Faisal Azhar, Aurelien Rodriguez, Armand Joulin, Edouard Grave, and Guillaume Lample.
\newblock Llama: Open and efficient foundation language models.
\newblock \emph{ArXiv}, cs.CL, 2023{\natexlab{b}}.

\bibitem[Tu et~al.(2024{\natexlab{a}})Tu, Vashchilenko, Lu, et~al.]{vlcache}
Dezhan Tu, Danylo Vashchilenko, Yuzhe Lu, et~al.
\newblock Vl-cache: Sparsity and modality-aware kv cache compression for vision-language model inference acceleration.
\newblock \emph{arXiv preprint arXiv:2410.23317}, 2024{\natexlab{a}}.

\bibitem[Tu et~al.(2024{\natexlab{b}})Tu, White, Kossaifi, Bonev, Kovachki, Pekhimenko, Azizzadenesheli, and Anandkumar]{MS-guaranteed2023}
Renbo Tu, Colin White, Jean Kossaifi, Boris Bonev, Nikola Kovachki, Gennady Pekhimenko, Kamyar Azizzadenesheli, and Anima Anandkumar.
\newblock Guaranteed approximation bounds for mixed-precision neural operators.
\newblock 2024{\natexlab{b}}.
\newblock URL \url{https://arxiv.org/abs/2307.15034}.

\bibitem[Tunstall et~al.(2023)Tunstall, von Werra, Han, Bakhtin, Scialom, Stiennon, Zettlemoyer, and Scao]{tunstall2023zephyr}
Lewis Tunstall, Leandro von Werra, Sheon Han, Anton Bakhtin, Thomas Scialom, Nisan Stiennon, Luke Zettlemoyer, and Teven~Le Scao.
\newblock Zephyr: Direct preference optimization as a reinforcement learning alternative for aligning language models.
\newblock \emph{arXiv preprint arXiv:2309.01375}, 2023.
\newblock URL \url{https://arxiv.org/abs/2309.01375}.

\bibitem[Vaswani et~al.(2017)Vaswani, Shazeer, Parmar, Uszkoreit, Jones, Gomez, Kaiser, and Polosukhin]{vaswani2017attention}
Ashish Vaswani, Noam Shazeer, Niki Parmar, Jakob Uszkoreit, Llion Jones, Aidan~N Gomez, Lukasz Kaiser, and Illia Polosukhin.
\newblock Attention is all you need.
\newblock \emph{Advances in neural information processing systems}, 30, 2017.

\bibitem[Vouitsis et~al.(2024)Vouitsis, Liu, Gorti, Villecroze, Cresswell, Yu, Loaiza-Ganem, and Volkovs]{GPU-MF}
Noël Vouitsis, Zhaoyan Liu, Satya~Krishna Gorti, Valentin Villecroze, Jesse~C. Cresswell, Guangwei Yu, Gabriel Loaiza-Ganem, and Maksims Volkovs.
\newblock Data-efficient multimodal fusion on a single gpu, 2024.
\newblock URL \url{https://arxiv.org/abs/2312.10144}.

\bibitem[Waleffe et~al.(2024)Waleffe, Byeon, Riach, Norick, Korthikanti, Dao, Gu, et~al.]{megatron2024mamba}
Roger Waleffe, Wonmin Byeon, Duncan Riach, Brandon Norick, Vijay Korthikanti, Tri Dao, Albert Gu, et~al.
\newblock An empirical study of mamba-based language models.
\newblock \emph{ArXiv}, cs.LG, 2024.

\bibitem[Wan et~al.(2023)Wan, Wang, Liu, Alam, Zheng, Liu, Qu, Yan, Zhu, Zhang, et~al.]{wan2023efficient}
Zhongwei Wan, Xin Wang, Che Liu, Samiul Alam, Yu~Zheng, Jiachen Liu, Zhongnan Qu, Shen Yan, Yi~Zhu, Quanlu Zhang, et~al.
\newblock Efficient large language models: A survey.
\newblock \emph{arXiv preprint arXiv:2312.03863}, 2023.

\bibitem[Wang et~al.(2023)Wang, Min, Deng, Shen, Wu, Zettlemoyer, and Sun]{wang-2023-CoT}
Boshi Wang, Sewon Min, Xiang Deng, Jiaming Shen, You Wu, Luke Zettlemoyer, and Huan Sun.
\newblock Towards understanding chain-of-thought prompting: An empirical study of what matters.
\newblock In Anna Rogers, Jordan Boyd-Graber, and Naoaki Okazaki, editors, \emph{Proceedings of the 61st Annual Meeting of the Association for Computational Linguistics (Volume 1: Long Papers)}, pages 2717--2739, Toronto, Canada, July 2023. Association for Computational Linguistics.
\newblock \doi{10.18653/v1/2023.acl-long.153}.
\newblock URL \url{https://aclanthology.org/2023.acl-long.153/}.

\bibitem[Wang et~al.(2024{\natexlab{a}})Wang, Wan, Hekmati, Zong, Alam, Zhang, and Krishnamachari]{wang2024iot}
Xin Wang, Zhongwei Wan, Arvin Hekmati, Mingyu Zong, Samiul Alam, Mi~Zhang, and Bhaskar Krishnamachari.
\newblock Iot in the era of generative ai: Vision and challenges.
\newblock \emph{arXiv preprint arXiv:2401.01923}, 2024{\natexlab{a}}.

\bibitem[Wang et~al.(2024{\natexlab{b}})Wang, Zheng, Wan, and Zhang]{wang2024svd}
Xin Wang, Yu~Zheng, Zhongwei Wan, and Mi~Zhang.
\newblock Svd-llm: Truncation-aware singular value decomposition for large language model compression.
\newblock \emph{arXiv preprint arXiv:2403.07378}, 2024{\natexlab{b}}.

\bibitem[Wang and Xiao(2024)]{wang2024loma}
Yumeng Wang and Zhenyang Xiao.
\newblock Loma: Lossless compressed memory attention.
\newblock \emph{arXiv preprint arXiv:2401.09486}, 2024.

\bibitem[Wei et~al.(2024)Wei, Zhu, Zhao, Cheng, Li, Lü, Cheng, Zhang, Zhang, Zeng, Wang, Ma, Hu, Yan, Fang, and Zhou]{wei2024skyworkmoe}
Tianwen Wei, Bo~Zhu, Liang Zhao, Cheng Cheng, Biye Li, Weiwei Lü, Peng Cheng, Jianhao Zhang, Xiaoyu Zhang, Liang Zeng, Xiaokun Wang, Yutuan Ma, Rui Hu, Shuicheng Yan, Han Fang, and Yahui Zhou.
\newblock Skywork-moe: A deep dive into training techniques for mixture-of-experts language models, 2024.
\newblock URL \url{https://arxiv.org/abs/2406.06563}.

\bibitem[Wen et~al.(2024)Wen, Cao, Fu, et~al.]{victor}
Yuxin Wen, Qingqing Cao, Qichen Fu, et~al.
\newblock Efficient vision-language models by summarizing visual tokens into compact registers.
\newblock \emph{arXiv preprint arXiv:2410.14072}, 2024.

\bibitem[Wu et~al.(2024)Wu, Waheed, Zhang, Abdul-Mageed, and Aji]{wu2023lamini}
Minghao Wu, Abdul Waheed, Chiyu Zhang, Muhammad Abdul-Mageed, and Alham~Fikri Aji.
\newblock Lamini-lm: A diverse herd of distilled models from large-scale instructions.
\newblock 2024.
\newblock URL \url{https://arxiv.org/abs/2304.14402}.

\bibitem[Wu et~al.(2023)Wu, Yao, and He]{wu2023zeroquant}
Xiaoxia Wu, Zhewei Yao, and Yuxiong He.
\newblock Zeroquant-fp: A leap forward in llms post-training w4a8 quantization using floating-point formats.
\newblock \emph{arXiv preprint arXiv:2307.09782}, 2023.

\bibitem[Wu et~al.(2025)Wu, Gao, and Wu]{GPU-SD}
Yize Wu, Ke~Gao, and Yanjun Wu.
\newblock Easyspec: Layer-parallel speculative decoding for efficient multi-gpu utilization, 2025.
\newblock URL \url{https://arxiv.org/abs/2502.02493}.

\bibitem[Xi et~al.(2024)Xi, Chen, Zhao, Teh, Chen, and Zhu]{xi2024jetfire}
Haocheng Xi, Yuxiang Chen, Kang Zhao, Kai~Jun Teh, Jianfei Chen, and Jun Zhu.
\newblock Jetfire: Efficient and accurate transformer pretraining with int8 data flow and per-block quantization.
\newblock \emph{arXiv preprint arXiv:2403.12422}, 2024.

\bibitem[Xia et~al.(2023{\natexlab{a}})Xia, Zheng, Li, Zhuang, Zhou, Qiu, Li, Lin, and Song]{P-flashllm}
Haojun Xia, Zhen Zheng, Yuchao Li, Donglin Zhuang, Zhongzhu Zhou, Xiafei Qiu, Yong Li, Wei Lin, and Shuaiwen~Leon Song.
\newblock Flash-llm: Enabling cost-effective and highly-efficient large generative model inference with unstructured sparsity.
\newblock \emph{arXiv preprint arXiv:2309.10285}, 2023{\natexlab{a}}.

\bibitem[Xia et~al.(2023{\natexlab{b}})Xia, Gao, Zeng, and Chen]{P-shearedllama}
Mengzhou Xia, Tianyu Gao, Zhiyuan Zeng, and Danqi Chen.
\newblock Sheared llama: Accelerating language model pre-training via structured pruning.
\newblock \emph{arXiv preprint arXiv:2310.06694}, 2023{\natexlab{b}}.

\bibitem[Xiao et~al.(2024{\natexlab{a}})Xiao, Tian, Chen, Han, and Lewis]{xiao2024}
G.~Xiao, Y.~Tian, B.~Chen, S.~Han, and M.~Lewis.
\newblock Efficient streaming language models with attention sinks.
\newblock In \emph{The Twelfth International Conference on Learning Representations}, 2024{\natexlab{a}}.

\bibitem[Xiao et~al.(2023)Xiao, Lin, Seznec, Wu, Demouth, and Han]{xiao2023smoothquant}
Guangxuan Xiao, Ji~Lin, Mickael Seznec, Hao Wu, Julien Demouth, and Song Han.
\newblock Smoothquant: Accurate and efficient post-training quantization for large language models.
\newblock In \emph{International Conference on Machine Learning}, pages 38087--38099. PMLR, 2023.

\bibitem[Xiao et~al.(2024{\natexlab{b}})Xiao, Tian, Chen, Han, and Lewis]{xiao2024efficientstreaminglanguagemodels}
Guangxuan Xiao, Yuandong Tian, Beidi Chen, Song Han, and Mike Lewis.
\newblock Efficient streaming language models with attention sinks.
\newblock 2024{\natexlab{b}}.
\newblock URL \url{https://arxiv.org/abs/2309.17453}.

\bibitem[Xin et~al.()Xin, Luo, Liu, Du, Zhou, Cheng, Lee, Du, Wang, Chen, et~al.]{citation-0}
Yi~Xin, Siqi Luo, Xuyang Liu, Yuntao Du, Haodi Zhou, Xinyu Cheng, Christina~Luoluo Lee, Junlong Du, Haozhe Wang, MingCai Chen, et~al.
\newblock V-petl bench: A unified visual parameter-efficient transfer learning benchmark.
\newblock In \emph{The Thirty-eight Conference on Neural Information Processing Systems Datasets and Benchmarks Track}.

\bibitem[Xing et~al.(2024)Xing, Huang, Dong, et~al.]{pyramiddrop}
Long Xing, Qidong Huang, Xiaoyi Dong, et~al.
\newblock Pyramiddrop: Accelerating your large vision-language models via pyramid visual redundancy reduction.
\newblock \emph{arXiv preprint arXiv:2410.17247}, 2024.

\bibitem[Xiong et~al.(2024{\natexlab{a}})Xiong, Liu, Huang, Wu, Wu, Mu, Yao, Shen, Wan, Huang, et~al.]{xiong2024autoregressive}
Jing Xiong, Gongye Liu, Lun Huang, Chengyue Wu, Taiqiang Wu, Yao Mu, Yuan Yao, Hui Shen, Zhongwei Wan, Jinfa Huang, et~al.
\newblock Autoregressive models in vision: A survey.
\newblock \emph{arXiv preprint arXiv:2411.05902}, 2024{\natexlab{a}}.

\bibitem[Xiong et~al.(2024{\natexlab{b}})Xiong, Shen, Ye, Tao, Wan, Lu, Wu, Zheng, Guo, Kong, et~al.]{xiong2024uncomp}
Jing Xiong, Jianghan Shen, Fanghua Ye, Chaofan Tao, Zhongwei Wan, Jianqiao Lu, Xun Wu, Chuanyang Zheng, Zhijiang Guo, Lingpeng Kong, et~al.
\newblock Uncomp: Uncertainty-aware long-context compressor for efficient large language model inference.
\newblock \emph{arXiv preprint arXiv:2410.03090}, 2024{\natexlab{b}}.

\bibitem[Xu et~al.(2024)Xu, Shao, Chen, Tang, Zhang, Gao, An, Qiao, and Luo]{P-BESA}
Peng Xu, Wenqi Shao, Mengzhao Chen, Shitao Tang, Kaipeng Zhang, Peng Gao, Fengwei An, Yu~Qiao, and Ping Luo.
\newblock Besa: Pruning large language models with blockwise parameter-efficient sparsity allocation.
\newblock \emph{arXiv preprint arXiv:2402.16880}, 2024.

\bibitem[Xu et~al.(2025)Xu, Bu, Zhang, and Barnett]{MS-MultimodalHessian}
Shiyun Xu, Zhiqi Bu, Yiliang Zhang, and Ian Barnett.
\newblock A hessian-informed hyperparameter optimization for differential learning rate.
\newblock 2025.
\newblock URL \url{https://arxiv.org/abs/2501.06954}.

\bibitem[Yang et~al.(2023)Yang, Zhang, Chen, and Li]{yang2023efficientdm}
Huanrui Yang, Zhenyu Zhang, Yiran Chen, and Hai Li.
\newblock Efficientdm: Robust and efficient quantization for diffusion models.
\newblock \emph{arXiv preprint arXiv:2303.09556}, 2023.
\newblock URL \url{https://arxiv.org/abs/2303.09556}.

\bibitem[Yang et~al.(2024{\natexlab{a}})Yang, Li, Xie, Zhu, Yu, and Li]{MS-MultimodalFusion}
Juncheng Yang, Zuchao Li, Shuai Xie, Weiping Zhu, Wei Yu, and Shijun Li.
\newblock Cross-modal adapter: Parameter-efficient transfer learning approach for vision-language models.
\newblock 2024{\natexlab{a}}.
\newblock URL \url{https://arxiv.org/abs/2404.12588}.

\bibitem[Yang et~al.(2024{\natexlab{b}})Yang, Kim, Bae, Kwon, Park, Yang, Kwon, and Lee]{MS-notokenleft2024}
June~Yong Yang, Byeongwook Kim, Jeongin Bae, Beomseok Kwon, Gunho Park, Eunho Yang, Se~Jung Kwon, and Dongsoo Lee.
\newblock No token left behind: Reliable kv cache compression via importance-aware mixed precision quantization.
\newblock 2024{\natexlab{b}}.
\newblock URL \url{https://arxiv.org/abs/2402.18096}.

\bibitem[Yang et~al.(2024{\natexlab{c}})Yang, Chen, Zhang, Liu, Zhang, Ma, Verma, Zhang, Zhou, King, and Ying]{MS-LowRankFusion}
Menglin Yang, Jialin Chen, Yifei Zhang, Jiahong Liu, Jiasheng Zhang, Qiyao Ma, Harshit Verma, Qianru Zhang, Min Zhou, Irwin King, and Rex Ying.
\newblock Low-rank adaptation for foundation models: A comprehensive review.
\newblock 2024{\natexlab{c}}.
\newblock URL \url{https://arxiv.org/abs/2501.00365}.

\bibitem[Yao et~al.(2024)Yao, Anthony, Shafi, Subramoni, and DK~Panda]{Yao2024exflow}
Jinghan Yao, Quentin Anthony, Aamir Shafi, Hari Subramoni, and Dhabaleswar~K. DK~Panda.
\newblock Exploiting inter-layer expert affinity for accelerating mixture-of-experts model inference.
\newblock In \emph{2024 IEEE International Parallel and Distributed Processing Symposium (IPDPS)}, pages 915--925, 2024.
\newblock \doi{10.1109/IPDPS57955.2024.00086}.

\bibitem[Yao et~al.(2023)Yao, Gholami, Mahoney, and Keutzer]{yao2023quantllm}
Zhewei Yao, Amir Gholami, Michael~W. Mahoney, and Kurt Keutzer.
\newblock Quant-llm: Post-training quantization for large language models.
\newblock \emph{arXiv preprint arXiv:2308.07339}, 2023.
\newblock URL \url{https://arxiv.org/abs/2308.07339}.

\bibitem[Yi et~al.(2023)Yi, Guo, Wei, Zhou, Wang, and Xu]{yi2023edgemoe}
Rongjie Yi, Liwei Guo, Shiyun Wei, Ao~Zhou, Shangguang Wang, and Mengwei Xu.
\newblock Edgemoe: Fast on-device inference of moe-based large language models, 2023.
\newblock URL \url{https://arxiv.org/abs/2308.14352}.

\bibitem[Yuan et~al.(2023)Yuan, Niu, Liu, Liu, Wang, Shang, Sun, Wu, Wu, and Wu]{yuan2023rptq}
Zhihang Yuan, Lin Niu, Jiawei Liu, Wenyu Liu, Xinggang Wang, Yuzhang Shang, Guangyu Sun, Qiang Wu, Jiaxiang Wu, and Bingzhe Wu.
\newblock Rptq: Reorder-based post-training quantization for large language models.
\newblock \emph{arXiv preprint arXiv:2304.01089}, 2023.

\bibitem[Zagoruyko and Komodakis(2017)]{zagoruyko2017paying}
Sergey Zagoruyko and Nikos Komodakis.
\newblock Paying more attention to attention: Improving the performance of convolutional neural networks via attention transfer.
\newblock In \emph{Proceedings of the International Conference on Learning Representations (ICLR)}, 2017.

\bibitem[Zhai et~al.(2023)Zhai, He, Ma, Zong, Zhang, and Zhai]{Zhai2023smartmoe}
Mingshu Zhai, Jiaao He, Zixuan Ma, Zan Zong, Runqing Zhang, and Jidong Zhai.
\newblock {SmartMoE}: Efficiently training {Sparsely-Activated} models through combining offline and online parallelization.
\newblock In \emph{2023 USENIX Annual Technical Conference (USENIX ATC 23)}, pages 961--975, Boston, MA, July 2023. USENIX Association.
\newblock ISBN 978-1-939133-35-9.
\newblock URL \url{https://www.usenix.org/conference/atc23/presentation/zhai}.

\bibitem[Zhang et~al.(2024{\natexlab{a}})Zhang, XiaolongShi, Sun, and Sun]{P-zhang2024structured}
Honghe Zhang, XiaolongShi XiaolongShi, Jingwei Sun, and Guangzhong Sun.
\newblock Structured pruning for large language models using coupled components elimination and minor fine-tuning.
\newblock In \emph{Findings of the Association for Computational Linguistics: NAACL 2024}, pages 1--12, 2024{\natexlab{a}}.

\bibitem[Zhang et~al.(2023{\natexlab{a}})Zhang, Zhang, Shi, Chu, and Li]{zhang2023lora}
Longteng Zhang, Lin Zhang, Shaohuai Shi, Xiaowen Chu, and Bo~Li.
\newblock Lora-fa: Memory-efficient low-rank adaptation for large language models fine-tuning.
\newblock \emph{arXiv preprint arXiv:2308.03303}, 2023{\natexlab{a}}.

\bibitem[Zhang et~al.(2023{\natexlab{b}})Zhang, Chen, Shen, Yang, Ou, Yu, and Zhuang]{E-zhang2023loraprune}
Mingyang Zhang, Hao Chen, Chunhua Shen, Zhen Yang, Linlin Ou, Xinyi Yu, and Bohan Zhuang.
\newblock Loraprune: Pruning meets low-rank parameter-efficient fine-tuning.
\newblock \emph{arXiv preprint arXiv:2305.18403}, 2023{\natexlab{b}}.

\bibitem[Zhang et~al.(2024{\natexlab{b}})Zhang, Chen, Shen, Yang, Ou, Yu, and Zhuang]{P-loraprune}
Mingyang Zhang, Hao Chen, Chunhua Shen, Zhen Yang, Linlin Ou, Xinyi Yu, and Bohan Zhuang.
\newblock Loraprune: Structured pruning meets low-rank parameter-efficient fine-tuning.
\newblock In \emph{Findings of the Association for Computational Linguistics ACL 2024}, pages 3013--3026, 2024{\natexlab{b}}.

\bibitem[Zhang et~al.(2024{\natexlab{c}})Zhang, Liu, Zhao, Cheng, Bao, Zhang, Mitra, and Chen]{P-Dpruner}
Nan Zhang, Yanchi Liu, Xujiang Zhao, Wei Cheng, Runxue Bao, Rui Zhang, Prasenjit Mitra, and Haifeng Chen.
\newblock Pruning as a domain-specific llm extractor.
\newblock \emph{arXiv preprint arXiv:2405.06275}, 2024{\natexlab{c}}.

\bibitem[Zhang et~al.(2024{\natexlab{d}})Zhang, Bai, Lin, Zhao, Hou, and Cannistraci]{P-RIA}
Yingtao Zhang, Haoli Bai, Haokun Lin, Jialin Zhao, Lu~Hou, and Carlo~Vittorio Cannistraci.
\newblock Plug-and-play: An efficient post-training pruning method for large language models.
\newblock In \emph{The Twelfth International Conference on Learning Representations}, 2024{\natexlab{d}}.

\bibitem[Zhang et~al.(2023{\natexlab{c}})Zhang, Sheng, Zhou, Chen, Zheng, Cai, Song, Tian, Ré, Barrett, Wang, and Chen]{H2O2023}
Zhenyu Zhang, Ying Sheng, Tianyi Zhou, Tianlong Chen, Lianmin Zheng, Ruisi Cai, Zhao Song, Yuandong Tian, Christopher Ré, Clark Barrett, Zhangyang Wang, and Beidi Chen.
\newblock H$_2$o: Heavy-hitter oracle for efficient generative inference of large language models.
\newblock 2023{\natexlab{c}}.
\newblock URL \url{https://arxiv.org/abs/2306.14048}.

\bibitem[Zhang et~al.(2023{\natexlab{d}})Zhang, Yang, Chen, and Li]{zhang2023dilatequant}
Zhenyu Zhang, Huanrui Yang, Yiran Chen, and Hai Li.
\newblock Dilatequant: Accurate and efficient diffusion quantization via weight dilation.
\newblock \emph{arXiv preprint arXiv:2306.03881}, 2023{\natexlab{d}}.
\newblock URL \url{https://arxiv.org/abs/2306.03881}.

\bibitem[Zhao et~al.(2021)Zhao, Hua, Shen, Lou, and Jin]{zhao2021automatic}
Changsheng Zhao, Ting Hua, Yilin Shen, Qian Lou, and Hongxia Jin.
\newblock Automatic mixed-precision quantization search of bert.
\newblock \emph{arXiv preprint arXiv:2112.14938}, 2021.

\bibitem[Zhao et~al.(2024{\natexlab{a}})Zhao, Zhao, Drozdov, Rozonoyer, Sultan, Lee, Iyyer, and McCallum]{zhao2024multistage}
Jiachen Zhao, Wenlong Zhao, Andrew Drozdov, Benjamin Rozonoyer, Md~Arafat Sultan, Jay-Yoon Lee, Mohit Iyyer, and Andrew McCallum.
\newblock Multistage collaborative knowledge distillation from a large language model for semi-supervised sequence generation.
\newblock 2024{\natexlab{a}}.
\newblock URL \url{https://arxiv.org/abs/2311.08640}.

\bibitem[Zhao et~al.(2023)Zhao, Lin, Zhu, Ye, Chen, Zheng, Ceze, Krishnamurthy, Chen, and Kasikci]{zhao2023atom}
Yilong Zhao, Chien-Yu Lin, Kan Zhu, Zihao Ye, Lequn Chen, Size Zheng, Luis Ceze, Arvind Krishnamurthy, Tianqi Chen, and Baris Kasikci.
\newblock Atom: Low-bit quantization for efficient and accurate llm serving.
\newblock \emph{arXiv preprint arXiv:2310.19102}, 2023.
\newblock URL \url{https://arxiv.org/abs/2310.19102}.

\bibitem[Zhao et~al.(2024{\natexlab{b}})Zhao, Lin, Zhu, Ye, Chen, Zheng, Ceze, Krishnamurthy, Chen, and Kasikci]{zhao2024atom}
Yilong Zhao, Chien-Yu Lin, Kan Zhu, Zihao Ye, Lequn Chen, Size Zheng, Luis Ceze, Arvind Krishnamurthy, Tianqi Chen, and Baris Kasikci.
\newblock Atom: Low-bit quantization for efficient and accurate llm serving.
\newblock \emph{Proceedings of Machine Learning and Systems}, 6:\penalty0 196--209, 2024{\natexlab{b}}.

\bibitem[Zheng et~al.(2024{\natexlab{a}})Zheng, Liu, Bian, Ma, Zhang, Wang, Guo, and Qin]{zheng2024bidm}
Xingyu Zheng, Xianglong Liu, Yichen Bian, Xudong Ma, Yulun Zhang, Jiakai Wang, Jinyang Guo, and Haotong Qin.
\newblock Bidm: Pushing the limit of quantization for diffusion models.
\newblock \emph{arXiv preprint arXiv:2412.05926}, 2024{\natexlab{a}}.

\bibitem[Zheng et~al.(2024{\natexlab{b}})Zheng, Liu, Bian, Ma, Zhang, Wang, Guo, and Qin]{zheng2024bidmpushinglimitquantization}
Xingyu Zheng, Xianglong Liu, Yichen Bian, Xudong Ma, Yulun Zhang, Jiakai Wang, Jinyang Guo, and Haotong Qin.
\newblock Bidm: Pushing the limit of quantization for diffusion models.
\newblock 2024{\natexlab{b}}.
\newblock URL \url{https://arxiv.org/abs/2412.05926}.

\bibitem[Zhou et~al.(2016)Zhou, Wu, Ni, Zhou, Wen, and Zou]{zhou2016dorefa}
Shuchang Zhou, Yuxin Wu, Zekun Ni, Xinyu Zhou, He~Wen, and Yuheng Zou.
\newblock Dorefa-net: Training low bitwidth convolutional neural networks with low bitwidth gradients.
\newblock \emph{arXiv preprint arXiv:1606.06160}, 2016.
\newblock URL \url{https://arxiv.org/abs/1606.06160}.

\bibitem[Zhou et~al.(2024)Zhou, Ning, Hong, Fu, Xu, Li, Lou, Wang, Yuan, Li, Yan, Dai, Zhang, Dong, and Wang]{e-infer-s}
Zixuan Zhou, Xuefei Ning, Ke~Hong, Tianyu Fu, Jiaming Xu, Shiyao Li, Yuming Lou, Luning Wang, Zhihang Yuan, Xiuhong Li, Shengen Yan, Guohao Dai, Xiao-Ping Zhang, Yuhan Dong, and Yu~Wang.
\newblock A survey on efficient inference for large language models.
\newblock 2024.
\newblock URL \url{https://arxiv.org/abs/2404.14294}.

\end{thebibliography}

\end{document}